\documentclass[12pt]{article}
\makeatletter \@addtoreset{equation}{section} \makeatother
\renewcommand{\theequation}{\thesection.\arabic{equation}}
\newcommand{\ba}{\begin{array}}
\newcommand{\ea}{\end{array}}
\newcommand{\beq}{\begin{equation}}
\newcommand{\eeq}{\end{equation}}
\newcommand{\bea}{\begin{eqnarray}}
\newcommand{\eea}{\end{eqnarray}}

\def\bce{\begin{center}}
\def\ece{\end{center}}

\def\nonu{\nonumber}

\def\pa{\partial}

\def\be{\beta}

\def\la{\lambda}

\def\eps6{{\displaystyle \mathop{\epsilon}^{6}}{}}
\def\g6{{\displaystyle \mathop{g}^{6}}{}}

\def\nab6{{\displaystyle \mathop{\nabla}^{6}}{}}

\def\0{{\sst{(0)}}}
\def\1{{\sst{(1)}}}
\def\2{{\sst{(2)}}}
\def\3{{\sst{(3)}}}
\def\4{{\sst{(4)}}}
\def\5{{\sst{(5)}}}
\def\6{{\sst{(6)}}}
\def\7{{\sst{(7)}}}
\def\8{{\sst{(8)}}}

\def\ba{\begin{array}}
\def\ea{\end{array}}
\def\beq{\begin{equation}}
\def\eeq{\end{equation}}
\def\be{\begin{equation}}
\def\ee{\end{equation}}

\def\la{\lambda}
\def\eps{\epsilon}

\def\ba{\begin{array}}
\def\ea{\end{array}}
\def\beq{\begin{equation}}
\def\eeq{\end{equation}}
\def\be{\begin{equation}}
\def\ee{\end{equation}}

\def\la{\lambda}
\def\eps{\epsilon}

\def\eps6{{\displaystyle \mathop{\epsilon}^{6}}{}}

\def\nab6{{\displaystyle \mathop{\nabla}^{6}}{}}

\newcommand{\tz}{\frac{\theta_{12}}{z_{12}}}
\newcommand{\tzb}{\frac{\bar{\theta}_{12}}{z_{12}}}
\newcommand{\tzzbb}{\frac{\theta_{12} \bar{\theta}_{12}}{z_{12}^2}}
\newcommand{\tzbb}{\frac{\theta_{12} \bar{\theta}_{12}}{z_{12}}}

\newcommand{\bean}{\begin{eqnarray*}}
\newcommand{\eean}{\end{eqnarray*}}

\begin{document}
\thispagestyle{empty} \addtocounter{page}{-1}
   \begin{flushright}
\end{flushright}

\vspace*{1.3cm}
  
\centerline{ \Large \bf   
The Operator Product Expansion of }
\vspace*{0.5cm}
\centerline{ \Large \bf  the Lowest Higher Spin Current 
at Finite $N$ }
\vspace*{1.5cm}
\centerline{{\bf Changhyun Ahn 
}} 
\vspace*{1.0cm} 
\centerline{\it 
Department of Physics, Kyungpook National University, Taegu
702-701, Korea} 
\vspace*{0.8cm} 
\centerline{\tt ahn@knu.ac.kr 
} 
\vskip2cm

\centerline{\bf Abstract}
\vspace*{0.5cm}
 
For the ${\cal N}=2$ Kazama-Suzuki(KS) model on ${\bf CP}^3$,
the lowest higher spin current with spins $(2, \frac{5}{2}, \frac{5}{2},3)$
is obtained from the generalized GKO coset construction.
By computing the operator product expansion of this current and
itself, the next higher spin current with spins $(3, \frac{7}{2},
\frac{7}{2}, 4)$ is also derived.
This is a realization of the ${\cal N}=2$ ${\cal W}_{N+1}$ algebra with
$N=3$ in the
supersymmetric WZW model.
By incorporating the self-coupling constant of lowest higher spin current 
which is known for the
general $(N,k)$,  
we present the complete {\it nonlinear} operator product expansion of the lowest
higher spin current with spins $(2, \frac{5}{2}, \frac{5}{2}, 3)$ in
the ${\cal N}=2$ KS model on ${\bf CP}^N$ space. 
This should coincide with the asymptotic symmetry of the higher spin 
$AdS_3$ supergravity at the quantum level.
The large $(N,k)$ 't Hooft limit and the corresponding classical
nonlinear algebra
are 
also discussed.  
 
\baselineskip=18pt
\newpage
\renewcommand{\theequation}
{\arabic{section}\mbox{.}\arabic{equation}}

\section{Introduction}

There are 
three approaches in the study of extended symmetries in conformal
field theory \cite{BS}.
Initiated by Zamolodchikov \cite{Zamolodchikov}, the approach $1$ is to propose the
number of extra currents with given spins and close the algebra. 
The direct construction of extended conformal algebra has been used
with the help of computer power. 
The associativity for the algebra should be checked.
On the other hand, developed by Fateev and Lukyanov \cite{FL1,FL2,FL3}, the
approach $2$ is to use the Drinfeld-Sokolov reduction based on the
classical Lie algebras(or Lie superalgebras). At the quantum level, 
the construction of the corresponding algebra is based on the
quantization of Miura transformation.      
The currents of the algebra in Miura basis are not primary nor
quasiprimary in general.
Furthermore, 
the approach $3$, found by \cite{BBSS1,BBSS2}, is to study the extended algebra
on the basis of the Casimir construction  for level $1$ WZW models for
simply laced Lie algebras. The difficult step is to identify the
complete set of independent generating currents. 
By construction, the associativity is satisfied automatically. 

The duality between the $W_N$ minimal model conformal field theories and 
the higher spin theory of Vasiliev on the $AdS_3$ 
has been proposed by Gaberdiel and Gopakumar in \cite{GG,GG1,GG2},
who claim that
the $W_N$ minimal model conformal field theory
is dual, in the 't Hooft $\frac{1}{N}$ expansion, 
to  the higher spin theory coupled to one complex scalar.
The ${\cal N}=2$ supersymmetric extension of this duality, 
the higher spin $AdS_3$ supergravity,
has been studied in \cite{CHR,CG,HP} where the dual conformal field theory is given by 
${\cal N}=2$ ${\bf CP}^N$ Kazama-Suzuki(KS) model \cite{KSNPB,KSPLB} in two dimensions.    

Recently,
for $N=4$ in the KS-model on ${\bf CP}^N$ space, 
the two higher spin currents with spins $(2, \frac{5}{2}, \frac{5}{2}, 3)$
and spins $(3,
\frac{7}{2}, \frac{7}{2}, 4)$ in terms of constrained WZW currents 
are constructed and the operator product
expansion of the lowest higher spin current and itself in
${\cal N}=2$ ${\cal W}_{N+1}$ algebra, at the linear level, 
is presented in \cite{Ahn1206}.
By taking the large $(N, k)$ limit on the various operator product 
expansions in components, at the linear order,
it leads to the corresponding operator product expansions in 
${\cal N}=2$ classical algebra describing the asymptotic symmetries of 
bulk theory found in \cite{HP}.

In this paper, 
we reconsider the KS coset model and find out the first nontrivial operator
product expansion
including the nonlinear terms. 
Although the two higher spin currents were found in \cite{Ahn1206},
it is rather complicated to describe the operator product expansions
for the various component fields explicitly. Therefore, 
we consider the $N=3$ case, which is the simplest one, in detail.  
For ${\cal N}=2$ ${\cal W}_{N+1}$ algebra, there exist $N$-multiplets
whose first component spins are $1$(corresponding to the stress energy
tensor), 
$2, \cdots, N$.
The complete algebra for the higher spin currents 
consists of $\frac{1}{2}N(N-1)$ operator product expansions.
Among them, we only compute the simplest operator product expansion between 
the lowest higher spin current and itself.
In components, this is equivalent to $16(=4\times 4)$ operator product expansions.
It turns out that this operator product expansion looks similar to the
one in ${\cal N}=2$ ${\cal W}_3$ algebra in the sense that 
the field contents are the same except the higher spin current with
spins $(3, \frac{7}{2}, \frac{7}{2}, 4)$ and its descendant fields are 
absent. 
Of course, the coefficient functions, i.e., 1) the central charge and
2) the
self-coupling constant of the lowest higher spin current, 
appearing in the right hand side
of operator product
expansion are replaced with their $N$-generalizations respectively.  

In section $2$, based on the construction of higher spin current with
spins $(2, \frac{5}{2}, \frac{5}{2}, 3)$ in the ${\bf CP}^3$ KS coset
model, we describe 
the complete nonlinear operator product expansion of the lowest
higher spin current with spins $(2, \frac{5}{2}, \frac{5}{2}, 3)$ in
the ${\cal N}=2$ KS model on ${\bf CP}^N$ coset space.
The next higher spin current $(3, \frac{7}{2}, \frac{7}{2}, 4)$ occurs
also in the right hand side of above operator product expansion.

In section $3$, we take the large $(N,k)$ limit of  the operator product 
expansion we
have found in section 2.

In section $4$, we compare the algebra we have described in section 2
with the corresponding classical algebra describing the asymptotic 
symmetries in the bulk theory.

In section $5$, we summarize what we have found in this paper and 
comment on some future directions.

In the Appendices $A$-$D$, 
we present some details which are necessary to
sections $2$, $3$, and $4$.

There are some partial relevant works \cite{BHMS}-\cite{Ahn1106} which deal
with the higher spin symmetry in different contexts.
In \cite{BHMS}, the three-dimensional Chern-Simons vector models are 
discussed.
The more general coset models with larger symmetry 
than $W_N$ symmetry are described in \cite{GHKSS}. 
In \cite{Vasiliev}, the higher spin holographic duality is discussed in various 
dimensions.
The states of black hole in the CFT are compared to those 
in the bulk in \cite{GHJ}.   
In \cite{HGPR}, the ${\cal N}=2$ asymptotic symmetry is described.
The nonsupersymmetric holographic dual 
with different group is described in \cite{Ahn2012}.
In \cite{CY}, the further stduy on $W_N$ minimal model is considered. 
The three-point correlation function with arbitrary spin 
is proposed in \cite{AKP}.
In \cite{CGGR}, the light primaries in the $W_N$ minimal model are described
at finite $N$.
The CFT computation for the three-point function with 
spin-$4$ current is described in \cite{Ahn2011}.
In \cite{PR}, further property of correlation functions is given. 
The structure constants of all classical $W_N$ algebra 
are explained in \cite{CFP}. 
In \cite{GV}, the minimal model holography with even spin is proposed.
For fixed 't Hooft coupling constant, the three-point 
function with general spin is proposed in \cite{CY1}. 
In \cite{GGHR}, the more precise duality property is checked by computing the 
partition function. 
The duality in the minimal model holography with orthogonal group is 
described in \cite{Ahn1106}.

In particular, there are some overlaps with the recent work by Candu
and Gaberdiel \cite{CG1}.
       
\section{The ${\cal N}=2$ quantum ${\cal W}_{N+1}$ algebra: the
  operator product expansion of lowest higher spin current }

\subsection{Review}

Let us consider the hermitian symmetric space
\bea
{\bf CP}^N = \frac{SU(N+1)}{SU(N) \times U(1)},
\label{cosetcpn}
\eea
where the structure constant of the coset is inherited from the 
complex structure of the numerator and denominator groups. 
Let $G=SU(N+1)$ be an even-dimensional Lie group and 
$H=SU(N) \times U(1)$ be an even-dimensional
subgroup. The $N$ should be even number.
We introduce a complex basis for the Lie algebra in which the complex
structure is diagonal.
The structure constants in this basis are determined by 
the commutation relations of the matrix generators 
of Appendix $A$.
It is
convenient to decompose the group $G$-indices into the subgroup
$H$-indices 
and the coset $\frac{G}{H}$-indices explicitly. 
The lower case middle roman indices $m, n, p, \cdots $, running from $1$
to $\frac{N^2}{2}$, refer to 
the Lie algebra of $H$ while the lower case top roman indices $a, b, c,
\cdots$, running from $\frac{N^2}{2}+1$ to 
$\frac{1}{2} \left[ (N+1)^2-1 \right]$,
refer to the remaining Lie algebra generators corresponding to the
coset $\frac{G}{H}$. 
We shall denote by $\bar{m}, \bar{n},
\bar{p}, \cdots $ and $\bar{a}, \bar{b}, \bar{c}, \cdots$ the complex
conjugate indices.
For odd $N$, we will describe the corresponding coset model in the
subsection \ref{cp3}.

The ten operator product expansions between the ${\cal N}=2$ 
$SU(N+1)$ WZW
currents  
are given in  \cite{Ahn1206} where both the subgroup index
structure and the remaining index structure are manifest.  

Then  the stress tensor $T(Z)$
for the supersymmetric coset model based on ${\cal N}=2$ ${\bf CP}^N$
model is obtained from \cite{HS}: quadratic and linear terms 
 \footnote{
The $Z$ stands for ${\cal N}=2$ superspace coordinates, one real
bosonic coordinate $z$, and pair of two conjugate Grassman coordinates
$\theta, \bar{\theta}$: $Z=(z, \theta, \bar{\theta})$.
The complex spinor covariant derivatives are 
$
D
=\frac{\partial}{\partial \theta}-\frac{1}{2} \overline {\theta}
\frac{\partial}{\pa z}$, and $ 
\overline{D} =\frac{\partial}{\partial \overline{\theta}}-\frac{1}{2}
\theta \frac{\partial}{\pa z}$. }
\bea
T(Z)
=-\frac{1}{(k+N+1)} \left[J^a J^{\bar{a}}-f_{\bar{m}
      \bar{a}}^{\,\,\,\,\,\,\,\,\bar{a}} 
D K^{\bar{m}}-f_{m \bar{a}}^{\,\,\,\,\,\,\,\, \bar{a}} 
\overline{D} K^{m}
\right](Z),
\label{stress}
\eea
where
$J^a(Z), J^{\bar{a}}(Z)$ live in the 
coset $\frac{G}{H}$ and 
$K^m(Z), K^{\bar{m}}(Z)$ live in the subgroup 
$H$. The positive integer $k$ is the level of WZW currents.
One obtains the standard operator product expansion of ${\cal N}=2$
superconformal algebra,
\bea
T (Z_{1}) \; T (Z_{2})=
\frac{1}{z^{2}_{12}}
\,\,\frac{c}{3}
  +  
\tzzbb \,\,T(Z_2) -\tz \,\,D T(Z_2) +\tzb \,\,\overline{D} T(Z_2) 
+\tzbb \,\, \partial T (Z_2),
\label{ttope}
\eea
where
the central charge is
\bea
c  =  \frac{3 N k}{N+k+1}.
\label{centralcharge}
\eea
The component result of (\ref{ttope}) is given by the Appendix $B$.
The superfield $T(Z)$ has 
no $U(1)$ charge because there is no $\frac{1}{z_{12}}$-term
in (\ref{ttope}). The coefficient of $\frac{\theta_{12}
  \bar{\theta}_{12}}{z_{12}^2}$ implies that the superspin of $T(Z)$
is equal to $1$ or its component spins are given by $(1, \frac{3}{2},
\frac{3}{2}, 2)$.

\subsection{The ${\bf CP}^3$ coset model}
\label{cp3}

Before we are going to discuss the $N=3$ case, let us remind the $N=2$ case where 
the ${\cal N}=2$ ${\cal W}_3$ algebra \cite{Romans,Odake} is realized.
The field contents are the stress tensor $T(Z)$ (\ref{stress}) 
and one higher spin current
$W(Z)$ with spins $(2, \frac{5}{2}, \frac{5}{2},3)$. They can be 
realized by the
constrained WZW currents \cite{Ahn94} based on the approach $3$. 
In ${\cal N}=2$ superspace, the only nontrivial nonlinear operator
product expansion of this algebra is
characterized by the operator product expansion between  $W(Z_1)$ and $W(Z_2)$. 
On the other hand, the coefficient functions on the right hand side are
completely fixed by the Jacobi identities  between these two currents
$T(Z)$ and $W(Z)$ along the line of the
approach $1$ explained in the introduction. The 
two approaches give the same result.  

We will consider the $N=3$ case, see how the algebra looks
different(compared to the $N=2$ case), 
and present the nonlinear algebra for general $N$ in next subsection.
Let us focus on the $N=3$ case where the $SU(4)$ generators in the
complex basis are given
by the Appendix ($A.1$).
The field contents of ${\cal N}=2$ ${\cal W}_4$
algebra are given by the stress tensor with spins $(1,
\frac{3}{2}, \frac{3}{2}, 2)$ and two primary higher spin currents with spins $(2,
\frac{5}{2}, \frac{5}{2}, 3)$, and $
(3, \frac{7}{2}, \frac{7}{2}, 4)$ \cite{BW,Wisskirchen}(based on the
approach $1$). 
Let us write them explicitly as follows \footnote{As in
  \cite{Ahn1206}, for the component fields, instead of introducing the
  different notations for the various fields, we stick to use the
  covariant spinor derivatives to describe the three components for
  given one superfield. As we specify the bosonic coordinate $z$ on
  them, one can tell them from the corresponding superfield. }:
\bea
T(Z) & = & 
T+ \theta \,\, D T|_{\theta=\bar{\theta}=0} + \bar{\theta} \,\, 
\overline{D} T|_{\theta=\bar{\theta}=0}+ 
\theta \bar{\theta}\,\, (-1) \frac{1}{2} [ D, \overline{D} ] 
T|_{\theta=\bar{\theta}=0},
\nonu \\
W(Z) & = & 
W+ \theta \,\, D W|_{\theta=\bar{\theta}=0} + 
\bar{\theta} \,\, 
\overline{D} W|_{\theta=\bar{\theta}=0}+ 
\theta \bar{\theta}\,\, (-1) \frac{1}{2} [ D, \overline{D} ] 
W|_{\theta=\bar{\theta}=0},
\nonu \\
V(Z) & = & 
V+ \theta \,\, D V|_{\theta=\bar{\theta}=0} + \bar{\theta} \,\, 
\overline{D} V|_{\theta=\bar{\theta}=0}+ 
\theta \bar{\theta}\,\, (-1) \frac{1}{2} [ D, \overline{D} ] 
V|_{\theta=\bar{\theta}=0}.
\label{twv}
\eea
The field $V(Z)$ satisfies the defining equation for the primary field
condition.
The coset can be described as ${\bf CP}^3$ by
introducing the extra $U(1)$'s in both the numerator and the denominator
in order to have even-dimensional groups
$G$ and $H$  as in (\ref{cosetcpn})
\bea
{\bf CP}^3 = \frac{SU(4) \times U(1)}{SU(3) \times U(1) \times U(1)}.
\label{cosetcpn1}
\eea 
In the Appendix $A$, we present the generators for each group factors.
The corresponding ${\cal N}=2$
current algebra with $SU(4)$ group in the complex basis can be
obtained by inserting the structure constants
and the metric. 
Then it is an immediate task to find the higher spin currents in terms of the fundamental
currents defined in the supersymmetric WZW model(in the context of the
approach $3$).
The stress tensor is given by (\ref{stress}).
It is nontrivial to find the extra higher spin currents in 
the Casimir
construction that includes the higher spin generators. 
There exist $16$ independent fundamental WZW currents, $K^m(Z),
K^{\bar{m}}(Z)$ 
and $J^a(Z)$, $J^{\bar{a}}(Z)$ where $m, \bar{m} =1,2,3,4,5$ and $a,
\bar{a} =6,7,8$.
These $16$ supercurrents are in fact constrained, e.g. the components
$D K^m(z)$ and $-\frac{1}{2} [ D, \overline{D}] K^m(z)$ 
of the fundamental WZW supercurrent $K^m(Z)$ are determined by the other $2$
components $K^m(z)$ and $\overline{D} K^m(z)$. The higher spin currents 
in the Casimir construction must be constructed out of the independent 
fundamental currents.

One way to write down the lowest higher spin
current $W(Z)$ with spins $(2, \frac{5}{2}, \frac{5}{2}, 3)$ is
to take into account of all the possible terms for given superspin $2$
in the WZW currents.   
Finally, it turns out, from the equation of $(2.24)$ in \cite{Ahn1206},  
that the correct higher spin current with spins
$(2, \frac{5}{2}, \frac{5}{2}, 3)$,
takes the form \footnote{All the composite fields are normal ordered
  from right to left, along the line of \cite{BBSS1,BBSS2}.}
\bea
W(Z) &=& 
\frac{A(k)}{(-8+7 k+3 k^2)} \left[  \, 
(-8+7 k+3 k^2)
f_{\bar{c} a}^{\;\;\;\;\bar{m}} f_{c\bar{b}}^{\;\;\;\;n} 
J^a J^{\bar{b}} K^m K^{\bar{n}} \right.
\nonu \\
&- & 2 (4+k) (-1+2 k) 
\, f_{\bar{c} a}^{\;\;\;\;n} f_{c\bar{b}}^{\;\;\;\;\bar{m}} 
 J^a J^{\bar{b}} K^m K^{\bar{n}}  \nonu \\
& + & 
\frac{1}{2} (-56+6 k+9 k^2+k^3) 
\, J^a J^b J^{\bar{a}} J^{\bar{b}} 
 + 
\frac{1}{2}(4+k) (-16+21 k+7 k^2)
 \,
f_{\bar{m} a}^{\;\;\;\;b}  J^{a} J^{\bar{b}} D K^{\bar{m}}
\nonu \\
& - & \frac{1}{2} k (28+11 k+k^2) 
( f_{a\bar{m}}^{\;\;\;\;b} J^a K^{\bar{m}} D J^{\bar{b}} 
- 
f_{m a}^{\;\;\;\;b} K^m \overline{D} J^{a} J^{\bar{b}}) 
\nonu \\
& + & \frac{1}{2} (-64+68 k+49 k^2+7 k^3) 
\, f_{m a}^{\;\;\;\;b}  J^{a} J^{\bar{b}} \overline{D} K^{m}
+ \frac{1}{6} k (4+k) (7+k)
\, f_{m n}^{\;\;\;\;p}  \overline{D} K^{m} K^n K^{\bar{p}}
\nonu \\
&- &  k (4+k)^2 (7+k)  \, \overline{D} J^a D J^{\bar{a}} 
+  2 (4+k)^2 (-1+2 k)   \,  \overline{D} K^m D
K^{\bar{m}}
\nonu \\
& + & \frac{1}{4} (-256-112 k+48 k^2+35 k^3+5 k^4)
 \, ( J^a \pa J^{\bar{a}}  - \pa J^a J^{\bar{a}})
\nonu \\ 
& - &  
(4+k)^2 (-1+2 k)
\, K^m \pa K^{\bar{m}} 
-\frac{1}{12} (192+160 k+24 k^2+3 k^3+k^4)
\, \pa K^m K^{\bar{m}}
\nonu \\
& -&  \frac{1}{4} k (4+k)^2 (7+k) 
( J^a [D, \overline{D}] J^{\bar{a}} 
+ [D, \overline{D}] J^a J^{\bar{a}})
\nonu \\
& + & (4+k)^2 (-1+2 k) 
\, K^m  [D, \overline{D}] K^{\bar{m}} 
  -\frac{1}{12} (4+k)^2 (12-17 k+k^2)   
\, [D, \overline{D}] K^m K^{\bar{m}}
\nonu \\
& - & 2 (4+k)^2 (-1+2 k)
 \, D K^m \overline{D} K^{\bar{m}}
 + 
\frac{1}{6} (4+k)^2 (6-5 k+k^2) 
\, f_{m\bar{n}}^{\;\;\;\;\bar{n}} \pa \overline{D} K^{m}
\nonu \\
& + &  (4+k)^2 (-1+2 k)
\, f_{\bar{m} \bar{n}}^{\;\;\;\;\bar{n}} \pa D K^{\bar{m}}
 - 
\frac{1}{3} (-8+k) (4+k)^2 
\, f_{\bar{m}\bar{a}}^{\;\;\;\;\bar{a}}
  f_{n \bar{b}}^{\;\;\;\;\bar{b}} D K^{\bar{m}} \overline{D} K^{n}
\nonu \\ 
& + &  
(-3+k) (4+k)^2 
(f_{\bar{m} \bar{b}}^{\;\;\;\;\bar{b}} J^a J^{\bar{a}} D
K^{\bar{m}}
+ f_{m \bar{b}}^{\;\;\;\;\bar{b}} J^a J^{\bar{a}} \overline{D}
K^{m})
\nonu \\
& + & 
\frac{1}{6} (4+k)^2 (7+k) 
( f_{\bar{m}\bar{a}}^{\;\;\;\;\bar{a}}
  f_{\bar{n}\bar{b}}^{\;\;\;\;\bar{b}} D K^{\bar{m}} D K^{\bar{n}} + 
 f_{m\bar{a}}^{\;\;\;\;\bar{a}}
  f_{n \bar{b}}^{\;\;\;\;\bar{b}} \overline{D} K^{m}
  \overline{D} K^{n}
)
\nonu \\
 & + &  \left.
\frac{1}{2} k (4+k)^2 (7+k)
\,
f_{\bar{m} a}^{\;\;\;\;a } \pa D K^{\bar{m}} \right](Z),
\label{superspin2}
\eea
where the overall constant is fixed as 
\bea
A(k)^2  & = & \frac{3 \left(-8+7 k+3 k^2\right)^2}{4 (-1+k) (2+k) (4+k)^6
  (7+k) (-1+2 k)} \nonu \\ 
& = & 
-\frac{(-9+c)^6 \left(-54+33 c+c^2\right)^2}{1088391168 (-21+c) (-1+c) (9+c) (-9+5 c)}.
\label{Ak}
\eea
We will see how we obtain this result  in a moment.
The central charge, from (\ref{centralcharge}),  is
\bea
c_{N=3} =\frac{9k}{k+4}.
\label{centralN3}
\eea
The superfield $W(Z)$ 
consists of quartic, cubic, quadratic and linear terms.
One realizes that the factor $(4+k)$ in (\ref{superspin2}) reflects the 
more general
$N$-dependent factor $(N+k+1)$. We have seen the factor $(3+k)$ in the
${\bf CP}^2$ coset model and the factor $(5+k)$ in the ${\bf CP}^4$
coset model \cite{Ahn1206}.  
If one considers the higher spin currents with spins $(2, \frac{5}{2},
\frac{5}{2}, 3)$ for the ${\bf CP}^N$ coset model, then the factor
$(N+k+1)$ should arise in various places for the composite fields.

The two conditions we use above, in order to determine the various
undetermined coefficients in $(2.24)$ of \cite{Ahn1206}, 
are the regularity condition
\bea
K^m(Z_1) \; W(Z_2) =0, \qquad K^{\bar{m}}(Z_1) \; W(Z_2)=0,
\label{KWvanishing}
\eea
and the primary field condition 
\bea
T (Z_{1}) \; W (Z_{2})=
 \tzzbb \,\, 2 W(Z_2) -\tz \,\, D W(Z_2) +\tzb \,\, \overline{D}
 W(Z_2) +\tzbb \,\, \partial 
W (Z_2).
\label{TW}
\eea
The component results for (\ref{TW}) are given by the Appendix $B$.
Notice that $W(Z)$ is uncharged.
All the $c$-dependent(or $k$-dependent) coefficient functions are completely fixed by
these two conditions except the overall constant $A(k)$.

Now we are ready to compute the operator product expansion $W(Z_1)
W(Z_2)$ from the realization of (\ref{superspin2}). We will use the
component approach by starting with the operator product expansion
between the lower spin components of $W(Z)$.

\subsubsection{ The operator product expansion $W(z)\; W(w)$}

Let us compute the spin-$2$ spin-$2$ 
operator product expansion $W(Z_1) \; W(Z_2)$
explicitly.
As explained in \cite{Ahn1206}, due to the limitation of ${\cal N}=2$
package \cite{KT}, one resorts to the original package \cite{Thielemans}. 
From the components (\ref{twv}), first of all, one can start with the operator
product expansion $W(z) \; W(w)$: the first component of $W(Z_1)$ and
the first component of $W(Z_2)$.  
The first component of $W(Z)$ is obtained by taking $\theta,
\bar{\theta}$ in (\ref{superspin2}) to zero.

$\bullet$ $\frac{1}{(z-w)^4}$-term

Let us consider the operator product expansion of
$W(z) \; W(w)$ for the highest singular term $\frac{1}{(z-w)^4}$. 
It turns out, by taking the explicit operator product expansion between
these two fields, that one obtains
$
\frac{6 (-1+k) k (2+k) (4+k)^5 (7+k) (-1+2 k)  }{\left(-8+7 k+3 k^2\right)^2}
A(k)^2$.
Normalizing it to be $\frac{c}{2}$ where the
central charge $c$ is given by (\ref{centralN3}), 
one sees the normalization factor (\ref{Ak}). 

$\bullet$ $\frac{1}{(z-w)^3}$-term

Let us move on the next singular term $\frac{1}{(z-w)^3}$. 
Such a term is not allowed by the symmetry of the operator product expansion 
$W(z)\; W(w)$.
Suppose that there is
such a singular term. After reversing the arguments
$z$ and $w$ and expanding around $w$, 
the same term will appear with an opposite sign.
This
implies that it is identically zero. 

$\bullet$ $\frac{1}{(z-w)^2}$-term

The next singular term $\frac{1}{(z-w)^2}$ contains many different
combinations of spin-$2$ fields in terms of WZW currents.    
The possible spin-$2$ fields(some of them are primary
and others are not primary) as $W(w), [D,
\overline{D}]T(w), T^2(w)$ and $\pa T(w)$. However, the last one is not
allowed because  it is a descendant of $T(w)$ 
and the latter did not appear in the more singular terms.
Then it is straightforward to rewrite this singular term in terms of
above three spin-$2$ fields by introducing three undetermined coefficients
which depend on the level $k$. 
The easiest way to determine the relative coefficients is to compute
the singular term using the representation (\ref{superspin2}) and equate 
the result to the ansatz above. 
Then the unknown three coefficients are
completely fixed. 
In particular, the coefficient of $W(w)$ turns out to be
$
\frac{4 (-3+k) (1+k) (4+k)^3 }{-8+7 k+3 k^2} A(k)
$ by explicit computations. 
In our notation, this is equal to $2\alpha$ and with
(\ref{Ak}), one gets the self-coupling constant, the coefficient
function appearing in the $W(w)$ term of the  right hand side in the
$W(z)\; W(w)$,  
\bea
\alpha_{N=3}^2 =\frac{3(-3+k)^2(1+k)^2}{(-1+k)(2+k)(7+k)(-1+2k)} =
-\frac{(27-7 c)^2 (3+c)^2}{2 (-21+c) (-1+c) (9+c) (-9+5 c)},
\label{alphan3}
\eea
where the level $k$ is replaced by the central charge $c$
(\ref{centralN3}) 
in the last expression. We will see that this self-coupling constant
is consistent with the general expression with arbitrary $N$
that can be obtained from the Jacobi identity \cite{CG1}.

$\bullet$ $\frac{1}{(z-w)}$-term

Let us describe the final lowest singular term $\frac{1}{(z-w)}$. 
The possible spin-$3$ terms are characterized by 
\bea
&& T^3(w), \,\, T W(w), \,\, V(w), \,\, [D, \overline{D}] W(w), \,\,T[D, \overline{D}] T(w),
\nonu \\
&& \overline{D} T D T(w), \,\, \pa^2 T(w), \,\, \pa W(w), \,\, \pa [D, \overline{D}]
T(w), \,\, \mbox{and}  \,\, \pa T T (w).
\label{spin3terms}
\eea 
For the first six independent terms, due to
the symmetry described in the $\frac{1}{(z-w)^3}$ term, we do not see
them in this singular term, and we do not see the $\pa^2 T(w)$-term
because there is no $T(w)$-term in the right hand side as above.    
Then, we are left with the last three independent terms.
They are exactly the corresponding descendant terms for the spin-$2$
fields $W(w)$, $[D, \overline{D}]T(w)$, and $T T (w)$.  
It turns out, by explicit computations, 
that all the coefficients are given by the half of the
coefficients given in the singular term $\frac{1}{(z-w)^2}$. Also note that 
$\pa T^2(w) = 2 \pa T T(w) = 2 T \pa T(w)$ with the operator product
expansion
$T(z)\; T(w)$ in the Appendix $B$.
The relative coefficient $\frac{1}{2}$ behavior is a consequence
of conformal invariance \cite{Bowcock,BS}.

Therefore, the operator
product expansion $W(z) W(w)$ is summarized in ($C.1$). 
Basically, this is the same as the one in $N=2$ case \cite{Ahn1206}.
Of course, the central charge and the self-coupling constant in this
case are different and given by (\ref{centralN3}) and (\ref{alphan3}).

\subsubsection{The operator product expansion $W(z) \; \overline{D} W(w)$}

Now let us 
consider the other operator product expansion.
In order to compute the spin-$2$ spin-$\frac{5}{2}$ operator product expansion 
$W(z) \; \overline{D} W(w)$, one should know the explicit form for the 
primary spin-$\frac{5}{2}$ field $\overline{D} W(w)$.
This can be read off from the operator product expansion $\overline{D}
T (z) \; W(w)$ given in ($B.2$). 
By looking at the singular term $\frac{1}{(z-w)}$,  it turns out the
spin-$\frac{5}{2}$ field $\overline{D} W(w)$ 
consists of $189$ WZW currents which will not
be written down here.

For given $ \overline{D} W(w)$(or one can obtain this in closed form
by acting $\overline{D}$ on the field $W(Z)$ (\ref{superspin2}) and putting $\theta,
\bar{\theta}$ to zero), by explicit computation of 
$W(z) \; \overline{D} W (w)$,
one can easily identify the highest singular term $\frac{1}{(z-w)^3}$ 
with $-3 D T(w)$.
The second order singular term can contain all spin-$\frac{5}{2}$ fields,
$ \overline{D} W(w)$, $\pa \overline{D} T(w)$, $T \overline{D}
T(w)$ and its conjugate fields. 
Because $\overline{D} W$ has charge $-1$, 
only the first three spin-$\frac{5}{2}$ fields appear on the right hand
side.
As before, we can write down an ansatz for this singular term in terms of
these spin-$\frac{5}{2}$ fields and solve it by comparing with the singular
term computed with the GKO approach.
The final first-order singular term can be written in terms of
following spin-$\frac{7}{2}$ fields:
\bea
&& \pa \overline{D} W(w), \,\,
\pa^2 \overline{D} T(w), \,\, T \overline{D} W(w), \,\, T T \overline{D}
T(w),\,\,
\overline{D} T W(w), \nonu \\
&& \overline{D} T [D, \overline{D}] T(w), 
\,\, \pa \overline{D} T T(w), \,\, \pa T \overline{D} T(w) \,\, \mbox{and}
\,\, \overline{D} V(w).
\label{spin7halffield}
\eea
Compared to the $N=2$ case, there exists an extra term $\overline{D} V(w)$ which
is the third component of superfield $V(Z_2)$ with spins $(3,
\frac{7}{2}, \frac{7}{2}, 4)$ (\ref{twv}). We will come to this issue after the
discussion of next operator product expansion where one realizes the
presence of $V(w)$, the first component of $V(Z_2)$.
Therefore, the remaining nonzero terms after subtracting the eight
terms with correct coefficient functions in this singular term are
really the third component of primary superfield $V(Z_2)$.
The final result is summarized by ($C.2$).

\subsubsection{The operator product expansion $W(z) \; D W(w)$}

Similarly, one can analyze the spin-$2$ spin-$\frac{5}{2}$ 
operator product expansion 
$W(z) \; D W(w)$. 
The explicit form for $D W(w)$ can be obtained from the operator product expansion $D
T(z) \; W(w)$ in ($B.2$) or by acting $D$ on the $W(Z)$ and
putting $\theta, \bar{\theta}$ to zero.
Since we know the form for $D W(w)$ explicitly, one can compute the
operator product expansion.
In this case, the spin-$\frac{7}{2}$ field $ D V(w)$
arises at the lowest singular term 
as well as the conjugated fields of (\ref{spin7halffield}).
This is also presented in ($C.2$) with explicit
coefficient functions.

\subsubsection{ The operator product expansion
$W(z) \; (-1) \frac{1}{2} [D, \overline{D}] W(w)$}

Now let us find out the explicit form for the spin-$3$ field $-\frac{1}{2} [D,
\overline{D}] W(z)$ in order to compute the spin-$2$ spin-$3$ operator product expansion
$W(z) \; (-1) \frac{1}{2} [D, \overline{D}] W(w)$. 
From the operator product expansion 
$D T(z) \; \overline{D} W(w)$ in ($B.2$), one can read off this
primary spin-$3$ field.
The singular term $\frac{1}{(z-w)}$ contains this spin-$3$ field as
well as $\frac{1}{2} \pa W(w)$. 
Then it is easy to obtain the spin-$3$ field which consists of $1065$
independent WZW currents(Also one can obtain this field from the 
superfield $W(Z)$
(\ref{superspin2}) with the derivatives $D$ and $\overline{D}$).

Let us consider the operator product expansion $W(z) \; (-1) \frac{1}{2}
[D, \overline{D}] W(w)$. See also the Appendix ($C.3$).
The fourth-order singular term has primary spin-$1$ field $T(w)$ and the coefficient
is given by $3$. There is no third-order singular term.
The second-order singular term contains the spin-$3$ fields. 
Among $10$ possible terms we described before in (\ref{spin3terms}), 
there are no $\pa T
T(w)$ and $\pa  W(w)$. This is consistent with the
fact that the spin-$2$ fields $T^2(w)$ and $W(w)$ do not occur in the
third-order singular term.  
After collecting $7$-independent terms with appropriate coefficient
functions, we are left with a nonzero spin-$3$ field which is denoted
by $3V(w)$ where
\bea
 V(w) & = & \frac{24 (-1+2 k) (11+2 k) \left(32-12 k-11 k^2+6 k^3\right)}
{(-1+k) (2+k) (4+k)^3 (-4+5 k) (8+5 k) (-16+11 k)} J^6 J^7 J^8
J^{\bar{6}} J^{\bar{7}} J^{\bar{8}}(w)
\nonu \\
& + & \frac{9(-1+2k)}{(-1+k)(2+k)(4+k)^4(7+k)} \left[ K^{2} K^{3} K^{4}
  K^{\bar{2}} K^{\bar{3}} K^{\bar{4}} - K^1 K^2 K^4 K^{\bar{1}}
  K^{\bar{2}} K^{\bar{4}} \right](w) \nonu \\
& + & \mbox{other 1282 terms},
\label{Vexpression}
\eea
where the abbreviated terms contain fifth-order, fourth-order, $\cdots$,
second-order 
and first-order terms in constrained WZW currents.  
Of course, it would be interesting to write down (\ref{Vexpression})
using the group theory structure constants as in (\ref{superspin2}) \footnote{
From the expressions (\ref{stress}), and (\ref{superspin2}), 
one can write down the above spin-$3$ fields (\ref{spin3terms}) except
$V(w)$ itself
in terms of the structure constants and WZW currents.
Now we introduce the arbitrary coefficients inside of each spin-$3$
field.
Then one can check whether the known spin-$3$ field
(\ref{Vexpression}) can be written in terms of the combinations of
spin-$3$ fields in (\ref{spin3terms}) or not. In other words, 
are the undetermined coefficients uniquely fixed? 
Eventhough there will be extra terms, it will not be so difficult to write them 
using the summation index structure.}. 
The overall scale factor of spin-$3$ field can be fixed
after computing the operator product expansion $V(z) V(w)$ which will
not be done in this paper.
For the first-order singular term, one should consider the possible
spin-$4$ fields. One expects that the descendant fields for the above
eight spin-$3$ fields(i.e., the derivative terms) appear. Moreover, the two
fields
$\overline{D} T D W(w)$ and $DT \overline{D} W(w)$ also occur. 

Since the spin-$3$ field $V(w)$ is found, one can compute the operator
product expansion $D T(z) \; V(w)$ and the first-order singular term
should be equal to $-D V(w)$ from the primary field condition in ($B.2$). 
One checks that the first-order singular term coincides with the
remaining terms described in the first-order singular term of the
operator product expansion $W(z) \; D W(w)$ as above. This confirms that
we have the correct normalization of components relative to each other.    
Similarly,  the operator
product expansion $\overline{D} T(z)\; V(w)$ provides the correct
expression for $\overline{D} V(w)$.

\subsubsection{ The operator product expansion $D W(z) \; \overline{D} W(w)$}

What about the spin-$4$ field $(-1) \frac{1}{2} [D, \overline{D}] V(z)$?
One can compute the covariant derivatives of a superfield only if
one knows all of its components. Or it can be determined as in the previous
analysis for  $(-1) \frac{1}{2} [D, \overline{D}] W(z)$.
From the operator product expansion 
$D T(z) \; \overline{D} V(w)$ in ($B.2$), one can read off this
primary spin-$4$ field.
The singular term $\frac{1}{(z-w)}$ contains this spin-$4$ field as
well as $\frac{1}{2} \pa V(w)$.   
This spin-$4$ field 
occurs in the operator product expansion  $D W(z) \; \overline{D} W(w)$. 
For example, the second component of $W(Z_1)$ and the third component
of $W(Z_2)$. The relevant singular term is given by the first-order
singular term.
One can exhaust all the possible uncharged spin-$4$ fields as follows:
\bea
&& \pa^2 W(w), \,\, \pa^2 [D, \overline{D}] T(w),  \,\, T [D, \overline{D}] W(w),
\,\,  T T [D, \overline{D}] T(w), \,\, T \overline{D} T D T(w),
\nonu \\
&& \overline{D} T D W(w), \,\,  D T \overline{D} W(w), \,\,
\pa 
\overline{D} T D T(w), \,\,   \pa D T 
\overline{D} T(w), \,\, 
[D, \overline{D}] T W(w), \nonu
\\
&&  [D, \overline{D}] T [D, 
\overline{D}] T(w), \,\,    \pa T \pa T(w), \,\, \pa^2 T T(w),
\,\, [D, \overline{D}] V(w), 
\,\, \pa [D, \overline{D}] W(w), \nonu \\
&& \pa^3 T(w), \,\,   \pa T  W(w), \,\, T \pa W(w), \,\,
\pa [D, \overline{D}] T  T(w), \,\, 
\pa T [D, \overline{D}] T(w), \nonu \\
&&   \pa T T T(w), \,\, \pa V(w), \,\, T^4(w), \,\, T T W(w), \,\, T
V(w), \,\, W^2(w).
\label{26terms}
\eea

Among these $26$-terms (\ref{26terms}), 
it turns out that the coefficients for the
last five terms are vanishing \footnote{ In the package
  \cite{Thielemans}, sometimes it takes too much time to compute the
first-order pole terms by using simply the command
``$\mbox{OPESimplify}[\mbox{OPEPole}[1][\mbox{ope}],\mbox{Factor}]$''.
Instead, one defines $\mbox{result} \equiv \mbox{OPEPole}[1][\mbox{ope}];$ 
and $L \equiv \mbox{GetOperators}[\mbox{result}];$ 
and then computes
``$\mbox{Sum}[\mbox{Simplify}[\mbox{Coefficients}[\mbox{result},L[[i]]]]*L[[i]],
\{i,1,\mbox{Length}[L]\}]$''. In other words, one gets the raw
expression by ``result'', gets the independent terms by $L$, 
simplifies the coefficient functions appearing in the list of $L$, and
then sums over the product between the independent fields and the
coefficient functions. This leads to the final
fist-order pole terms with simplified coefficient functions. }.
It is rather complicated to fix all the coefficients explicitly using the 
GKO representation and make 
sure that 
the remaining terms after subtracting everybody besides
$ [D, \overline{D}] V$ 
can be interpreted as the ${\bf CP}^4$
coset field of spin-$4$. 
In other words, this new field is primary field of spin
$4$ under
the stress tensor. Moreover, the fact that this is coset field implies
that one should have the regularity conditions $ K^m(z) \; [D,
\overline{D} ] V(w)=0$, $\overline{D} K^m(z)  \; [D,
\overline{D} ] V(w)=0$,
$K^{\bar{m}}(z) \; [D, \overline{D}] V(w)=0$, and $ D K^{\bar{m}}(z)
\; [D,
\overline{D} ] V(w)=0$. 
We have checked these identities explicitly.   

So far, we have checked the following operator product expansions,
\bea
W(z) W(w), \,\, W(z) D W(w), \,\, W(z) \overline{D} W(w), \,\, 
W(z) (-1) \frac{1}{2} [D, \overline{D}] W(w), \,\, D W(z) \overline{D} W(w).
\label{doneope}
\eea
One can proceed further with the remaining operator product
expansions. However, it is rather complicated procedure to check all
the nontrivial operator product expansions.
Instead, by resorting to ${\cal N}=2$ supersymmetry, one can reexpress 
without any ambiguity the 
above operator product expansions (\ref{doneope}) using ${\cal N}=2$ superspace
formalism.
Now one can write one single operator product expansion $W(Z_1)
\;W(Z_2)$ explicitly which will be presented in next subsection. 
Then the unchecked operator product expansions can be read off from
this and they will appear in the Appendix $C$ completely.

In summary, one concludes, compared to the $N=2$ case \cite{Ahn94,Ahn93},  
that the operator product expansion of the
lowest higher spin current with spins $(2,\frac{5}{2}, \frac{5}{2}, 3)$ in
${\cal N}=2$ KS coset model on ${\bf CP}^3$ (\ref{cosetcpn1}) has the next higher spin
current
with spins $(3, \frac{7}{2}, \frac{7}{2}, 4)$
(\ref{Vexpression}). 
One can rewrite all the structure constants in terms of $c$ and the 
self-coupling constant (\ref{alphan3}). 
All dependence on $N$ is contained in these 
two parameters, provided the higher spin field $(3,\frac{7}{2},\frac{7}{2},4)$
is normalized as in equation (\ref{openolimit}).

\subsection{The ${\bf CP}^N$ coset model}

So far, we have considered the $N=2$ case in \cite{Ahn94}, the $N=3$
case in this paper, and the $N=4$ case in \cite{Ahn1206}.
As we increase the $N (> 4)$, we expect to have the similar features.  

1) The self-coupling constant, that appears in the $W(w)$-dependent terms, 
changes according to $N$ but its expression is known for general $N$ \cite{BW}, 

2) the higher spin currents, whose normalization should be fixed, occur and

3) the $c$-dependent coefficient functions in terms of $c$, that appear in
$W(w)$-independent and -dependent terms in the right hand side, do not change. 
Of course, the central charge itself depends on $N$.

For example, for $N=4$ the self-coupling
constant is known and 
one could take the same $c$-dependent coefficient functions
as in the $N=3$ case. 

What about the presence of an extra higher spin current?
If we denote the higher spin current $X(Z)$ 
with spins $(4, \frac{9}{2},\frac{9}{2},5)$ which is one of the field
contents in ${\cal N}=2$ ${\cal W}_5$ algebra, 
by dimensional analysis, the spin-$4$ field $X(Z_2)$ can appear in the
singular term $
\frac{\theta_{12}\bar{\theta}_{12}}{z_{12}}$ but it seems that this is
not the case. Actually, this is direct consequence of the 
Jacobi identities \cite{CG1}. 
Eventually, this higher spin current will appear in other operator
product expansions between the higher spin currents whose spins are
greater than the spin of lowest higher spin current.

The final operator product expansion of the superspin $2$ current and
itself which is the main result of this paper, 
from the component results of (\ref{doneope}), 
can be written as
\bea
&& W(Z_1) \; W(Z_2)  =  \frac{1}{z_{12}^4} \,\, \frac{c}{2} 
 +\frac{\theta_{12}
  \bar{\theta}_{12}}{z_{12}^4}
\,\, 3 T(Z_2) +\frac{\bar{\theta}_{12}}{z_{12}^3} \,\, 3 \overline{D} T(Z_2)
-\frac{\theta_{12}}{z_{12}^3} \,\, 3 D T(Z_2) +\frac{\theta_{12} 
\bar{\theta}_{12}}{z_{12}^3} \,\, 3 \pa T(Z_2)
\nonu \\
&& + \frac{1}{z_{12}^2} \left[ 2 \alpha \, W - \frac{c}{(-1+c)} \,\,
[D,
  \overline{D}] T -\frac{3}{(-1+c)} \,\, T^2 \right](Z_2) \nonu \\
&& +\frac{\bar{\theta}_{12}}{z_{12}^2}
 \left[ \alpha \, \overline{D} W +\frac{(-3+2c)}{(-1+c)} 
  \,\, \pa \overline{D} T -\frac{3}{(-1+c)} \,\, T \overline{D} T \right](Z_2)
\nonu \\
&& +  \frac{\theta_{12}}{z_{12}^2}
 \left[ \alpha \, D W -\frac{(-3+2c)}{(-1+c)} 
  \,\, \pa D T -\frac{3}{(-1+c)} \,\, T D T \right](Z_2) 
\nonu \\
&& + 
\frac{\theta_{12} \bar{\theta}_{12}}{z_{12}^2} \left[ 
- \frac{3(-8+c)}{2(-12+5c)} \,\, \alpha \, [D, \overline{D} ] W  -   
\frac{9c(-12+5c)}{4(-1+c)(6+c)(-3+2c)} \,\, \pa [D, \overline{D}] T \right.
\nonu
\\
&&  +
\frac{3(18-15c+2c^2+2c^3)}{2(-1+c)(6+c)(-3+2c)} \,\, \pa^2 T + 3 V
 + \frac{42}{(-12+5c)} \,\, \alpha \, T W
\nonu \\
&& -\frac{3(36-9c+8c^2)}{2(-1+c)(6+c)(-3+2c)} \,\,
T [D, \overline{D}] T -\frac{9(3+4c)}{(-1+c)(6+c)(-3+2c)} \,\, T^3
\nonu \\
&& \left. - \frac{9c(-12+5c)}{(-1+c)(6+c)(-3+2c)} 
\,\, \overline{D} T D T \right](Z_2)
\nonu \\
&& +  \frac{1}{z_{12}} \left[  
\alpha \, \pa  W -\frac{c}{2(-1+c)} \,\, \pa [D,
  \overline{D}] T -\frac{3}{(-1+c)} \,\, \pa T T \right](Z_2) \nonu \\
&& + 
\frac{\bar{\theta}_{12}}{z_{12}}
 \left[ \frac{3(-6+c)(-1+c)}{(3+c)(-12+5c)} \,\, \alpha \, \pa \overline{D} W +
\frac{3c(9+3c+2c^2)}{4(-1+c)(6+c)(-3+2c)} \,\, 
  \pa^2 \overline{D} T   + \overline{D} V \right. \nonu \\
&&  - \frac{6(-15+c)}{(3+c)(-12+5c)} \,\, \alpha \, T \overline{D} W  
-\frac{9(3+4c)}{(-1+c)(6+c)(-3+2c)} \,\, T T \overline{D} T 
\nonu \\
&& + \frac{54(-1+c)}{(3+c)(-12+5c)} \,\, \alpha \, \overline{D} T W
-\frac{27c}{2(6+c)(-3+2c)} \,\, \overline{D} T [D, \overline{D}] T 
\nonu \\
&& \left. -\frac{3(-18+24c+c^2)}{(-1+c)(6+c)(-3+2c)} \,\, \pa 
\overline{D} T T -\frac{9(-6+c)}{2(6+c)(-3+2c)} \,\, \pa T \overline{D} 
T \right](Z_2)
\nonu \\
&& + 
\frac{\theta_{12}}{z_{12}}
 \left[ \frac{3(-6+c)(-1+c)}{(3+c)(-12+5c)} \,\, \alpha \, \pa D W -
\frac{3c(9-3c+c^2)}{2(-1+c)(6+c)(-3+2c)} 
  \,\, \pa^2 D T - D V \right. \nonu \\
&&  +\frac{6(-15+c)}{(3+c)(-12+5c)} \,\, \alpha \, T D W 
 +\frac{9(3+4c)}{(-1+c)(6+c)(-3+2c)} \,\, T T D T  \nonu \\
&&  +
\frac{27c}{2(6+c)(-3+2c)} \,\, [D, \overline{D}] T D T
-\frac{54(-1+c)}{(3+c)(-12+5c)} \,\, \alpha \, D T W  \nonu \\
&& \left. -\frac{3(-18+24c+c^2)}{(-1+c)(6+c)(-3+2c)} \,\, \pa D T T 
-\frac{9(-6+c)}{2(6+c)(-3+2c)} \,\, \pa T D T  \right](Z_2)
\nonu \\
&& + 
\frac{\theta_{12}\bar{\theta}_{12}}{z_{12}}
 \left[- \frac{(-15+c)c}{(3+c)(-12+5c)} \,\, \alpha \, \pa [D,
   \overline{D}] W  +\frac{(-18-3c-2c^2+2c^3)}{2(-1+c)(6+c)(-3+2c)} 
  \,\, \pa^3 T + 2 \pa V \right. \nonu \\
&& +
\frac{18(6+c)}{(3+c)(-12+5c)} \,\, \alpha \, T \pa W  
+ \frac{12(3+4c)}{(3+c)(-12+5c)} \,\, \alpha \, \pa T W
\nonu \\
&& \left.  + \frac{6}{(3+c)} \,\, \alpha \, \overline{D} T D W 
-\frac{6c(-12+5c)}{(-1+c)(6+c)(-3+2c)} \,\, \pa \overline{D} T D T
\right. \nonu \\
&&  -\frac{6(9-3c+c^2)}{(-1+c)(6+c)(-3+2c)} \,\, \pa [D, \overline{D}] T T
 + \frac{6}{(3+c)} \,\, \alpha \, D T \overline{D} W  \nonu \\
&& +\frac{6c(-12+5c)}{(-1+c)(6+c)(-3+2c)} \,\, \pa D T \overline{D} T
-\frac{3c(3+4c)}{(-1+c)(6+c)(-3+2c)} \,\, \pa T [D, \overline{D}] T 
\nonu \\
&& \left.
-\frac{18(3+4c)}{(-1+c)(6+c)(-3+2c)} \,\, \pa T T T  \right](Z_2) +\cdots.
\label{openolimit}
\eea
Here \footnote{The coupling constant $c_{22,3}$ in front of $V(Z_2)$ in 
(\ref{openolimit}) was found in \cite{CG1}. 
From the equations $(2.18)$, $(2.19)$, $(2.20)$ and $(3.30)$ in \cite{CG1},
one can write down the square of this coupling constant as 
$c_{22,3}^2 =\frac{3 (3+c) (-12+5 c)}{(6+c) (-3+2 c)}+
\frac{6 (-15+c) (-1+c) \alpha^2}{(3+c) (-12+5 c)}$ with (\ref{selfcoupling}). 
This implies that the correct normalization for this field should  
be $\hat{V}(Z_2)$ with $c_{22,3} \hat{V}(Z_2) = 3 V(Z_2)$.
We thank the referee for pointing out this.}
the central charge and self-coupling constant are  given by
\bea
 c & = &   \frac{3 N k}{ N +k +1}, \label{twoconsts} \\
 \alpha^2  & = &  
\frac{3 (1+k)^2 (k-N)^2 (1+N)^2}{(-1+k) (-1+N) (1+2 k+N) (1+k+2 N)
  (-1-N+k (-1+3 N))}
\nonu \\
& = & \frac{(3+c)^2 
\left(c+2 c N-3 N^2\right)^2}{(-1+c) (3-c+6 N) (-1+N) (c+3 N) (-3 N+c (2+N))}.
\label{selfcoupling}
\eea
The $N$- and $k$-dependences in the right hand side occur through
these two values. 
The full quantum operator product expansion is characterized by two
free parameters, the central charge and the self-coupling constant.
For $N=2$ case, one easily sees that the above operator product
expansion leads to the previous results in \cite{Romans} by putting
the spin-$3$ field $V(w)$(and its descendant fields) and
spin-$\frac{5}{2}$ fields $D V(w)$ and $\overline{D} V(w)$ to vanish. 
For fixed $N$, the central charge and the self-coupling constant
depend on only the level $k$.
For $N=3$ case, as we explained before, the general expression for
(\ref{openolimit})
reduces to the findings in previous subsection. One also sees that 
$\alpha$ becomes the one in (\ref{alphan3}) when $N=3$.  
In the $\frac{\theta_{12} \bar{\theta}_{12}}{z_{12}}$-singular term,
we do not see the possible terms with spins $(4, \frac{9}{2},
\frac{9}{2}, 5)$ like as $T^4(Z_2)$, $T T W(Z_2)$, $W^2(Z_2)$ or $T
V(Z_2)$ \footnote{
We have noticed this fact in previous section in the context of
component results.
These terms appear in the different singular term of the operator product expansion of 
$W(Z_1) \; T W(Z_2)$. This implies that 
one should modify the fourth component of the higher spin current with 
spins $(2, \frac{5}{2}, \frac{5}{2}, 3)$ by adding the first component
of the higher spin current $T W(Z)$ 
with spins $(3, \frac{7}{2},\frac{7}{2}, 4)$. 
We will see this feature in detail soon.
Moreover, the possible $10$ terms, among (\ref{26terms}),  
with same spins, coming from $T(Z_2)$
and $W(Z_2)$ with the appropriate derivatives, 
are not present in the term $\frac{\theta_{12}
 \bar{\theta}_{12}}{z_{12}}$ 
of (\ref{openolimit}).}.

Note that the $29$ nonlinear terms in (\ref{openolimit}) which appear
in the last entries of each singular term  are the new we observe 
and the remaining $26$
linear terms are found in \cite{Ahn1206}.
We present the $16$ component operator product expansions in the
Appendix $C$.

In order to compare the bulk theory computations, one should 
also consider the higher spin currents in non-primary basis. 
By subtracting(or excising) the spin-$1$ current, one can construct the following
spin-$2$ current \cite{Romans} by adding the quadratic piece in $T(z)$:
\bea
-\frac{1}{2} \widetilde{[D, \overline{D}] T}(z)
=-\frac{1}{2} [D, \overline{D}] T(z) -\frac{3}{2c} T^2(z).
\label{mod}
\eea
The corresponding central charge is given by $(c-1)$ and the operator
product expansion satisfies
\bea
(-1) \frac{1}{2} \widetilde{[D, \overline{D}] T}(z) \;
(-1) \frac{1}{2} \widetilde{[D, \overline{D}] T}(w)
& = & \frac{1}{(z-w)^4} \,\, \frac{(-1+c)}{2} +\frac{1}{(z-w)^2}\,\,
 2 (-1) \frac{1}{2} \widetilde{[D, \overline{D}] T}(w) 
\nonu \\
&+ &   \frac{1}{(z-w)} \,\, (-1) \frac{1}{2} \pa 
\widetilde{[D, \overline{D}] T}(w) +\cdots,
\label{opestress}
\eea
corresponding to $(3.21a)$ of \cite{HP}.
Then there are no any singular terms with spin-$1$ current
\bea
(-1) \frac{1}{2} \widetilde{[D, \overline{D}] T}(z)
\; T(w) =0.
\label{opetstress}
\eea
This corresponds to $(3.21b)$ of \cite{HP}.
This comes from the coefficient $\frac{3}{2c}$ in the second term of (\ref{mod}).
Furthermore, the previous spin-$\frac{3}{2}$ currents in primary basis
are not primary
fields, due to the extra term in (\ref{mod}), 
but the corresponding operator product expansions contain the
nonlinear terms 
\bea
&& (-1) \frac{1}{2} \widetilde{[D, \overline{D}] T}(z)
\; (D T \pm \overline{D} T)(w) 
=\frac{1}{(z-w)^2} \left[\frac{3}{2}-\frac{3}{2c}\right] \left[ D T \pm
\overline{D} T \right](w)
\nonu \\
 && +\frac{1}{(z-w)} \left[\pa (D T \pm \overline{D} T) -
\frac{3}{c} T ( D T \mp \overline{D} T) \right](w) +\cdots.
\label{spin3half}
\eea
Even in the $c \rightarrow \infty$ limit, the nonlinear terms 
$\frac{3}{c} T ( D T \mp \overline{D} T)$ arise while $c$-dependent
linear term vanishes \cite{BW1,AKS} where the only 
$c$-independent factor can survive in the linear field term
and the $\frac{1}{c}$-dependent factor can survive in the quadratic field term. 
Strictly speaking, the currents 
$D T(w)$ and $\overline{D} T(w)$ are {\it not} primary fields of
spin $\frac{3}{2}$ under
the stress energy tensor (\ref{mod}) even at the linear order for the
finite $c$. They become primary fields $c \rightarrow \infty$ limit
at the linear order.
One realizes that the corresponding bulk
expressions are given by $(3.21d)$ and $(3.21e)$ of \cite{HP}.
One also checks the other relations from the operator product
expansions ($B.1$). In particular, as soon as the spin-$2$ term 
$[D, \overline{D}] T(w)$ in the right hand side of the operator
product expansion appears, one should rewrite it using the relation (\ref{mod}). 
This implies that the operator product expansions in the
spin-$\frac{3}{2}$ and spin-$\frac{3}{2}$ contain the nonlinear term
$T^2(w)$
as in $(3.21h)$ and $(3.21j)$ of \cite{HP} \footnote{For the
  $(3.21c)$ of \cite{HP}, 
one has $T(z) \; T(w)$ of the Appendix $C$. For the
  $(3.21f)$ and $(3.21g)$ of \cite{HP}, 
one has $T(z) \; (D T  \mp \overline{D} T)(w)$ which gives 
$(D T \pm \overline{D} T)(w)$ in the right hand side of the operator
product expansion. For $(3.21i)$ of \cite{HP}, one has $(D T
-\overline{D} T)(z) \; (D T + \overline{D} T)(w)$ of the Appendix $C$. }. 
Therefore, the four component fields $T(z)$, $D T(z)$, $\overline{D}
T(z)$,
and $-\frac{1}{2} \widetilde{[D, \overline{D}] T}(z)$ of
${\cal N}=2$ superconformal algebra, with an appropriate classical
limit, coincide with those quantities in the bulk theory \cite{HP}.  
Their operator product expansions with stress energy tensor are given
by (\ref{opestress}), (\ref{opetstress}), and (\ref{spin3half}).

What about the higher spin currents? 
Due to this modification of stress energy tensor, one should also 
add the extra term in the spin-$3$ current, in order to preserve the
primary field condition,  as follows:
\bea
-\frac{1}{2} \widetilde{[D, \overline{D}] W}(z)
=
-\frac{1}{2} [D, \overline{D}] W(z) -\frac{6}{c} T W(z).
\label{twoccur}
\eea
One can easily check the standard primary field condition with spin
$3$, 
\bea
(-1) \frac{1}{2} \widetilde{[D, \overline{D}] T}(z) \;
(-1) \frac{1}{2} \widetilde{[D, \overline{D}] W}(w)
& = & \frac{1}{(z-w)^2} 3 (-1) \frac{1}{2} \widetilde{[D, \overline{D}] W}(w) 
\nonu \\
& + & \frac{1}{(z-w)}  (-1) \frac{1}{2} \pa \widetilde{[D, \overline{D}] W}(w)
+\cdots,
\label{pripri}
\eea
corresponding to $(3.26a)$ of \cite{HP}.
For the 
spin-$2$ current $W(w)$, the primary field condition does not
change because the extra term $T^2(z)$ in (\ref{mod}) does not produce
any extra singular structure, i.e., $T(z) \; W(w) =0$,  
\bea
&& (-1) \frac{1}{2} \widetilde{[D, \overline{D}] T}(z) \; W(w) 
=\frac{1}{(z-w)^2} 2 W(w) +\frac{1}{(z-w)} \pa W(w) +\cdots.
\label{wpri}
\eea
For the spin-$\frac{5}{2}$ currents, one sees the similar behavior as
in (\ref{spin3half})
\bea
&& (-1) \frac{1}{2} \widetilde{[D, \overline{D}] T}(z)
\;(D W \pm \overline{D} W)(w) 
=\frac{1}{(z-w)^2} \left[\frac{5}{2}-\frac{3}{2c}\right] \left[ D W \pm
\overline{D} W \right](w)
\nonu \\
 && +\frac{1}{(z-w)} \left[\pa (D W \pm \overline{D} W) -
\frac{3}{c} T ( D W \mp \overline{D} W) \right](w) +\cdots.
\label{spin5half}
\eea
In this case, the nonlinear terms are $\frac{3}{c} T ( D W \mp
\overline{D} W)$ which do not vanish in the $c \rightarrow \infty$ limit.
This corresponds to the equation $(3.26b)$ of \cite{HP}.
As before, these spin-$\frac{5}{2}$ fields are not primary field.
One can easily see that there is no singular term in the operator
product expansion between $T(z)$ and the field (\ref{twoccur}) which corresponds
to $(3.26c)$ of \cite{HP}.
For the operator product expansions between $ (D T \pm \overline{D}
T)(z)$ with the current (\ref{twoccur}) in the Appendix $B$, 
one has the nonlinear terms from the equations($(3.26f)$ and
$(3.26j)$). 
Similarly, from the
operator product expansions between $(D W \pm \overline{D} W)$ and 
$(D T \pm \overline{D} T)$ in the Appendix $B$, 
the nonlinear terms in the right hand side
are the same as the ones($(3.26h)$ and $(3.26m)$) in \cite{HP}
\footnote{For the linear terms, the relations $(3.26d)$ and $(3.26e)$
  in \cite{HP} correspond to the operator
  product expansions $T(z) \; (D W \pm \overline{D} W)(w)$ in the Appendix
$B$.  The equations $(3.26g)$ and $(3.26i)$ correspond to the operator
  product expansions $(D T- \overline{D} T)(z) \; W (w)$ and 
$(D T -\overline{D} T)(z) \; (D W +\overline{D} W)(w)$. 
Furthermore,  the equations $(3.26k)$ and $(3.26l)$ correspond to the operator
  product expansions $(D T + \overline{D} T)(z) \; W(w)$ and 
 $(D T + \overline{D} T)(z) \; (D W -\overline{D} W)(w)$ in the Appendix
$B$. }.
Therefore,  
the four component fields $W(z)$, $D W(z)$, $\overline{D}
W(z)$,
and $-\frac{1}{2} \widetilde{[D, \overline{D}] W}(z)$ of
${\cal N}=2$ lowest higher spin current 
coincide with those quantities in the bulk theory \cite{HP} after the
$c \rightarrow \infty$. Their operator product expansions with stress
energy tensor are given by (\ref{pripri}), (\ref{wpri}), and (\ref{spin5half}).

For the comparison with the classical 
asymptotic symmetry algebra  in  the bulk theory, one should consider
the modified two quantities (\ref{mod}) and (\ref{twoccur}). In other
words, the old quantities for the spin-$2$ and spin-$3$ currents have
the extra terms.
Then one expects that the composite operators containing the fields 
$T^2(z)$ and $T W(z)$ and the various derivative terms(coming from the
normal ordering procedure) occur in the right
hand side. 
In ${\cal N}=2$ superspace, one should also consider the operator
product expansions $T(Z_1) \;T W(Z_2)$, $ T^2(Z_1) \;T W(Z_2)$, 
$T^2(Z_1) \;W(Z_2)$, and $W(Z_1) \;T W(Z_2)$.
We will come to this issue in next sections.

\section{The large $(N,k)$ 't Hooft limit of ${\cal N}=2$ quantum ${\cal
    W}_{N+1}$ algebra}

The large $(N,k)$ limit with fixed 't Hooft coupling constant 
is characterized by \cite{GG}
\bea
c(N,k) = \frac{3N k}{N+k+1} \longrightarrow 3(1-\la) N, \qquad \la \equiv \frac{N}{N+k}.
\label{limit}
\eea
For this limit, the self coupling constant behaves as follows \cite{Ahn1206}:
\bea
\alpha(N,k)^2 \longrightarrow 
-\frac{(-1+2 \lambda )^2}{(-2+\lambda ) (1+\lambda )}.
\label{alphalimit}
\eea
One can think of other limit \cite{GG1} where for fixed $N$, one takes
the large $c$ limit. 
Then, from the finite $(N,c)$ expression for the self-coupling
constant (\ref{selfcoupling}), one obtains 
$\alpha(N,k)^2 \rightarrow -\frac{(2N+1)^2}{(N-1)(N+2)}$. 

Then it is easy to see that  under the large $(N,k)$ limit
the operator product expansion (\ref{openolimit}) becomes the
following classical limit
\bea
&& W(Z_1) \; W(Z_2)  =  \frac{1}{z_{12}^4} \,\, \frac{c}{2} 
 +\frac{\theta_{12}
  \bar{\theta}_{12}}{z_{12}^4}
\,\, 3 T(Z_2) +\frac{\bar{\theta}_{12}}{z_{12}^3} \,\, 3 \overline{D} T(Z_2)
\nonu \\
&& -\frac{\theta_{12}}{z_{12}^3} \,\, 3 D T(Z_2) +\frac{\theta_{12} 
\bar{\theta}_{12}}{z_{12}^3} \,\, 3 \pa T(Z_2)
 + \frac{1}{z_{12}^2} \left[ 2 \alpha \, W -  \,\,
[D,
  \overline{D}] T -\frac{3}{c} \,\, T^2 \right](Z_2) \nonu \\
&& +\frac{\bar{\theta}_{12}}{z_{12}^2}
 \left[ \alpha \, \overline{D} W + 2 
  \,\, \pa \overline{D} T -\frac{3}{c} \,\, T \overline{D} T \right](Z_2)
 +  \frac{\theta_{12}}{z_{12}^2}
 \left[ \alpha \, D W -2 
  \,\, \pa D T -\frac{3}{c} \,\, T D T \right](Z_2) 
\nonu \\
&& + 
\frac{\theta_{12} \bar{\theta}_{12}}{z_{12}^2} \left[ 
- \frac{3}{10} \,\, \alpha \, [D, \overline{D} ] W 
  +
\frac{3}{2} \,\, \pa^2 T + 3 V
 + \frac{42}{5c} \,\, \alpha \, T W
-\frac{6}{c} \,\,
T [D, \overline{D}] T \right.
\nonu \\
&& \left. -\frac{18}{c^2} \,\, T^3
 - \frac{45}{2c} 
\,\, \overline{D} T D T \right](Z_2)
 +  \frac{1}{z_{12}} \left[  
\alpha \, \pa  W -\frac{1}{2} \,\, \pa [D,
  \overline{D}] T -\frac{3}{c} \,\, \pa T T \right](Z_2) \nonu \\
&& + 
\frac{\bar{\theta}_{12}}{z_{12}}
 \left[ \frac{3}{5} \,\, \alpha \, \pa \overline{D} W +
\frac{3}{4} \,\, 
  \pa^2 \overline{D} T   + \overline{D} V   
- \frac{6}{5c} \,\, \alpha \, T \overline{D} W  
-\frac{18}{c^2} \,\, T T \overline{D} T 
 + \frac{54}{5c} \,\, \alpha \, \overline{D} T W
\right. \nonu \\
&& \left. -\frac{27}{4c} \,\, \overline{D} T [D, \overline{D}] T 
 -\frac{3}{2c} \,\, \pa 
\overline{D} T T -\frac{9}{4c} \,\, \pa T \overline{D} 
T \right](Z_2)
\nonu \\
&& + 
\frac{\theta_{12}}{z_{12}}
 \left[ \frac{3}{5} \,\, \alpha \, \pa D W -
\frac{3}{4} 
  \,\, \pa^2 D T - D V   +\frac{6}{5c} \,\, \alpha \, T D W 
 +\frac{18}{c^2} \,\, T T D T  
\right. \nonu \\
&& \left. + \frac{27}{4c} \,\, [D, \overline{D}] T D T
-\frac{54}{5c} \,\, \alpha \, D T W   -\frac{3}{2c} \,\, \pa D T T 
-\frac{9}{4c} \,\, \pa T D T  \right](Z_2)
\nonu \\
&& + 
\frac{\theta_{12}\bar{\theta}_{12}}{z_{12}}
 \left[- \frac{1}{5} \,\, \alpha \, \pa [D,
   \overline{D}] W  +\frac{1}{2} 
  \,\, \pa^3 T + 2 \pa V  +
\frac{18}{5c} \,\, \alpha \, T \pa W  
+ \frac{48}{5c} \,\, \alpha \, \pa T W \right.
\nonu \\
&&   + \frac{6}{c} \,\, \alpha \, \overline{D} T D W 
-\frac{15}{c} \,\, \pa \overline{D} T D T
 -\frac{3}{c} \,\, \pa [D, \overline{D}] T T
 + \frac{6}{c} \,\, \alpha \, D T \overline{D} W  \nonu \\
&& \left. +\frac{15}{c} \,\, \pa D T \overline{D} T
-\frac{6}{c} \,\, \pa T [D, \overline{D}] T 
-\frac{36}{c^2} \,\, \pa T T T  \right](Z_2) +\cdots.
\label{opelimit}
\eea
The term $\pa [D, \overline{D}] T(w)$ in the $\frac{\theta_{12}
\bar{\theta}_{12}}{z_{12}^2}$ term goes away \cite{BW1,AKS,AIKS}. See also the previous
work \cite{AKS} for the reduction of classical algebra from the
quantum algebra where the precise limiting procedure is given. 
It is known that there exists an ${\cal N}=2$ supersymmetric version
of classical $W_3$ algebra in ${\cal N}=2$ superspace \cite{IK}
corresponding to the component approach \cite{LPRetal}.

One sees the relative coefficients appearing in
$\frac{\theta_{12}}{z_{12}}$ term, 
$\frac{27}{4c}$, $\frac{3}{2c}$, and $\frac{9}{4c}$ 
are consistent with those in \cite{IK} where the corresponding
coefficients are given by $\frac{36}{c_{IK}}$,
$\frac{8}{c_{IK}}$, and $\frac{12}{c_{IK}}$ in spin-$\frac{7}{2}$ field.
One can find that the relative coefficients 
$\frac{6}{5c}$ and $\frac{54}{5c}$ correspond to 
the values $\frac{8}{c_{IK}}$ and $\frac{72}{c_{IK}}$.
We do not have to worry about 
the ordering between the fields in the classical expression. 
For example, $D T W(Z_2)$ in the above singular term 
is the same as $W D T(Z_2)$.
One can interchange any fields in the classical algebra
(\ref{opelimit}).
For the fermionic fields, there exists minus sign between the
interchanging of any two fermionic fields. 
For example, $\overline{D} T D T (Z_2) = - D T \overline{D} T (Z_2)$.
In the package \cite{Thielemans}, this can be done by using
``SetOPEOptions[OPEMethod, ClassicalOPEs]''. 
Moreover,
the relative coefficients appearing in
$\frac{\theta_{12} \bar{\theta}_{12}}{z_{12}^2}$ term, 
$\frac{6}{c}$, and $\frac{45}{2c}$ 
correspond to the coefficients $\frac{32}{c_{IK}}$,
and $\frac{120}{c_{IK}}$ in spin-$3$ field. 

Of course, the component approach 
results are read off from the Appendix $C$
with an appropriate limit for the central charge.
One can easily check that the above result (\ref{opelimit}) contains 
the operator product expansion \cite{IK} with an appropriate normalization.

\section{ The ${\cal N}=2$ classical ${\cal W}_{\infty}^{\rm{cl}}[\la]$ algebra
in the bulk theory }

In this section, we would like to see the operator product expansions
appearing in the asymptotic symmetry of the higher spin $AdS_3$
supergravity theory from our findings in section $2$. 
In the Appendix $C$, we have our component
operator product expansions completely. 
As we observed, the spin-$3$ field is modified via (\ref{twoccur}).
Then from primary basis to nonprimary basis, one should recompute the
operator product expansions containing this spin-$3$ field. 
Furthermore, one should change the old fields appearing in the right
hand side into the new fields, with the defining equations (\ref{mod})
and (\ref{twoccur}).

Let us consider the three operator product expansions containing the
spin-$2$ current. Other remaining operator product expansions will be
given in the Appendix $D$.

In this section, we take the large $(N,k)$ 't Hooft limit with 
(\ref{limit}) and 
(\ref{alphalimit})
on the algebra we have found in previous section or in the Appendix $C$. 

\subsection{The operator product expansion $W(z) \; W(w)$}

For example, the spin-$2$ and spin-$2$
operator product expansion  can be written as
\bea
&& W(z) \; W(w)  =  
\frac{1}{(z-w)^4} \,\, \frac{c}{2}  +\frac{1}{(z-w)^2} \left[
  2 \alpha\, W -\frac{c}{(-1+c)} \,\, \widetilde{[D, \overline{D}] T}
   \right](w)   \nonu \\
&& +  \frac{1}{(z-w)} \left[ 
\alpha \, \pa W -\frac{c}{2(-1+c)} \,\, \pa \widetilde{[D, \overline{D}] T}
\right](w) 
 +   \cdots
\label{quantumww} \\
&& \nonu \\
&&  \longrightarrow 
\frac{1}{(z-w)^4} \,\, \frac{c}{2}  +\frac{1}{(z-w)^2} \left[
  2 \alpha\, W - \widetilde{[D, \overline{D}] T}
   \right](w)   \nonu \\
&& +  \frac{1}{(z-w)} \left[ 
\alpha \, \pa W -\frac{1}{2} \,\, \pa \widetilde{[D, \overline{D}] T}
\right](w) 
 +   \cdots,
\label{classicalww}
\eea
where the old stress energy tensor is replaced by the new stress
energy tensor (\ref{mod}).
In the classical $c \rightarrow \infty$ limit, one finds the coincidence with
the equation $(3.27)$ of \cite{HP}.
Note that the central charge $c$ depends on $(N,k)$ and the
self-coupling constant $\alpha$ depends on $(N,k)$ also. 
At the quantum level, one should use (\ref{quantumww}) rather than (\ref{classicalww}).
The difference between (\ref{quantumww}) and (\ref{classicalww}) is
the fact that the two parameters 
$c$ and $\alpha$ are replaced by their limiting values (\ref{limit})
and (\ref{alphalimit}) respectively.
The field contents do not change.
In general, one expects that there exist some fields in the quantum
operator product expansion which will disappear in the classical limit.
Note that the relative coefficient $\frac{1}{2}$ between the
second-order and first-order singular terms(i.e., the latter can be
written as $\frac{1}{2}$ times the total derivative of the former) is well-known
numerical factor which can be observed from the conformal invariance.

\subsection{The operator product expansion $W(z) \; ( D W \pm
  \overline{D}  W) (w)$}

Let us consider the spin-$2$ and spin-$\frac{5}{2}$ operator product
expansions. Again, from the operator product expansion ($C.2$),
one obtains
\bea
&& W(z) \; (D W \pm \overline{D} W)(w)  =  
\frac{1}{(z-w)^3} 3 \left[ D T \mp \overline{D} T \right](w)
\nonu \\
&&  +\frac{1}{(z-w)^2} \left[
\alpha \, (D W \pm \overline{D} W) +  \frac{c}{(-1+c)} 
\,\, 
\pa (D T \mp \overline{D} T) -\frac{3}{(-1+c)} \,\, T ( D T \pm  \overline{D} T ) \right](w) 
\nonu \\
&& + \frac{1}{(z-w)} \left[ ( D V \mp \overline{D} V) +
\frac{2(-27+12c+c^2)}{(3+c)(-12+5c)} \,\,
  \alpha \,
\pa (D W  \pm \overline{D} W) \right. \nonu \\
&&  \mp \frac{(-3+c)(3+c)(-9+2c)}{4(-1+c)(6+c)(-3+2c)} \,\, 
\pa^2 \overline{D} T
- \frac{6(-15+c)}{(3+c)(-12+5c)} \,\, \alpha \, T ( D W \mp  \overline{D} W) 
\nonu \\
&&  +\frac{9(-15+c)}{2(-1+c)(6+c)(-3+2c)} \,\, T T (D T \mp  \overline{D}
  T) +\frac{54(-1+c)}{(3+c)(-12+5c)} \,\, \alpha \, (D T  \mp \overline{D}
  T) W  \nonu \\
&&  - \frac{27c}{2(6+c)(-3+2c)} \,\, ( \widetilde{[D, \overline{D}]T} D T  \mp
\overline{D} T \widetilde{[D, \overline{D}] T })
  +\frac{(-15+c)(-3+c)c}{2(-1+c)(6+c)(-3+2c)} \,\, \pa^2 D T \nonu \\
&&  -\frac{3(-15+c)c}{(-1+c)(6+c)(-3+2c)} \,\, (\pa D T T
  + \frac{1}{2} \pa T D T)  \nonu \\
&& \left. \mp 
\frac{3(-54+39c+c^2)}{2(-1+c)(6+c)(-3+2c)}  \,\, \pa T  \overline{D} T  
\mp \frac{3(-27+12c+c^2)}{(-1+c)(6+c)(-3+2c)} \,\, \pa \overline{D} T T  \right] (w)
+  \cdots
\nonu \\
&& \nonu \\
&& \longrightarrow 
\frac{1}{(z-w)^3} 3 \left[ D T \mp \overline{D} T \right](w)
\nonu \\
&&  +\frac{1}{(z-w)^2} \left[
\alpha \, (D W \pm \overline{D} W) +   
\pa (D T \mp \overline{D} T) -\frac{3}{c} \,\, T ( D T \pm  \overline{D} T ) \right](w) 
\nonu \\
&& + \frac{1}{(z-w)} \left[ ( D V \mp \overline{D} V) +
\frac{2}{5} \,\,
  \alpha \,
\pa (D W  \pm \overline{D} W) 
- \frac{6}{5c} \,\, \alpha \, T ( D W \mp  \overline{D} W) 
\right. \nonu \\
&&  -\frac{9}{4c^2} \,\, T T (D T \mp  \overline{D}
  T) +\frac{54}{5c} \,\, \alpha \, (D T  \mp \overline{D}
  T) W   - \frac{27}{4c} \,\,  \widetilde{[D, \overline{D}]T} ( D T
  \mp \overline{D} T)
  \nonu \\
&& \left. +\frac{1}{4} \,\, \pa^2 ( D T \mp \overline{D} T ) 
 -\frac{3}{2c} \,\, (\pa D T  
  \pm \pa  \overline{D} T) T   - \frac{3}{4c}  \,\, \pa T ( D T \pm
  \overline{D} T) 
\right] (w)
+  \cdots.
\label{class1}
\eea
The $14$ nonlinear terms in the classical expression 
are exactly the same as the ones in \cite{HP}.
They are denoted by $B F_{\frac{5}{2},\frac{5}{2}}$ and $B
F_{\frac{5}{2}, 3}$ respectively. The ordering in the composite
operators is important at the quantum level.  This will give rise to the
different coefficient function in the derivative terms. For example, 
$(D T  \mp \overline{D}
  T) W(w)$, 
$\mp \overline{D} T \widetilde{[D, \overline{D}] T}(w)$ and $(\pa D T  
  \pm \pa  \overline{D} T) T(w)$. However, in the  classical limit
we do not have to worry about the ordering because the different
ordering gives the same result, as we described in previous section. 
Note that the relative coefficients $-\frac{3}{c}$, $-\frac{3}{2c}$,
and $-\frac{3}{4c}$ correspond to $4, 2, 1$ in the $a_{\frac{3}{2}} \,
\psi_{\frac{3}{2}}$($a_{\frac{3}{2}} \,
\psi_{2}$ in the second sign) terms \cite{HP}. 
One realizes that the relative coefficients $\frac{54}{5c}$ and
$-\frac{6}{5c}(= -\frac{1}{9} \times \frac{54}{5c})$ 
in the $\alpha$-dependent terms give the correct
values in \cite{HP}.

The relative coefficients $\frac{1}{3}$ and $\frac{1}{12}$
appearing in the descendant fields for $(D T \mp \overline{D} T)(w)$
can be obtained the formula \cite{Bowcock,BS}, 
$\frac{1}{n!} \times \frac{\Gamma(h_i-h_j+h_k+n)}
{\Gamma(h_i-h_j+h_k)} \times \frac{\Gamma(2h_k)}{\Gamma(2h_k+n)}$
where $h_i, h_j$, and $h_k$ are the conformal dimensions in the
operator product expansion $\phi^i(z) \phi^j(w) \sim \phi^k(w)$ and
$n$ is the number of derivatives of $\phi^k(w)$. Let us compute for
the case of the field $(D T \pm \overline{D} T)(w)$ and its descendant
field.
One sees that $h_i =2, h_j =\frac{5}{2}$, and $h_k=\frac{3}{2}$ in the
primary basis. Then
the coefficient of $\pa (D T \pm \overline{D} T)(w)$ can be read off
from
$\frac{1}{1!} \times \frac{\Gamma(2-\frac{5}{2}+\frac{3}{2}+1)}
{\Gamma(2-\frac{5}{2}+\frac{3}{2})} \times
\frac{\Gamma(2\times \frac{3}{2})}{\Gamma(2\times \frac{3}{2}+1)}=
\frac{1}{3}$ where $n=1$ 
and the coefficient of $\pa^2 (D T \pm \overline{D}
T)(w)$
is given by $\frac{1}{2!} \times \frac{\Gamma(2-\frac{5}{2}+\frac{3}{2}+2)}
{\Gamma(2-\frac{5}{2}+\frac{3}{2})} \times
\frac{\Gamma(2\times \frac{3}{2})}{\Gamma(2\times \frac{3}{2}+2)}=
\frac{1}{12}$ where $n=2$. 
For given coefficient $3$ in the $(D T \pm
\overline{D} T)$, the former becomes $3 \times \frac{1}{3}=1$ and the
latter becomes $3 \times \frac{1}{12}=\frac{1}{4}$ as in (\ref{class1}). 

\subsection{The operator product expansion $W(z) \;
\widetilde{(-1)\frac{1}{2} [D, \overline{D}] W}(w) $}

In this case, one should compute the extra operator product expansion 
between  $W(z)$ and $T W(w)$.
Let us describe the operator product expansion between the spin-$2$
and spin-$3$, by adding this extra contribution to ($C.3$),   
\bea
&& W(z) \; \widetilde{(-1)\frac{1}{2} [D, \overline{D}] W}(w)  =  
\frac{1}{(z-w)^2} \left[-\frac{3(-8+c)}{2(-12+5c)} \,\, \alpha \, \widetilde{[D, 
\overline{D} ] W}  \right. \nonu \\
&& -\frac{9c(-12+5c)}{4(-1+c)(6+c)(-3+2c)} \,\, \pa \widetilde{[D,
\overline{D}] T} +\frac{27(-12+5c)}{(-1+c)(6+c)(-3+2c)} \,\,
\pa T T  \nonu \\
&& +
\frac{27(-12+5c)}{2(-1+c)(6+c)(-3+2c)} \,\, T \widetilde{[D, \overline{D}]T}  
 -\frac{54(-12+5c)}{(-1+c)c(6+c)(-3+2c)} \,\, T^3 \nonu \\
&& \left. -
\frac{9c(-12+5c)}{(-1+c)(6+c)(-3+2c)} \,\,
\overline{D} T D T  -\frac{3c(-12+5c)}{2(-1+c)(6+c)(-3+2c)} \,\, \pa^2 T
+ 3 V \right](w) 
\nonu \\
&& +  
\frac{1}{(z-w)} \left[ -\frac{(-72+15c+c^2)}{2(3+c)(-12+5c)} 
\,\, \alpha \, \pa \widetilde{[D, 
\overline{D} ] W} + \frac{18}{c(3+c)}
\,\, \alpha \, T \pa W  \right. \nonu \\
&&  -\frac{6}{(3+c)} \,\, \alpha \, ( \overline{D} T D W +  D T \overline{D} W )-
\frac{3c(-12+5c)}{(-1+c)(6+c)(-3+2c)} \,\, ( \pa \overline{D} T D T - 
\pa D T \overline{D} T  ) \nonu \\
&& +\frac{9(-12+5c)}{2(-1+c)(6+c)(-3+2c)} \,\, \pa \widetilde{[D, \overline{D}] T}
T  +
\frac{36}{c(3+c)} \,\, \alpha \, \pa T W \nonu \\
&&   
+\frac{9(-12+5c)}{2(-1+c)(6+c)(-3+2c)} \,\, \pa T \widetilde{[D, \overline{D}] 
T} -\frac{54(-12+5c)}{(-1+c)c(6+c)(-3+2c)} \,\, \pa T T T  \nonu \\
&& \left. -\frac{(-12+5c)}{4(-1+c)(-3+2c)} \,\, \pa^3 T 
+ \pa V \right](w) 
+\cdots
\nonu \\
&& \nonu \\
 && \longrightarrow
\frac{1}{(z-w)^2} \left[-\frac{3}{10} \,\, \alpha \, \widetilde{[D, 
\overline{D} ] W}  -
\frac{45}{2c} \,\,
\overline{D} T D T 
+ 3 V \right](w) 
\nonu \\
&& +  
\frac{1}{(z-w)} \left[ -\frac{1}{10} 
\,\, \alpha \, \pa \widetilde{[D, 
\overline{D} ] W}  
-\frac{6}{c} \,\, \alpha \, ( \overline{D} T D W +  D T \overline{D} W )-
\frac{15}{2c} \,\, ( \pa \overline{D} T D T - 
\pa D T \overline{D} T  ) 
\right. \nonu \\
&& \left. + \pa V \right](w) 
+\cdots.
\label{class2}
\eea
There is no fourth-order singular term due to the particular
combination of spin-$3$ field in (\ref{twoccur}). See the equation 
($C.3$) where the fourth-order singular term occurs.
Compared to the previous two cases, many terms in the quantum operator
product expansion vanish in the classical limit.
In the $\alpha$-independent nonlinear term for the classical limit, the
relative coefficients give the correct values and also the
coefficients $-\frac{15}{2c}$ and $\frac{15}{2c}$ are consistent with
those in the bulk theory \cite{HP}. Note that there is an extra term
$\overline{D} T D T(w)$ in the second singular term and this doesn't
appear in the classical algebra.  
One can rewrite the last two nonlinear terms in the first-order
singular term as $-\frac{15}{2c} \pa (\overline{D} T D
T)(w)$. Therefore, the presence of $-\frac{45}{2c} \overline{D} T D
T(w)$ generates its above descendant field. The relative coefficient
$\frac{1}{3}$ is true from the conformal invariance.  

We will present the other remaining operator product expansions in the
Appendix $D$ given by ($D.1$), ($D.2$), ($D.3$), and
($D.4$).
As in (\ref{class2}), for the operator product expansions containing 
the new spin-$3$ field, one should recombine the contributions from $T
W(z)$. Furthermore, one should write down the old fields using the two
relations (\ref{mod}) and (\ref{twoccur}).  
 
\section{Conclusions and outlook }

We have found the complete ``nonlinear'' operator product expansion,
characterized by (\ref{openolimit}) together with (\ref{twoconsts}) 
and (\ref{selfcoupling}), of the lowest
higher spin current with spins $(2, \frac{5}{2}, \frac{5}{2}, 3)$ in
the ${\cal N}=2$ KS model on ${\bf CP}^N$ space.
In component approach, they are given in
($C.1$)-($C.10$) in primary basis.
The asymptotic symmetry of the higher spin 
$AdS_3$ supergravity, at the quantum level, should preserve the ${\cal
N}=2$ ${\cal W}_{N+1}$ algebra which contains the above nonlinear
operator product expansion \footnote{The nonlinear structure of the 
't Hooft limit of (\ref{openolimit}) is in principle fixed by 
${\cal N}=2$ conformal invariance. Therefore, these nonlinear terms are bound 
to match with the ones 
obtained from the Drinfeld-Sokolov 
reduction of supersymmetric higher spin algebra 
if the linear 
ones do. We thank the referee for pointing this out.}.

$\bullet$ It is an open problem to describe the 
quantum analysis of $AdS_3$ supergravity(or gravity in the context of
\cite{GH}) 
bulk theory. 
As noticed in (\ref{spin3half}) and (\ref{spin5half}), one should take
into account the quantum behavior(or normal ordered product in the
composite operators). We expect that there are additional
$c$-dependent terms in the variations of various generators with
different spins in the asymptotic
symmetry algebra.

$\bullet$ The field contents for the ${\cal N}=2$ ${\cal W}_{N+1}$
algebra are given by $T(Z)$, $W(Z)$, $V(Z)$, $X(Z)$, a current with
spins $(5,\frac{11}{2}, \frac{11}{2}, 6)$, $\cdots$, a current with
spins $(N, N+\frac{1}{2}, N+\frac{1}{2}, N+1)$. It is an open problem to
construct these currents in terms of ${\cal N}=2$ WZW constrained 
currents. The
first step is how to generalize the current (\ref{superspin2}), 
satisfying the conditions (\ref{KWvanishing}) and (\ref{TW}), 
to arbitrary $N$.   
Once this current is found, then in principle, one can proceed to
compute the operator product expansion, by hand, to determine other
higher spin currents. 

$\bullet$ So far, we have considered the operator product expansion of
$W(Z_1) \; W(Z_2)$. It would be interesting to see 
the remaining operator product expansions for the higher
spin currents. For example, $W(Z_1)\; V(Z_2)$ and $V(Z_1)\; V(Z_2)$ for
general $N$. As in previous paragraph, 
in order to obtain these, it is necessary to find out the $W(Z)$
for general $N$. 
For ${\cal N}=2 $ ${\cal W}_4$ algebra where $N=3$, these are
constructed in \cite{BW}, along the line of the approach $1$,
 but the explicit results are still missing. 

$\bullet$ As described in the introduction, the approach $2$ is based
on the quantum Miura transformation. Some of the currents in the 
Fateev-Lukyanov construction are found in \cite{Ito,Ozer,HP}.
It is an open problem to compute all the other currents
systematically and see
whether these satisfy the above operator product expansion (\ref{openolimit}).
In this case, in general, the fields are not primary. Therefore, 
the nontrivial task is to find the right primary fields with
$N$-dependent coefficient functions. 

$\bullet$ One can also consider the bosonic subalgebra of
(\ref{openolimit}) for particular $N$.
For $N=2$, this analysis was done in \cite{Romans}.
For $N=3$, the subalgebra should contain the bosonic $W_4$
algebra. In the notation of \cite{BS}, one denotes it as ${\cal
  W}(2,3,4)$. 
The field contents of this
algebra are given by the stress tensor of spin $2$, one higher spin
current of
spin $3$, and other higher spin current of spin $4$. 
It would be interesting to obtain the algebra found in \cite{BFKNRV,KW}.
The main thing is to identify the correct primary fields for given 
fields coming from $T(Z)$, $W(Z)$ and $V(Z)$.
Similarly, the $N=4$ case corresponds to the bosonic $W_5$ algebra
denoted by ${\cal W}(2,3,4,5)$. It is an open problem to obtain the
algebra in \cite{Hornfeck1,Hornfeck2} from our result. 

\vspace{.7cm}

\centerline{\bf Acknowledgments}

CA would like to thank A. Wisskirchen for sending the paper \cite{Wisskirchen}
and thank the referee for pointing out many comments which 
are very helpful to improve the draft. 
This work was supported by the Mid-career Researcher Program through
the National Research Foundation of Korea (NRF) grant 
funded by the Korean government (MEST) (No. 2012045385).
CA acknowledges warm hospitality from 
the School of  Liberal Arts (and Institute of Convergence Fundamental
Studies), Seoul National University of Science and Technology.

\appendix

\renewcommand{\thesection}{\large \bf \mbox{Appendix~}\Alph{section}}
\renewcommand{\theequation}{\Alph{section}\mbox{.}\arabic{equation}}

\section{The generators and structure constants of $U(1) \times 
SU(4)$ in complex basis}

Let us describe the $5$ generators $T_m$ where $m=1, 2, \cdots, 5$
and the remaining $3$ generators $T_a$ where $a=6, 7, 8$ as follows:
\bea
T_1 & = &
\left(
\begin{array}{cccc}
0 & 0 &0 &0    \\ 
1 & 0 &0 &0   \\
0 & 0 &0 &0  \\
0 & 0 &0 & 0   \\
\end{array} \right),
\,\,\,
T_2 =
\left(
\begin{array}{cccc}
0 & 0 &0 &0    \\ 
0 & 0 &0 &0   \\
1 & 0 &0 &0  \\
0 & 0 &0 & 0   \\
\end{array} \right), 
\,\,\ T_3 =
\left(
\begin{array}{cccc}
0 & 0 &0 &0    \\ 
0 & 0 &0 &0   \\
0 & 1 &0 &0  \\
0 & 0 &0 & 0   \\
\end{array} \right),
\nonu \\
T_4  & = &
\left(
\begin{array}{cccc}
\frac{i}{2}+\frac{1}{\sqrt{12}} & 0 &0 &0    \\ 
0 & -\frac{i}{2} +\frac{1}{\sqrt{12}} &0 &0   \\
0 & 0 &-\frac{2}{\sqrt{12}} &0  \\
0 & 0 &0 & 0   \\
\end{array} \right),
\nonu \\
\,\,\,
T_5  &= &
\left(
\begin{array}{cccc}
\frac{i}{\sqrt{24}}+\frac{\sqrt{5}}{\sqrt{40}}  & 0 &0 &0    \\ 
0 & \frac{i}{\sqrt{24}}+\frac{\sqrt{5}}{\sqrt{40}} &0 &0   \\
0 & 0 &\frac{i}{\sqrt{24}}+\frac{\sqrt{5}}{\sqrt{40}} &0  \\
0 & 0 &0 & -\frac{3i}{\sqrt{24}}+\frac{\sqrt{5}}{\sqrt{40}}   \\
\end{array} \right), 
\nonu \\
T_6 & = &
\left(
\begin{array}{cccc}
0 & 0 &0 &0    \\ 
0 & 0 &0 &0   \\
0 & 0 &0 &0  \\
1 & 0 &0 & 0   \\
\end{array} \right),
\,\,\,
T_{7} =
\left(
\begin{array}{cccc}
0 & 0 &0 &0    \\ 
0 & 0 &0 &0   \\
0 & 0 &0 &0  \\
0 & 1 &0 & 0   \\
\end{array} \right), 
\,\,\,
T_{8} =
\left(
\begin{array}{cccc}
0 & 0 &0 &0    \\ 
0 & 0 &0 &0   \\
0 & 0 &0 &0  \\
0 & 0 &1 & 0   \\
\end{array} \right).
\label{su4generators}
\eea
Note that $T_4= i H_1 + H_2$ and $T_5 = i H_3 + H_4$ where the three
Cartan generators are given 
$H_1 = \frac{1}{2} \mbox{diag}(1,-1,0,0)$, $H_2 =
\frac{1}{\sqrt{12}} \mbox{diag}(1,1,-2,0)$, $H_3 =
\frac{1}{\sqrt{24}} 
\mbox{diag}(1,1,1,-3)$  and $H_4 = 
\frac{\sqrt{5}}{\sqrt{40}} \mbox{diag}(1,1,1,1)$.
The generators have been normalized so that 
$\mbox{tr} \, T_a T_{\bar{b}} =\delta_{a b}$ and $\mbox{tr} \, T_m T_{\bar{n}}=
\delta_{mn}$.
The conjugated generators $T_{\bar{m}}$ and $T_{\bar{a}}$ can be
obtained from $T_{\bar{m}} =T_m^{\dagger}$ and
$T_{\bar{a}}=T_a^{\dagger}$. 
One can easily see that the $8$ generators $T_m, T_{\bar{m}}$ where $m=1,
2, \cdots, 4$ consist of 
the $SU(3)$ subgroup generators. The remaining diagonal generators 
$(T_5 +T_{\bar{5}})$ and $(T_5 -T_{\bar{5}})$ 
correspond to the two $U(1)$'s in the subgroup $H$ of
the coset model $(2.6)$ and the generator 
$(T_5 +T_{\bar{5}})$
 corresponds to the $U(1)$ in the 
numerator.

\section{The operator product expansions $(2.3)$ and  
$(2.11)$ in the component  approach }

Practically, it is often useful to compute the operator product
expansions in component approach. Now we would like to rewrite 
the ${\cal N}=2$ superspace formula $(2.3)$ 
as follows: 
\bea
T(z) \; T(w) & = & \frac{1}{(z-w)^2} \,\,\frac{c}{3} + \cdots,
\nonu \\
T(z) \; D T(w) & = & \frac{1}{(z-w)} \,\, D T(w) + \cdots,
\nonu \\
T(z) \; \overline{D} T(w) & = & -\frac{1}{(z-w)} \,\, \overline{D} T(w) + \cdots,
\nonu \\
T(z) \; [ D, \overline{D} ] T(w) & = & -\frac{1}{(z-w)^2} \,\, 2 T(w) + \cdots,
\nonu \\
D T(z)  \; T(w) & = & -\frac{1}{(z-w)} \,\, D T(w) + \cdots,
\nonu \\
D T(z) \; \overline{D} T(w) & = & \frac{1}{(z-w)^3} \,\, \frac{c}{3}
+\frac{1}{(z-w)^2}  \,\, T(w) \nonu \\
&+ & 
\frac{1}{(z-w)} \,\, 
\frac{1}{2} \left[ -[ D, \overline{D} ] T +\pa T \right](w) + \cdots,
\nonu \\
D T(z) \; [D, \overline{D}] T(w) & = & - \frac{1}{(z-w)^2} \,\,3 D T(w) -
\frac{1}{(z-w)} \,\, \pa D T(w) + \cdots,
\nonu \\
\overline{D} T(z)  \; T(w) & = & \frac{1}{(z-w)} \,\, \overline{D} T(w) + \cdots,
\nonu \\
\overline{D} T(z) \; D T(w) & = & \frac{1}{(z-w)^3} \,\, 
\frac{c}{3} - \frac{1}{(z-w)^2}  \,\, T(w) \nonu \\
& - &
\frac{1}{(z-w)} \,\,
\frac{1}{2} \left[ [ D, \overline{D} ] T +\pa T \right](w) + \cdots,
\nonu \\
\overline{D} T(z) \; [ D, \overline{D} ] T(w) & = & - \frac{1}{(z-w)^2}
\,\, 3 
\overline{D} T(w) -
\frac{1}{(z-w)} \,\, \pa \overline{D} T(w) + \cdots,
\nonu \\
(-1) \frac{1}{2} \left[ D, \overline{D} \right] T(z) \; T(w) & = &  
\frac{1}{(z-w)^2} \,\, 
 T(w) +
\frac{1}{(z-w)}  \,\, \pa  T(w) + \cdots,
\nonu \\
(-1) \frac{1}{2} \left[ D, \overline{D} \right] T(z) \; D T(w) & = &  
\frac{1}{(z-w)^2} \,\, \frac{3}{2} 
D T(w) +
\frac{1}{(z-w)}  \,\, \pa D T(w) + \cdots,
\nonu \\
(-1) \frac{1}{2} \left[ D, \overline{D} \right] T(z) \; \overline{D} T(w)
& = & 
 \frac{1}{(z-w)^2} \,\, \frac{3}{2} 
\overline{D} T(w) +
\frac{1}{(z-w)}  \,\, \pa \overline{D} T(w) + \cdots,
\nonu \\
(-1) \frac{1}{2} 
\left[ D, \overline{D} \right] T(z) \; [ D, \overline{D} ] T(w) & = &  
- \frac{1}{(z-w)^4} c + \frac{1}{(z-w)^2} \,\, 2 
 [ D, \overline{D} ] T(w) \nonu \\
& + &
\frac{1}{(z-w)}  \,\, \pa [ D, \overline{D} ] T(w) + \cdots.
\label{opett}
\eea
Similarly, the operator product expansions between the stress energy
tensor and the higher spin current $\Phi$ with spins $(\Delta, 
\Delta +\frac{1}{2},\Delta+\frac{1}{2}, \Delta+1)$
can be written as 
\bea
T(z) \; D \Phi(w) & = & \frac{1}{(z-w)} \,\, D \Phi(w) + \cdots,
\nonu \\
T(z) \; \overline{D} \Phi(w) & = & -\frac{1}{(z-w)} \,\, \overline{D} \Phi(w) + \cdots,
\nonu \\
T(z) \; [ D, \overline{D} ] \Phi(w) & = & -\frac{1}{(z-w)^2} \,\, 2\Delta 
\Phi(w) + \cdots,
\nonu \\
D T(z)  \; \Phi(w) & = & -\frac{1}{(z-w)} \,\, D \Phi (w) + \cdots,
\nonu \\
D T(z) \; \overline{D} \Phi(w) & = & \frac{1}{(z-w)^2} \,\, \Delta \Phi(w) +
\frac{1}{(z-w)} \,\,
\frac{1}{2} \left[ -[ D, \overline{D} ] \Phi +\pa \Phi \right](w) + \cdots,
\nonu \\
D T(z) \; [D, \overline{D}] \Phi(w) & = & - \frac{1}{(z-w)^2} \,\, (2\Delta+1) 
D \Phi(w) -
\frac{1}{(z-w)} \,\, \pa D \Phi(w) + \cdots,
\nonu \\
\overline{D} T(z)  \; \Phi(w) & = & \frac{1}{(z-w)} \,\, \overline{D} \Phi(w) + \cdots,
\nonu \\
\overline{D} T(z) \; D \Phi(w) & = & - \frac{1}{(z-w)^2} \,\, \Delta \Phi(w) -
\frac{1}{(z-w)} \,\,
\frac{1}{2} \left[ [ D, \overline{D} ] \Phi +\pa \Phi \right](w) + \cdots,
\nonu \\
\overline{D} T(z) \; [ D, \overline{D} ] \Phi(w) & = & - \frac{1}{(z-w)^2}
\,\, (2\Delta+1) 
\overline{D} \Phi(w) -
\frac{1}{(z-w)} \,\, \pa \overline{D} \Phi(w) + \cdots,
\nonu \\
(-1) \frac{1}{2} \left[ D, \overline{D} \right] T(z) \; \Phi(w) & = &
\frac{1}{(z-w)^2} \,\, \Delta 
 \Phi(w) +
\frac{1}{(z-w)} \,\, \pa  \Phi(w) + \cdots,
\nonu \\
(-1) \frac{1}{2} \left[ D, \overline{D} \right] T(z) \; D \Phi(w) & = &  
\frac{1}{(z-w)^2} \,\, (\Delta+\frac{1}{2}) 
D \Phi(w) +
\frac{1}{(z-w)} \,\, \pa D \Phi(w) + \cdots,
\nonu \\
(-1) \frac{1}{2} \left[ D, \overline{D} \right] T(z) \; \overline{D} \Phi(w)
& = & 
 \frac{1}{(z-w)^2} \,\, (\Delta+\frac{1}{2})
\overline{D} \Phi(w) +
\frac{1}{(z-w)}  \,\, \pa \overline{D} \Phi(w) + \cdots,
\nonu \\
(-1) \frac{1}{2} 
\left[ D, \overline{D} \right] T(z) \; [ D, \overline{D} ] \Phi(w) & = &
\frac{1}{(z-w)^2} \,\,
(\Delta+1) [ D, \overline{D} ] \Phi(w) +
\frac{1}{(z-w)}  \,\, \pa [ D, \overline{D} ] \Phi(w) \nonu \\
& + & \cdots.
\label{opetv}
\eea
All the component fields except the fourth component of $T(Z)$ in
$(2.5)$
are primary, from (\ref{opett}) and (\ref{opetv}),  
with respect to the stress energy tensor $(-1) \frac{1}{2}
[D, \overline{D}] T(z)$. 
As we described in section $2$, under the modified stress energy
tensor $(2.21)$, some of them are not primary.
In general, once we know the stress energy tensor and the first
component of any primary field, then other three components of this
primary field are determined by the primary field condition. We used
this property all the time in this paper.

\section{The operator product expansions of ${\cal N}=2$ quantum 
${\cal W}_{N+1}$ algebra in component approach: Primary basis }

The operator product expansion of spin-$2$ and spin-$2$, from the
relation $(2.18)$,  can be
written as \footnote{With the ${\cal N}=2$ package [40], one
obtains the following  component operator product expansions
automatically by using the command ``N2OPEToComponents''. }
\bea
&& W(z) \; W(w)  =  
\frac{1}{(z-w)^4} \,\, \frac{c}{2} \nonu \\
&& +\frac{1}{(z-w)^2} \left[
  2 \alpha\, W -\frac{c}{(-1+c)} \,\, [D, \overline{D}] T
  -\frac{3}{(-1+c)} \,\, T^2 \right](w)   \nonu \\
&& +  \frac{1}{(z-w)} \left[ 
\alpha \, \pa W -\frac{c}{2(-1+c)} \,\, \pa [D, \overline{D}] T
-\frac{3}{(-1+c)} \,\, \pa T T \right](w) 
 +   \cdots
\nonu \\
&& \nonu \\
&& \longrightarrow 
\frac{1}{(z-w)^4} \,\, \frac{c}{2} 
 +\frac{1}{(z-w)^2} \left[
  2 \alpha\, W - \,\, [D, \overline{D}] T
  -\frac{3}{c} \,\, T^2 \right](w)   \nonu \\
&& +  \frac{1}{(z-w)} \left[ 
\alpha \, \pa W -\frac{1}{2} \,\, \pa [D, \overline{D}] T
-\frac{3}{c} \,\, \pa T T \right](w) 
 +   \cdots.
\label{opeww} 
\eea
Also we take the large $c$ limit. In section $2$, we introduced a new
stress energy tensor by looking at the second-order pole.
One easily sees the relative coefficient $\frac{1}{2}$ between the second-order
and the first-order singular terms from the conformal invariance.

The spin-$2$ and spin-$\frac{5}{2}$ operator product expansions are 
\bea
&& W(z) \; (D W \pm \overline{D} W)(w)  =  
\frac{1}{(z-w)^3} 3 \left[ D T \mp \overline{D} T \right](w)
\nonu \\
&&  +\frac{1}{(z-w)^2} \left[
\alpha \, (D W \pm \overline{D} W) +  \frac{c}{(-1+c)} 
\,\, 
\pa (D T \mp \overline{D} T) -\frac{3}{(-1+c)} \,\, T ( D T \pm  \overline{D} T ) \right](w) 
\nonu \\
&& + \frac{1}{(z-w)} \left[ ( D V \mp \overline{D} V) +
\frac{2(-27+12c+c^2)}{(3+c)(-12+5c)} \,\,
  \alpha \,
\pa (D W  \pm \overline{D} W) \right. \nonu \\
&&  \mp \frac{c(63-9c+2c^2)}{4(-1+c)(6+c)(-3+2c)} \,\, 
\pa^2 \overline{D} T
- \frac{6(-15+c)}{(3+c)(-12+5c)} \,\, \alpha \, T ( D W \mp  \overline{D} W) 
\nonu \\
&&  -\frac{9(3+4c)}{(-1+c)(6+c)(-3+2c)} \,\, T T (D T \mp  \overline{D}
  T) +\frac{54(-1+c)}{(3+c)(-12+5c)} \,\, \alpha \, (D T  \mp \overline{D}
  T) W  \nonu \\
&&  - \frac{27c}{2(6+c)(-3+2c)} \,\, ( [D, \overline{D}]T D T  \mp
\overline{D} T [D, \overline{D}]
  T)
  +\frac{(-15+c)(-3+c)c}{2(-1+c)(6+c)(-3+2c)} \,\, \pa^2 D T \nonu \\
&&  -\frac{3(-15+c)c}{(-1+c)(6+c)(-3+2c)} \,\, (\pa D T 
  \pm \pa \overline{D} T) T \nonu \\
&& \left. -\frac{3(-54+39c+c^2)}{2(-1+c)(6+c)(-3+2c)}  \,\, \pa T (
  D T \pm  \overline{D} T ) \right] (w)
+  \cdots
\nonu \\
&& \nonu \\
&& \longrightarrow
\frac{1}{(z-w)^3} 3 \left[ D T \mp \overline{D} T \right](w)
\nonu \\
&&  +\frac{1}{(z-w)^2} \left[
\alpha \, (D W \pm \overline{D} W) +  
\,\, 
\pa (D T \mp \overline{D} T) -\frac{3}{c} \,\, T ( D T \pm  \overline{D} T ) \right](w) 
\nonu \\
&& + \frac{1}{(z-w)} \left[ ( D V \mp \overline{D} V) +
\frac{2}{5} \,\,
  \alpha \,
\pa (D W  \pm \overline{D} W) 
  \mp \frac{1}{4} \,\, 
\pa^2 \overline{D} T
- \frac{6}{5c} \,\, \alpha \, T ( D W \mp  \overline{D} W) 
\right. \nonu \\
&&  -\frac{18}{c^2} \,\, T T (D T \mp  \overline{D}
  T) +\frac{54}{5c} \,\, \alpha \, (D T  \mp \overline{D}
  T) W  
  - \frac{27}{4c} \,\, ( [D, \overline{D}]T D T  \mp
\overline{D} T [D, \overline{D}]
  T)
  \nonu \\
&&  \left. -\frac{3}{2c} \,\, (\pa D T 
  \pm \pa \overline{D} T) T  -\frac{3}{4c}  \,\, \pa T (
  D T \pm  \overline{D} T ) + \frac{1}{4} \pa^2 D T \right] (w)
+  \cdots.
\label{ope2fivehalf} 
\eea
One realizes that the symmetry behind the two types of combination in 
the second- and third-components
of $W(w)$ occurs in the right hand side except some derivative terms
but at the classical level these derivative terms also can be combined.

The operator product expansion between the spin-$2$ and spin-$3$ is
given by
\bea
&& W(z) \; (-1)\frac{1}{2} [D, \overline{D}] W(w)  =  
\frac{1}{(z-w)^4} 3 T(w) \nonu \\
&& +\frac{1}{(z-w)^2} \left[-\frac{3(-8+c)}{2(-12+5c)} \,\, \alpha \, [D, 
\overline{D} ] W -\frac{9c(-12+5c)}{4(-1+c)(6+c)(-3+2c)} \,\, \pa [D,
\overline{D}] T \right.
\nonu \\
&&  +\frac{42}{(-12+5c)} \,\, \alpha \, T W-
\frac{3(36-9c+8c^2)}{2(-1+c)(6+c)(-3+2c)} \,\, T [D, \overline{D}]T  
\nonu \\
&&  -\frac{9(3+4c)}{(-1+c)(6+c)(-3+2c)} \,\, T^3 -
\frac{9c(-12+5c)}{(-1+c)(6+c)(-3+2c)} \,\,
\overline{D} T D T \nonu \\
&& \left. -\frac{3c(-12+5c)}{2(-1+c)(6+c)(-3+2c)} \,\, \pa^2 T
+ 3 V \right](w) 
\nonu \\
&& +  
\frac{1}{(z-w)} \left[ -\frac{(-72+15c+c^2)}{2(3+c)(-12+5c)} 
\,\, \alpha \, \pa [D, 
\overline{D} ] W + \frac{6(3+4c)}{(3+c)(-12+5c)}
\,\, \alpha \, T \pa W  \right. \nonu \\
&&  -\frac{6}{(3+c)} \,\, \alpha \, ( \overline{D} T D W +  D T \overline{D} W )-
\frac{3c(-12+5c)}{(-1+c)(6+c)(-3+2c)} \,\, ( \pa \overline{D} T D T - 
\pa D T \overline{D} T  ) \nonu \\
&& -\frac{3c(3+4c)}{2(-1+c)(6+c)(-3+2c)} \,\, \pa [D, \overline{D}] T
T  -
\frac{6(-15+c)}{(3+c)(-12+5c)} \,\, \alpha \, \pa T W \nonu \\
&&   
+\frac{9(-12+5c)}{2(-1+c)(6+c)(-3+2c)} \,\, \pa T [D, \overline{D}] 
T -\frac{9(3+4c)}{(-1+c)(6+c)(-3+2c)} \,\, \pa T T T  \nonu \\
&& \left. -\frac{c(-6+13c)}{4(-1+c)(6+c)(-3+2c)} \,\, \pa^3 T 
+ \pa V \right](w) 
+\cdots
\nonu \\
&& \nonu \\
&& \longrightarrow 
\frac{1}{(z-w)^4} 3 T(w) \nonu \\
&& +\frac{1}{(z-w)^2} \left[-\frac{3}{10} \,\, \alpha \, [D, 
\overline{D} ] W    +\frac{42}{5c} \,\, \alpha \, T W-
\frac{6}{c} \,\, T [D, \overline{D}]T  
  -\frac{18}{c^2} \,\, T^3 -
\frac{45}{2c} \,\,
\overline{D} T D T 
+ 3 V \right](w) 
\nonu \\
&& +  
\frac{1}{(z-w)} \left[ -\frac{1}{10} 
\,\, \alpha \, \pa [D, 
\overline{D} ] W + \frac{24}{5c}
\,\, \alpha \, T \pa W    -\frac{6}{c} \,\, \alpha \, ( \overline{D} T
D W +  
D T \overline{D} W ) \right. \nonu \\
&& \left. -
\frac{15}{2c} \,\, ( \pa \overline{D} T D T - 
\pa D T \overline{D} T  )  -\frac{3}{c} \,\, \pa [D, \overline{D}] T
T  -
\frac{6}{5c} \,\, \alpha \, \pa T W    
 -\frac{18}{c^2} \,\, \pa T T T   + \pa V \right](w) 
\nonu \\
&& +\cdots.
\label{ope23} 
\eea
Note that the spin-$3$ field $V(w)$ and its descendant field $\pa V(w)$ arise here. 
In the $c \rightarrow \infty$ limit, the terms 
$\pa [D, \overline{D}] T(w)$, $\pa^2 T(w)$, $\pa T [D, \overline{D}]
T(w)$, and $\pa^3 T(w)$ are vanishing with an appropriate limiting procedure. 

The operator product expansion of spin-$\frac{5}{2}$ and
spin-$\frac{5}{2}$
occurs as follows:
\bea
&& D W(z)  \; D W(w)  = \frac{1}{(z-w)} \left[-\frac{12}{3+c} \,\, \alpha \,
  D T D W -\frac{6}{(-1+c)} \,\, \pa D T D T \right](w) +\cdots
\nonu \\
&& \nonu \\
&& \longrightarrow
 \frac{1}{(z-w)} \left[-\frac{12}{c} \,\, \alpha \,
  D T D W -\frac{6}{c} \,\, \pa D T D T \right](w) +\cdots.
\label{dwdw}
\eea
There are no any linear terms in the right hand side.
At the linear level, there were no singular terms [16].
Similarly, one has the following operator product expansion
\bea
&& D W(z) \; \overline{D} W (w) +\overline{D} W(z) \; D W(w)  =  
\frac{1}{(z-w)^5} 2 c \nonu \\
&& +\frac{1}{(z-w)^3} \left[ 4 \alpha \, W -
\frac{(-3+5c)}{(-1+c)} \,\, [D,
  \overline{D}]T -\frac{6}{(-1+c)} \,\, T^2 
\right](w) \nonu \\
&& +\frac{1}{(z-w)^2} \left[ 2 \alpha \, \pa W -
  \frac{(-3+5c)}{2(-1+c)} \,\, \pa [D,
  \overline{D}] T -\frac{6}{(-1+c)} \,\, \pa T T 
\right](w) \nonu \\
&& +\frac{1}{(z-w)} \left[\frac{3(-6+c)(-1+c)}{(3+c)(-12+5c)} \,\, \alpha
  \, \pa^2 W -
  \frac{3c(9-3c+c^2)}{2(-1+c)(6+c)(-3+2c)} 
\,\, \pa^2 [D, \overline{D}] T  \right.  \nonu \\
&& + \frac{6(-15+c)}{(3+c)(-12+5c)} \,\, \alpha \, T [D, \overline{D}] W
+ \frac{9(3+4c)}{(-1+c)(6+c)(-3+2c)} \,\, T T [D, \overline{D}] T  
\nonu \\
&& -\frac{36(3+4c)}{(-1+c)(6+c)(-3+2c)} \,\, T \overline{D} T D T
+\frac{12(3+4c)}{(3+c)(-12+5c)} \,\, \alpha ( \overline{D} T D W -  D T \overline{D} 
W)
\nonu \\
&& + \frac{3(18-51c+5c^2)}{(-1+c)(6+c)(-3+2c)} \,\, ( \pa 
\overline{D} T D T  + \pa D T 
\overline{D} T ) -\frac{54(-1+c)}{(3+c)(-12+5c)} \,\, \alpha \, 
[D, \overline{D}] T W \nonu \\
&& + \frac{27c}{2(6+c)(-3+2c)} \,\, [D, \overline{D}] T [D, 
\overline{D}] T -\frac{9(3+4c)}{(-1+c)(6+c)(-3+2c)} 
\,\, \pa [D, \overline{D}] T  T \nonu \\
&& -\frac{9(-6+c)}{2(6+c)(-3+2c)} \,\, \pa T \pa T -
\frac{3(-18+24c+c^2)}{(-1+c)(6+c)(-3+2c)} \,\, \pa^2 T T \nonu \\
&& \left. -\frac{3(3+4c)}{(-1+c)(6+c)(-3+2c)} \,\, \pa^3 T - [D, \overline{D}] V
\right](w)  +\cdots
\nonu \\
&& \nonu \\
&& \longrightarrow 
\frac{1}{(z-w)^5} 2 c \nonu \\
&& +\frac{1}{(z-w)^3} \left[ 4 \alpha \, W -
5 \,\, [D,
  \overline{D}]T -\frac{6}{c} \,\, T^2 
\right](w)  +\frac{1}{(z-w)^2} \left[ 2 \alpha \, \pa W -
  \frac{5}{2} \,\, \pa [D,
  \overline{D}] T -\frac{6}{c} \,\, \pa T T 
\right](w) \nonu \\
&& +\frac{1}{(z-w)} \left[\frac{3}{5} \,\, \alpha
  \, \pa^2 W -
  \frac{3}{4} 
\,\, \pa^2 [D, \overline{D}] T  
 + \frac{6}{5c} \,\, \alpha \, T [D, \overline{D}] W
+ \frac{18}{c^2} \,\, T T [D, \overline{D}] T  
\right.
\nonu \\
&& -\frac{72}{c^2} \,\, T \overline{D} T D T
+\frac{48}{5c} \,\, \alpha ( \overline{D} T D W -  D T \overline{D} 
W)
+ \frac{15}{2c} \,\, ( \pa 
\overline{D} T D T  + \pa D T 
\overline{D} T ) \nonu \\
&& -\left. \frac{54}{5c} \,\, \alpha \, 
[D, \overline{D}] T W + \frac{27}{4c} \,\, [D, \overline{D}] T [D, 
\overline{D}] T   -\frac{9}{4c} \,\, \pa T \pa T -
\frac{3}{2c} \,\, \pa^2 T T   - [D, \overline{D}] V
\right](w)  +\cdots,
\label{ope351} 
\eea
where the spin-$4$ field occurs in the right hand side.
In the $c \rightarrow \infty$ limit, the terms 
$\pa [D, \overline{D}] T T(w)$ and $\pa^3 T(w)$ are vanishing. 
One can easily analyze the relative coefficients appearing in the
spin-$2$ fields and its descendant fields.

The different combination between the spin-$\frac{5}{2}$ fields 
leads to
\bea
&& D W(z)  \; \overline{D} W (w) -\overline{D} W(z) \; D W(w)  =  
\frac{1}{(z-w)^4} \,\, 6 T(w) +\frac{1}{(z-w)^3} \,\, 3 \pa T(w)  
\nonu \\
&& +\frac{1}{(z-w)^2} \left[ \frac{2(6+c)}{(-12+5c)} \,\, \alpha \, [D,
  \overline{D}] W 
+ \frac{2c(9-3c+c^2)}{(-1+c)(6+c)(-3+2c)} \,\,
\pa^2 T \right. \nonu \\
&&  -\frac{3(-3+c)(-6+13c)}{2(-1+c)(6+c)(-3+2c)} \,\, \pa [D, 
\overline{D}] T  + \frac{84}{(-12+5c)} \,\, \alpha \, T W  \nonu \\
&&  -
\frac{6(9+5c^2)}{(-1+c)(6+c)(-3+2c)} \,\, T [D, 
\overline{D}] T  -\frac{18(3+4c)}{(-1+c)(6+c)(-3+2c)}
\,\, T^3 \nonu \\
&& \left. - \frac{6(-3+c)(-6+13c)}{(-1+c)(6+c)(-3+2c)} 
\,\, \overline{D} T D T  +6 V 
\right](w) \nonu \\
&& +\frac{1}{(z-w)} \left[ \frac{(6+c)}{(-12+5c)} 
\,\, \alpha \, \pa [D, \overline{D}] W +
  \frac{(-18+9c-13c^2+c^3)}{2(-1+c)(6+c)(-3+2c)} 
\,\, \pa^3 T \right. \nonu \\
&&  + \frac{42}{(-12+5c)} \,\, \alpha \, \pa (T  W)
  -\frac{3(-3+c)(-6+13c)}{(-1+c)(6+c)(-3+2c)}\,\,
(\pa \overline{D} T D T -\pa D T \overline{D} T)  \nonu \\
&&  -\frac{3(9+5c^2)}{(-1+c)(6+c)(-3+2c)} \,\,
( \pa [D, \overline{D}] T  T +\pa T [D, \overline{D}] T )    \nonu \\
&&  \left. -\frac{27(3+4c)}
{(-1+c)(6+c)(-3+2c)} \,\, \pa T T T   + 3 \pa V \right](w)  +\cdots
\nonu \\
&& \nonu \\
&& \longrightarrow 
\frac{1}{(z-w)^4} \,\, 6 T(w) +\frac{1}{(z-w)^3} \,\, 3 \pa T(w)  
\nonu \\
&& +\frac{1}{(z-w)^2} \left[ \frac{2}{5} \,\, \alpha \, [D,
  \overline{D}] W 
+  \,\,
\pa^2 T    + \frac{84}{5c} \,\, \alpha \, T W    -
\frac{15}{c} \,\, T [D, 
\overline{D}] T  -\frac{36}{c^2}
\,\, T^3 \right. \nonu \\
&& \left. - \frac{39}{c} 
\,\, \overline{D} T D T  +6 V 
\right](w) \nonu \\
&& +\frac{1}{(z-w)} \left[ \frac{1}{5} 
\,\, \alpha \, \pa [D, \overline{D}] W +
  \frac{1}{4} 
\,\, \pa^3 T + \frac{42}{5c} \,\, \alpha \, \pa (T  W)
  -\frac{39}{2c}\,\,
(\pa \overline{D} T D T -\pa D T \overline{D} T) \right.  \nonu \\
&&  \left. -\frac{15}{2c} \,\,
( \pa [D, \overline{D}] T  T +\pa T [D, \overline{D}] T )  -\frac{54}
{c^2} \,\, \pa T T T   + 3 \pa V \right](w)  +\cdots.
\label{ope352}
\eea
In this case, 
the spin-$3$ field $V(w)$ and its descendant field $\pa V(w)$ appear.
In the $c \rightarrow \infty$ limit, the term 
$\pa [D, \overline{D}] T(w)$ vanishes. 
By adding or subtracting the two equations (\ref{ope351}) and (\ref{ope352}),
one obtains $D W(z) \;\overline{D} W(w)$ and $\overline{D} W(z) \; D W(w)$ independently.

One also has the following operator product expansion without any
linear terms
\bea
&& \overline{D} W(z)  \;  \overline{D} W(w)  
= \frac{1}{(z-w)} \left[\frac{12}{3+c} \,\, \alpha \,
  \overline{D} T \overline{D} W -
\frac{6}{(-1+c)} \,\, \pa \overline{D} T \overline{D} T \right](w)
+\cdots
\nonu \\
&& \nonu \\
&& \longrightarrow 
  \frac{1}{(z-w)} \left[\frac{12}{c} \,\, \alpha \,
  \overline{D} T \overline{D} W -
\frac{6}{c} \,\, \pa \overline{D} T \overline{D} T \right](w)
+\cdots.
\label{dwdw1}
\eea
There were no singular terms at the linear level [16].

The next operator product expansion between 
the spin-$\frac{5}{2}$ and the spin-$3$ leads to
\bea
&& (D W  \pm \overline{D} W)(z) \; (-1)\frac{1}{2} [D, \overline{D}] W(w)  =
\frac{1}{(z-w)^4} \,\, \frac{15}{2} \left[ DT  \pm \overline{D} T \right](w)
\nonu \\
&& +\frac{1}
{(z-w)^3} \left[ \alpha \, (D W  \mp \overline{D} W)  + 
\frac{(-3+5c)}{2(-1+c)} \,\, \pa ( D T
  \pm \overline{D} T) -\frac{3}{(-1+c)} \,\, T (  D T \mp  \overline{D} T) 
\right](w) \nonu \\
&& + \frac{1}{(z-w)^2} \left[ \frac{(-9+21c+2c^2)}{(3+c)(-12+5c)}
\alpha \, \pa (D W \mp \overline{D} W)
  + 
\frac{(-3+c)(18-114c+5c^2)}{4(-1+c)(6+c)(-3+2c)} \,\, \pa^2  D T
   \right. \nonu \\
&& \pm \frac{(108+423c-93c^2+10c^3)}{8(-1+c)(6+c)(-3+2c)} \,\, \pa^2 
\overline{D} T +
\frac{3(57+13c)}{(3+c)(-12+5c)} \,\, \alpha \, T ( D W \pm  \overline{D} W )  
\nonu \\
&& -\frac{63(3+4c)}{2(-1+c)(6+c)(-3+2c)} \,\, ( T T D T \pm
T T \overline{D} T) + \frac{3(33+23c)}{(3+c)(-12+5c)} \,\, \alpha \, ( 
D T  \pm \overline{D} T) W
\nonu \\
&& -\frac{3(72-99c+55c^2)}{4(-1+c)(6+c)(-3+2c)} \,\,
( [D, \overline{D}] T 
D T  \pm  \overline{D} T [D, \overline{D}] T)  \nonu \\
&& -\frac{3(-36-87c+11c^2)}{2(-1+c)(6+c)(-3+2c)} \,\, 
(\pa D T  \mp \pa \overline{D} T) T \nonu \\
&& \left. + \frac{3(90-129c+25c^2)}{4(-1+c)(6+c)(-3+2c)} \,\, \pa T 
(D T \mp  \overline{D} T)   +\frac{7}{2} (D V \pm \overline{D} V) \right](w)
\nonu \\
&& +
\frac{1}{(z-w)} \left[  \frac{(6+c)}{2(-12+5c)} \,\,
\alpha \, \pa^2 (D W  \mp \overline{D} W)
  + 
\frac{(-15+c)(-3+c)c}{4(-1+c)(6+c)(-3+2c)} \,\,
\pa^3 ( D T
  \pm \overline{D} T) \right. \nonu \\
&& +\frac{21}{(-12+5c)} \,\, \alpha  \, T (D W \pm
\pa \overline{D} W) +\frac{3}{(3+c)} \,\, \alpha ( D T  
\mp \overline{D} T ) [D, \overline{D}] W  \nonu \\
&& +\frac{9(3+4c)}{(3+c)(-12+5c)} \,\, \alpha \, ( D T 
  \pm \overline{D} T) \pa W  +\frac{21}{(-12+5c)} \,\, \alpha \, ( \pa 
D T  \pm \pa \overline{D} T) W  \nonu \\
&& -\frac{3(72-63c+19c^2)}{4(-1+c)(6+c)(-3+2c)} \,\, (\pa 
D T  \pm \pa \overline{D} T) [D, \overline{D}] T 
\nonu \\
&& -\frac{27(3+4c)}{2(-1+c)(6+c)(-3+2c)} \,\, (\pa D T  
\pm \pa \overline{D} T) T T  +\frac{3(3+7c)}{4(-1+c)(-3+2c)} 
\,\, (\pa D T  \mp \pa  \overline{D} T) \pa T \nonu \\
&& -
\frac{3(9+5c^2)}{2(-1+c)(6+c)(-3+2c)} \,\, (\pa^2 D T  \mp
\pa^2 \overline{D} T) T - \frac{3}{(3+c)} \,\, \alpha \, [D, \overline{D}] T
(D W \mp  \overline{D} W) \nonu \\
&& \nonu \\
&& -\frac{81c}{4(6+c)(-3+2c)}  \,\, \pa [D, \overline{D}] T ( D T
\pm  \overline{D} T) + \frac{3(33+2c)}{(3+c)(-12+5c)} \,\, \alpha \, (\pa T D W \pm
\pa T \overline{D} W) \nonu \\
&& -\frac{27(3+4c)}{(-1+c)(6+c)(-3+2c)} \,\, \pa T T (D T \pm
 \overline{D} T) \nonu \\
&& \left. +\frac{27(-6+c)}{4(6+c)(-3+2c)}  \,\, \pa^2 T (D T \mp
 \overline{D} T) + \frac{3}{2} \,\, \pa (D V \pm \overline{D}
 V)\right](w) +\cdots
\nonu \\
&& \nonu \\
&& \longrightarrow 
\frac{1}{(z-w)^4} \,\, \frac{15}{2} \left[ DT  \pm \overline{D} T \right](w)
\nonu \\
&& +\frac{1}
{(z-w)^3} \left[ \alpha \, (D W  \mp \overline{D} W)  + 
\frac{5}{2} \,\, \pa ( D T
  \pm \overline{D} T) -\frac{3}{c} \,\, T (  D T \mp  \overline{D} T) 
\right](w) \nonu \\
&& + \frac{1}{(z-w)^2} \left[ \frac{2}{5}
\alpha \, \pa (D W \mp \overline{D} W)
  + 
\frac{5}{8} \,\, \pa^2  D T
   \pm \frac{5}{8} \,\, \pa^2 
\overline{D} T +
\frac{39}{5c} \,\, \alpha \, T ( D W \pm  \overline{D} W )  
\right.
\nonu \\
&& -\frac{63}{c^2} \,\, ( T T D T \pm
T T \overline{D} T) + \frac{69}{5c} \,\, \alpha \, ( 
D T  \pm \overline{D} T) W
-\frac{165}{8c} \,\,
( [D, \overline{D}] T 
D T  \pm  \overline{D} T [D, \overline{D}] T)  \nonu \\
&& \left. -\frac{33}{4c} \,\, 
(\pa D T  \mp \pa \overline{D} T) T  + \frac{75}{8c} \,\, \pa T 
(D T \mp  \overline{D} T)   +\frac{7}{2} (D V \pm \overline{D} V) \right](w)
\nonu \\
&& +
\frac{1}{(z-w)} \left[  \frac{1}{10} \,\,
\alpha \, \pa^2 (D W  \mp \overline{D} W)
  + 
\frac{1}{8} \,\,
\pa^3 ( D T
  \pm \overline{D} T) +\frac{21}{5c} \,\, \alpha  \, T (D W \pm
\pa \overline{D} W) \right. \nonu \\
&& +\frac{3}{c} \,\, \alpha ( D T  
\mp \overline{D} T ) [D, \overline{D}] W   +\frac{36}{5c} \,\, \alpha \, ( D T 
  \pm \overline{D} T) \pa W  +\frac{21}{5c} \,\, \alpha \, ( \pa 
D T  \pm \pa \overline{D} T) W  \nonu \\
&& -\frac{57}{8c} \,\, (\pa 
D T  \pm \pa \overline{D} T) [D, \overline{D}] T 
-\frac{27}{c^2} \,\, (\pa D T  
\pm \pa \overline{D} T) T T  +\frac{21}{8c} 
\,\, (\pa D T  \mp \pa  \overline{D} T) \pa T \nonu \\
&& -
\frac{15}{4c} \,\, (\pa^2 D T  \mp
\pa^2 \overline{D} T) T - \frac{3}{c} \,\, \alpha \, [D, \overline{D}] T
(D W \mp  \overline{D} W) 
 -\frac{81}{8c}  \,\, \pa [D, \overline{D}] T ( D T
\pm  \overline{D} T) \nonu \\
&& + \frac{6}{5c} \,\, \alpha \, (\pa T D W \pm
\pa T \overline{D} W) -\frac{54}{c^2} \,\, \pa T T (D T \pm
 \overline{D} T) \nonu \\
&& \left. +\frac{27}{8c}  \,\, \pa^2 T (D T \mp
 \overline{D} T) + \frac{3}{2} \,\, \pa (D V \pm \overline{D}
 V)\right](w) +\cdots.
\label{otherope}
\eea
The spin-$\frac{7}{2}$ field and its descendant field appear in the
right hand side. 
At the classical limit, the symmetry between $D$ and $\overline{D}$
which act on the fields is manifest at both sides.
At the quantum level, the two derivative terms do not have common factors. 

Also one has, from (\ref{otherope}), 
\bea
&&   (-1)\frac{1}{2} [D, \overline{D}] W(z) \; (D W  \pm \overline{D} W)(w) =
\frac{1}{(z-w)^4} \,\, \frac{15}{2} \left[ DT  \pm \overline{D} T \right](w)
\nonu \\
&& +\frac{1}
{(z-w)^3} \left[ -\alpha \, (D W  \mp \overline{D} W)  + 
\frac{(-6+5c)}{(-1+c)} \,\, \pa ( D T
  \pm \overline{D} T) +\frac{3}{(-1+c)} \,\, T (  D T \mp  \overline{D} T) 
\right](w) \nonu \\
&& + \frac{1}{(z-w)^2} \left[ -\frac{3(-9-6c+c^2)}{(3+c)(-12+5c)}
\alpha \, \pa (D W \mp \overline{D} W)
  + 
\frac{3c(81-30c+5c^2)}{4(-1+c)(6+c)(-3+2c)} \,\, \pa^2  D T
   \right. \nonu \\
&& \pm \frac{3(72+63c-5c^2+10c^3)}{8(-1+c)(6+c)(-3+2c)} \,\, \pa^2 
\overline{D} T +
\frac{3(57+13c)}{(3+c)(-12+5c)} \,\, \alpha \, T ( D W \pm  \overline{D} W )  
\nonu \\
&& -\frac{63(3+4c)}{2(-1+c)(6+c)(-3+2c)} \,\, ( T T D T \pm
T T \overline{D} T) + \frac{3(33+23c)}{(3+c)(-12+5c)} \,\, \alpha \, ( 
D T  \pm \overline{D} T) W
\nonu \\
&& -\frac{3(72-99c+55c^2)}{4(-1+c)(6+c)(-3+2c)} \,\,
( [D, \overline{D}] T 
D T  \pm  \overline{D} T [D, \overline{D}] T)  \nonu \\
&& -\frac{21(-15+c)c}{2(-1+c)(6+c)(-3+2c)} \,\, 
(\pa D T  \mp \pa \overline{D} T) T \nonu \\
&& \left. + \frac{9(6-31c+11c^2)}{4(-1+c)(6+c)(-3+2c)} \,\, \pa T 
(D T \mp  \overline{D} T)   +\frac{7}{2} (D V \pm \overline{D} V) \right](w)
\nonu \\
&& +
\frac{1}{(z-w)} \left[  -\frac{(-15+c) c}{(3+c)(-12+5c)} \,\,
\alpha \, \pa^2 (D W  \mp \overline{D} W)
  + 
\frac{c(9-3c+c^2)}{(-1+c)(6+c)(-3+2c)} \,\,
\pa^3 ( D T
  \pm \overline{D} T) \right. \nonu \\
&& +\frac{18(6+c)}{(3+c)(-12+5c)} \,\, \alpha  \, T (D W \pm
\pa \overline{D} W) -\frac{3}{(3+c)} \,\, \alpha ( D T  
\mp \overline{D} T ) [D, \overline{D}] W  \nonu \\
&& +\frac{3(24+11c)}{(3+c)(-12+5c)} \,\, \alpha \, ( D T 
  \pm \overline{D} T) \pa W  +\frac{12(3+4c)}{(3+c)(-12+5c)} \,\, \alpha \, ( \pa 
D T  \pm \pa \overline{D} T) W  \nonu \\
&& -\frac{27c}{(6+c)(-3+2c)} \,\, (\pa 
D T  \pm \pa \overline{D} T) [D, \overline{D}] T 
\nonu \\
&& -\frac{18(3+4c)}{(-1+c)(6+c)(-3+2c)} \,\, (\pa D T  
\pm \pa \overline{D} T) T T  +\frac{3(3+c)}{(-1+c)(-3+2c)} 
\,\, (\pa D T  \mp \pa  \overline{D} T) \pa T \nonu \\
&& -
\frac{6(9-3c+c^2)}{(-1+c)(6+c)(-3+2c)} \,\, (\pa^2 D T  \mp
\pa^2 \overline{D} T) T + \frac{3}{(3+c)} \,\, \alpha \, [D, \overline{D}] T
(D W \mp  \overline{D} W) \nonu \\
&& \nonu \\
&& -\frac{3(18-18c+7c^2)}{(-1+c)(6+c)(-3+2c)}  \,\, \pa [D, \overline{D}] T ( D T
\pm  \overline{D} T) + \frac{3(24+11c)}{(3+c)(-12+5c)} \,\, \alpha \, (\pa T D W \pm
\pa T \overline{D} W) \nonu \\
&& -\frac{36(3+4c)}{(-1+c)(6+c)(-3+2c)} \,\, \pa T T (D T \pm
 \overline{D} T) \nonu \\
&& \left. +\frac{3c(-12+5c)}{(-1+c)(6+c)(-3+2c)}  \,\, \pa^2 T (D T \mp
 \overline{D} T) + 2 \,\, \pa (D V \pm \overline{D}
 V)\right](w) +\cdots
\nonu \\
&& \nonu \\
&& \longrightarrow 
\frac{1}{(z-w)^4} \,\, \frac{15}{2} \left[ DT  \pm \overline{D} T \right](w)
\nonu \\
&& +\frac{1}
{(z-w)^3} \left[ -\alpha \, (D W  \mp \overline{D} W)  + 
5 \,\, \pa ( D T
  \pm \overline{D} T) +\frac{3}{c} \,\, T (  D T \mp  \overline{D} T) 
\right](w) \nonu \\
&& + \frac{1}{(z-w)^2} \left[ -\frac{3}{5}
\alpha \, \pa (D W \mp \overline{D} W)
  + 
\frac{15}{8} \,\, \pa^2  D T
   \pm \frac{15}{8} \,\, \pa^2 
\overline{D} T +
\frac{39}{5c} \,\, \alpha \, T ( D W \pm  \overline{D} W )  
\right.
\nonu \\
&& -\frac{63}{c^2} \,\, ( T T D T \pm
T T \overline{D} T) + \frac{69}{5c} \,\, \alpha \, ( 
D T  \pm \overline{D} T) W
-\frac{165}{8c} \,\,
( [D, \overline{D}] T 
D T  \pm  \overline{D} T [D, \overline{D}] T)  \nonu \\
&& \left. -\frac{21}{4c} \,\, 
(\pa D T  \mp \pa \overline{D} T) T  + \frac{99}{8c} \,\, \pa T 
(D T \mp  \overline{D} T)   +\frac{7}{2} (D V \pm \overline{D} V) \right](w)
\nonu \\
&& +
\frac{1}{(z-w)} \left[  -\frac{1}{5} \,\,
\alpha \, \pa^2 (D W  \mp \overline{D} W)
  + 
\frac{1}{2} \,\,
\pa^3 ( D T
  \pm \overline{D} T) +\frac{18}{5c} \,\, \alpha  \, T (D W \pm
\pa \overline{D} W) \right. \nonu \\
&& -\frac{3}{c} \,\, \alpha ( D T  
\mp \overline{D} T ) [D, \overline{D}] W   +\frac{33}{5c} \,\, \alpha \, ( D T 
  \pm \overline{D} T) \pa W  +\frac{48}{5c} \,\, \alpha \, ( \pa 
D T  \pm \pa \overline{D} T) W  \nonu \\
&& -\frac{27}{2c} \,\, (\pa 
D T  \pm \pa \overline{D} T) [D, \overline{D}] T 
-\frac{36}{c^2} \,\, (\pa D T  
\pm \pa \overline{D} T) T T  +\frac{3}{2c} 
\,\, (\pa D T  \mp \pa  \overline{D} T) \pa T \nonu \\
&& -
\frac{3}{c} \,\, (\pa^2 D T  \mp
\pa^2 \overline{D} T) T + \frac{3}{c} \,\, \alpha \, [D, \overline{D}] T
(D W \mp  \overline{D} W) 
 -\frac{21}{2c}  \,\, \pa [D, \overline{D}] T ( D T
\pm  \overline{D} T) \nonu \\
&& + \frac{33}{5c} \,\, \alpha \, (\pa T D W \pm
\pa T \overline{D} W) -\frac{72}{c^2} \,\, \pa T T (D T \pm
 \overline{D} T) \nonu \\
&& \left. +\frac{15}{2c}  \,\, \pa^2 T (D T \mp
 \overline{D} T) + 2 \,\, \pa (D V \pm \overline{D}
 V)\right](w) +\cdots.
\label{otherope1}
\eea

Finally, 
the spin-$3$ and spin-$3$ operator product expansion has the following form
\bea
&& (-1)\frac{1}{2} [D, \overline{D}] W(z) \;
 (-1)\frac{1}{2} [D, \overline{D}] W(w) 
=
 \frac{1}{(z-w)^6} \,\, \frac{5}{2} c \nonu \\
&& +\frac{1}{(z-w)^4} \left[  3 \alpha
  \, W
  -\frac{3(-4+5c)}{2(-1+c)} \,\, [D, \overline{D}] T
  -\frac{9}{2(-1+c)} \,\, T^2 \right](w) 
\nonu \\
&& + \frac{1}{(z-w)^3} \left[ \frac{3}{2} \alpha \, \pa W
  -\frac{3(-4+5c)}{4(-1+c)} \,\, \pa [D, \overline{D}] T -
\frac{9}{2(-1+c)} \pa T T \right](w) \nonu \\
&& +
\frac{1}{(z-w)^2} \left[ \frac{3(-12-19c+3c^2)}{4(3+c)(-12+5c)} \alpha
  \, \pa^2 W
  -\frac{9c(30-11c+2c^2)}{8(-1+c)(6+c)(-3+2c)} \,\, 
\pa^2 [D, \overline{D}] T \right. \nonu \\
&& -\frac{18(6+c)}{(3+c)(-12+5c)} \,\, \alpha \, T [D, \overline{D}] W +
\frac{18(3+4c)}{(-1+c)(6+c)(-3+2c)} \,\, ( T T [D, \overline{D}] T - \pa [D,
 \overline{D}] T T)
\nonu \\
&& -\frac{72(3+4c)}{(-1+c)(6+c)(-3+2c)} \,\, T \overline{D} T D T
+ \frac{6(24+11c)}{(3+c)(-12+5c)} \,\,
\alpha ( \overline{D} T D W - D T \overline{D} W ) \nonu \\
&& + \frac{42(-6+c)c}{(-1+c)(6+c)(-3+2c)} \,\, 
(\pa \overline{D} T D T+  \pa D T \overline{D} T ) -
\frac{12(3+4c)}{(3+c)(-12+5c)} \,\, \alpha  \, [D, \overline{D}] T W  \nonu \\
&& +\frac{3(18-27c+16c^2)}{2(-1+c)(6+c)(-3+2c)} \,\, [D, \overline{D}] T 
[D, \overline{D}] T 
\nonu \\
&& -\frac{9(6-19c+6c^2)}{4(-1+c)(6+c)(-3+2c)} \,\, \pa T
\pa T + \frac{9(6-43c+2c^2)}{4(-1+c)(6+c)(-3+2c)} \,\, \pa^2 T T  \nonu \\
&& \left.-\frac{6(3+4c)}{(-1+c)(6+c)(-3+2c)} \,\,  
\pa^3 T -2 [D, \overline{D}] V  \right](w)
\nonu \\
&& + \frac{1}{(z-w)} \left[  \frac{(-15+c)c}{2(3+c)(-12+5c)} \,\, \alpha \,
  \pa^3 W
  -\frac{c(30-11c+2c^2)}{4(-1+c)(6+c)(-3+2c)} \,\,
\pa^3 [D, \overline{D}] T  \right. \nonu \\
&& -\frac{9(6+c)}{(3+c)(-12+5c)} \,\, \alpha \, \pa (T  
[D, \overline{D}] W) +\frac{3(24+11c)}{(3+c)(-12+5c)} \,\, \alpha \, 
(\overline{D} T \pa D W +  \pa \overline{D} T D W) \nonu \\
&& 
 -\frac{36(3+4c)}{(-1+c)(6+c)(-3+2c)} \,\, ( \pa \overline{D} T 
T D T -  \pa D T T \overline{D} T) \nonu \\
&& +\frac{3(18-18c+7c^2)}{(-1+c)(6+c)(-3+2c)} \,\, \pa^2 \overline{D} T
D T -\frac{6(3+4c)}{(3+c)(-12+5c)} \,\, \alpha \, \pa ([D, \overline{D}] T  W )
 \nonu \\
&&   +\frac{3(18-27c+16c^2)}{2(-1+c)(6+c)(-3+2c)} \,\,
\pa [D, \overline{D}] T [D, \overline{D}] T 
-\frac{3(24+11c)}{(3+c)(-12+5c)} \,\, \alpha \, \pa (D T  \overline{D} W )
\nonu \\
&& +\frac{9(3+4c)}{(-1+c)(6+c)(-3+2c)} \,\,(\pa [D, \overline{D}] T T T -
\pa [D, \overline{D}] T \pa T)
\nonu  
 \\
&& +\frac{36(3+4c)}{(-1+c)(6+c)(-3+2c)} \,\, \pa D T T \overline{D} T
+ \frac{3(18-18c+7c^2)}{(-1+c)(6+c)(-3+2c)} \,\, \pa^2 D T \overline{D} 
T  \nonu \\
&&  
+ \frac{18(3+4c)}{(-1+c)(6+c)(-3+2c)} \,\, \pa T T [D, \overline{D}] 
T 
-\frac{9(6-c+2c^2)}{2(-1+c)(6+c)(-3+2c)} \,\, \pa^2 T \pa T
\nonu \\
&&  \left. +\frac{3(9-3c+c^2)}{(-1+c)(6+c)(-3+2c)} \,\, \pa^3 T T  
-\frac{3(3+4c)}{2(-1+c)(6+c)(-3+2c)} \,\, \pa^4 T  
 - \pa [D, \overline{D}] V
\right](w) +\cdots
\nonu \\
&& \nonu \\
&& \longrightarrow
 \frac{1}{(z-w)^6} \,\, \frac{5}{2} c +\frac{1}{(z-w)^4} \left[  3 \alpha
  \, W
  -\frac{15}{2} \,\, [D, \overline{D}] T
  -\frac{9}{2c} \,\, T^2 \right](w) 
\nonu \\
&& + \frac{1}{(z-w)^3} \left[ \frac{3}{2} \alpha \, \pa W
  -\frac{15}{4} \,\, \pa [D, \overline{D}] T -
\frac{9}{2c} \pa T T \right](w) \nonu \\
&& +
\frac{1}{(z-w)^2} \left[ \frac{9}{20} \alpha
  \, \pa^2 W
  -\frac{9}{8} \,\, 
\pa^2 [D, \overline{D}] T -\frac{18}{5c} \,\, \alpha \, T [D, \overline{D}] W +
\frac{36}{c^2} \,\, ( T T [D, \overline{D}] T - \pa [D,
 \overline{D}] T T) \right.
\nonu \\
&& -\frac{144}{c^2} \,\, T \overline{D} T D T
+ \frac{66}{5c} \,\,
\alpha ( \overline{D} T D W - D T \overline{D} W ) + \frac{21}{c} \,\, 
(\pa \overline{D} T D T+  \pa D T \overline{D} T ) -
\frac{48}{5c} \,\, \alpha  \, [D, \overline{D}] T W  \nonu \\
&& \left. +\frac{12}{c} \,\, [D, \overline{D}] T 
[D, \overline{D}] T 
-\frac{27}{4c} \,\, \pa T
\pa T + \frac{9}{4c} \,\, \pa^2 T T  -2 [D, \overline{D}] V  \right](w)
\nonu \\
&& + \frac{1}{(z-w)} \left[  \frac{1}{10} \,\, \alpha \,
  \pa^3 W
  -\frac{1}{4} \,\,
\pa^3 [D, \overline{D}] T   -\frac{9}{5c} \,\, \alpha \, \pa (T  
[D, \overline{D}] W) +\frac{33}{5c} \,\, \alpha \, 
(\overline{D} T \pa D W +  \pa \overline{D} T D W) \right. \nonu \\
&& 
 -\frac{72}{c^2} \,\, ( \pa \overline{D} T 
T D T -  \pa D T T \overline{D} T) +\frac{21}{2c} \,\, \pa^2 \overline{D} T
D T -\frac{24}{5c} \,\, \alpha \, \pa ([D, \overline{D}] T  W )
 \nonu \\
&&   +\frac{12}{c} \,\,
\pa [D, \overline{D}] T [D, \overline{D}] T 
-\frac{33}{5c} \,\, \alpha \, \pa (D T  \overline{D} W )
 +\frac{18}{c^2} \,\,(\pa [D, \overline{D}] T T T -
\pa [D, \overline{D}] T \pa T)
\nonu 
\\
&& +\frac{72}{c^2} \,\, \pa D T T \overline{D} T
+ \frac{21}{2c} \,\, \pa^2 D T \overline{D} 
T  + \frac{36}{c^2} \,\, \pa T T [D, \overline{D}] 
T 
-\frac{9}{2c} \,\, \pa^2 T \pa T
\nonu \\
&&  \left. +\frac{3}{2c} \,\, \pa^3 T T  
 - \pa [D, \overline{D}] V
\right](w) +\cdots.
\label{otherope2}
\eea
The spin-$4$ field and its descendant field appear in the right hand side.
In the $c \rightarrow \infty$ limit, the term 
$\pa^3 T(w)$ vanishes. 

Of course, the classical limits (\ref{opeww})-(\ref{otherope2}) where 
the central charge is given by $(3.1)$ and the self-coupling
constant is given by $(3.2)$ 
can be combined into $(3.3)$. For the fixed $N$ with large 
$c$ limit, the corresponding self-coupling constant behaves as before.
In next Appendices, we list some relevant operator product expansions
which will be necessary to transform the operator product expansions
in the primary basis of this Appendix $C$
into those in the non-primary basis of [13].

\section{The remaining operator product expansions in the $AdS_3$ bulk
  theory:
Nonprimary basis}

We have seen that 
the above relations $(4.2)$, $(4.3)$ and $(4.4)$
 provide the quantum and classical operator product
expansions in the notation of [13].
In this Appendix $D$, we would like to complete the remaining ones.  
 
The operator product expansion between the spin-$\frac{5}{2}$ and
spin-$\frac{5}{2}$(this does not change eventhough one introduces the
modified stress energy tensor $(2.21)$), 
from (\ref{dwdw}), (\ref{ope351}), and (\ref{dwdw1}),
is given by
\bea
&& (D W + \overline{D} W) (z) \,\, ( D W + \overline{D} W )(w)  =  
\frac{1}{(z-w)^5} 2 c \nonu \\
&& +\frac{1}{(z-w)^3} \left[ 4 \alpha \, W -
\frac{(-3+5c)}{(-1+c)} \,\, \widetilde{[D,
  \overline{D}]T} +\frac{9}{c} \,\, T^2 
\right](w) \nonu \\
&& +\frac{1}{(z-w)^2} \left[ 2 \alpha \, \pa W -
  \frac{(-3+5c)}{2(-1+c)} \,\, \pa \widetilde{[D,
  \overline{D}] T} +\frac{9}{c} \,\, \pa T T 
\right](w) \nonu \\
&& +\frac{1}{(z-w)} \left[\frac{3(-6+c)(-1+c)}{(3+c)(-12+5c)} \,\, \alpha
  \, \pa^2 W -
  \frac{3c(9-3c+c^2)}{2(-1+c)(6+c)(-3+2c)} 
\,\, \pa^2 \widetilde{[D, \overline{D}] T}  \right.  \nonu \\
&& + \frac{6(-15+c)}{(3+c)(-12+5c)} \,\, \alpha \, T \widetilde{[D, \overline{D}] W}
- \frac{9(-12+5c)}{(-1+c)(6+c)(-3+2c)} \,\, T T \widetilde{[D, \overline{D}] T}  
\nonu \\
&& +\frac{18(51+5c)}{c(3+c)(-12+5c)} \,\, \alpha T T W
+\frac{27(-15+c)}{2(-1+c)c(6+c)(-3+2c)} \,\, T^4
\nonu \\
&& -\frac{36(3+4c)}{(-1+c)(6+c)(-3+2c)} \,\, T \overline{D} T D T
+\frac{12(3+4c)}{(3+c)(-12+5c)} \,\, \alpha ( \overline{D} T D W -  D T \overline{D} 
W)
\nonu \\
&& + \frac{3(18-51c+5c^2)}{(-1+c)(6+c)(-3+2c)} \,\, ( \pa 
\overline{D} T D T  + \pa D T 
\overline{D} T ) -\frac{54(-1+c)}{(3+c)(-12+5c)} \,\, \alpha \, 
\widetilde{[D, \overline{D}] T} W \nonu \\
&& + \frac{27c}{2(6+c)(-3+2c)} \,\, \widetilde{[D, \overline{D}] T} \widetilde{ [D, 
\overline{D}] T} -\frac{9(3+4c)}{(-1+c)(6+c)(-3+2c)} 
\,\, \pa \widetilde{[D, \overline{D}] T}  T \nonu \\
&& -\frac{9(-6+c)}{2(6+c)(-3+2c)} \,\, \pa T \pa T -
\frac{3(63-57c+c^2)}{(-1+c)(6+c)(-3+2c)} \,\, \pa^2 T T \nonu \\
&&  -\frac{12}{(3+c)} \,\, \alpha D T D W -\frac{6}{(-1+c)} \,\, \pa D
T D T  \nonu \\
&&    +\frac{12}{(3+c)} \,\, \alpha \overline{D} T \overline{D}
  W 
- \frac{6}{(-1+c)} \,\, \pa \overline{D}
T \overline{D} T \nonu \\
&& \left. +\frac{54(3+4c)}{(-1+c)c(6+c)(-3+2c)} \,\, \pa T T T  - [D, \overline{D}] V
\right](w)  +\cdots
\nonu \\
&& \nonu \\
&& \longrightarrow 
\frac{1}{(z-w)^5} 2 c +\frac{1}{(z-w)^3} \left[ 4 \alpha \, W -
5 \widetilde{[D,
  \overline{D}]T} +\frac{9}{c} \,\, T^2 
\right](w) \nonu \\
&& +\frac{1}{(z-w)^2} \left[ 2 \alpha \, \pa W -
  \frac{5}{2} \,\, \pa \widetilde{[D,
  \overline{D}] T} +\frac{9}{2c} \,\, \pa T^2  
\right](w) \nonu \\
&& +\frac{1}{(z-w)} \left[\frac{3}{5} \,\, \alpha
  \, \pa^2 W -
  \frac{3}{4} 
\,\, \pa^2 \widetilde{[D, \overline{D}] T}   + 
\frac{6}{5c} \,\, \alpha \, T \widetilde{[D, \overline{D}] W}
- \frac{45}{2c^2} \,\, T T \widetilde{[D, \overline{D}] T}  
\right. \nonu \\
&& +\frac{18}{c^2} \,\, \alpha T T W
+\frac{27}{4c^3} \,\, T^4
-\frac{72}{c^2} \,\, T \overline{D} T D T
+\frac{48}{5c} \,\, \alpha ( \overline{D} T D W -  D T \overline{D} 
W)
\nonu \\
&& + \frac{15}{2c} \,\, ( \pa 
\overline{D} T D T  + \pa D T 
\overline{D} T ) -\frac{54}{5c} \,\, \alpha \, 
\widetilde{[D, \overline{D}] T} W + \frac{27}{4c} \,\, 
\widetilde{[D, \overline{D}] T} \widetilde{ [D, 
\overline{D}] T} \nonu \\
&&  -\frac{9}{4c} \,\, \pa T \pa T -
\frac{3}{2c} \,\, \pa^2 T T    - [D, \overline{D}] V
-\frac{12}{c} \,\, \alpha D T D W -\frac{6}{c} \,\, \pa D
T D T  
 \nonu \\
&& \left.   +\frac{12}{c} \,\, \alpha \overline{D} T \overline{D}
  W 
- \frac{6}{c} \,\, \pa \overline{D}
T \overline{D} T
\right](w)  +\cdots.
\label{rem1}
\eea
Except the terms $T \overline{D} T D T(w)$, $\overline{D} T D W(w)$,
$D T \overline{D} W$, $D T D W)(w)$, and $\overline{D} T \overline{D}
W(w)$, one can easily check that all the remaining fields appear in
the bulk theory computations [13].
The relative coefficients in the spin-$2$ field and its descendant
fields can be obtained from the formula [48],
$\frac{\Gamma[2+1]}{\Gamma[2]} \times \frac{\Gamma[4]}{\Gamma[4+1]}=
\frac{1}{2}$ and $\frac{1}{2!}\frac{\Gamma[2+2]}{\Gamma[2]} \times
\frac{\Gamma[4]}
{\Gamma[4+2]}=\frac{3}{20}$. That is, the coefficient $2$ in the $\pa
W(w)$
is obtained from $4 \times \frac{1}{2}=2$ and the coefficient
$\frac{3}{5}$ in $\pa^2 W(w)$ term
is coming from $4 \times \frac{3}{20}=\frac{3}{5}$.

From the equations (\ref{dwdw}), (\ref{ope351}), and (\ref{dwdw1}),
other type of spin-$\frac{5}{2}$ and spin-$\frac{5}{2}$ operator
product expansion is 
\bea
&& ( D W + \overline{D} W) (z) \,\,(D W  -\overline{D} W) (w)  =  
-\frac{1}{(z-w)^4} \,\, 6 T(w) -\frac{1}{(z-w)^3} \,\, 3 \pa T(w)  
\nonu \\
&& -\frac{1}{(z-w)^2} \left[ \frac{2(6+c)}{(-12+5c)} \,\, \alpha \, \widetilde{[D,
  \overline{D}] W} 
+ \frac{2c(9-3c+c^2)}{(-1+c)(6+c)(-3+2c)} \,\,
\pa^2 T \right. \nonu \\
&&  -\frac{3(-3+c)(-6+13c)}{2(-1+c)(6+c)(-3+2c)} \,\, \pa \widetilde{[D, 
\overline{D}] T}  + \frac{12}{c} \,\, \alpha \, T W  \nonu \\
&&  -
\frac{6(9+5c^2)}{(-1+c)(6+c)(-3+2c)} \,\, T \widetilde{[D, 
\overline{D}] T}  +\frac{18(9-3c+c^2)}{(-1+c)c(6+c)(-3+2c)}
\,\, T^3 \nonu \\
&& \left. - \frac{6(-3+c)(-6+13c)}{(-1+c)(6+c)(-3+2c)} 
\,\, \overline{D} T D T  + \frac{9(-3+c)(-6+13c)}{(-1+c)c(6+c)(-3+2c)} 
\,\, \pa T T  +6 V 
\right](w) \nonu \\
&& -\frac{1}{(z-w)} \left[ \frac{(6+c)}{(-12+5c)} 
\,\, \alpha \, \pa \widetilde{[D, \overline{D}] W} +
  \frac{c(9-3c+c^2)}{2(-1+c)(6+c)(-3+2c)} 
\,\, \pa^3 T \right. \nonu \\
&&  + \frac{6}{c} \,\, \alpha \, \pa (T  W)
  -\frac{3(-3+c)(-6+13c)}{(-1+c)(6+c)(-3+2c)}\,\,
(\pa \overline{D} T D T -\pa D T \overline{D} T)  \nonu \\
&&  -\frac{3(9+5c^2)}{(-1+c)(6+c)(-3+2c)} \,\,
( \pa [D, \overline{D}] T  T +\pa T [D, \overline{D}] T )    \nonu \\
&&   +\frac{27(9-3c+c^2)}
{(-1+c)c(6+c)(-3+2c)} \,\, \pa T T T   + 3 \pa V \nonu \\
&&  -\frac{12}{(3+c)} \,\, \alpha D T D W -\frac{6}{(-1+c)} \,\, \pa D
T D T  \nonu \\
&&  \left.  -\frac{12}{(3+c)} \,\, \alpha \overline{D} T \overline{D}
  W 
+ \frac{6}{(-1+c)} \,\, \pa \overline{D}
T \overline{D} T \right](w)  +\cdots
\nonu \\
&& \nonu \\
&& \longrightarrow
-\frac{1}{(z-w)^4} \,\, 6 T(w) -\frac{1}{(z-w)^3} \,\, 3 \pa T(w)  
\nonu \\
&& -\frac{1}{(z-w)^2} \left[ \frac{2}{5} \,\, \alpha \, \widetilde{[D,
  \overline{D}] W} 
+
\pa^2 T  + \frac{12}{c} \,\, \alpha \, T W    -
\frac{15}{c} \,\, T \widetilde{[D, 
\overline{D}] T}  +\frac{9}{c^2}
\,\, T^3 \right. \nonu \\
&& \left. - \frac{39}{c} 
\,\, \overline{D} T D T    +6 V 
\right](w) \nonu \\
&& -\frac{1}{(z-w)} \left[ \frac{1}{5} 
\,\, \alpha \, \pa \widetilde{[D, \overline{D}] W} +
  \frac{1}{4} 
\,\, \pa^3 T + \frac{6}{c} \,\, \alpha \, \pa (T  W)
  -\frac{39}{2c}\,\,
\pa (\overline{D} T D T) \right.  \nonu \\
&&   -\frac{15}{2c} \,\,
\pa (\widetilde{T [D, \overline{D}] T} ) +\frac{9}
{2c^2} \,\, \pa T^3   + 3 \pa V 
-\frac{12}{c} \,\, \alpha D T D W -\frac{6}{c} \,\, \pa D
T D T  
\nonu \\
&& \left.   -\frac{12}{c} \,\, \alpha \overline{D} T \overline{D}
  W 
+ \frac{6}{c} \,\, \pa \overline{D}
T \overline{D} T
\right](w)  +\cdots.
\label{rem2}
\eea
The relative coefficients $\frac{12}{c}$ and $\frac{6}{c}$ in the
$\alpha$-dependent terms provide the correct values in [13]. 
Also $\frac{15}{c}$ in the second order singular term and $\frac{15}{2c}$ in
the first order singular term have the correct behaviors. 
Furthermore, the coefficient $\frac{9}{c^2}$ in $T^3(w)$ term and the
coefficient $\frac{27}{2c^2}$ in its derivative term behave correctly
as in [13]. 
There are also the terms $\overline{D} T D T(w)$, $DT D W$, and
$\overline{D} T \overline{D} W$.
The relative coefficients $6, 3, 1$,
and $\frac{1}{4}$ in the spin-$1$ field and its descendant
fields can be obtained from the coefficient formula in [48,1]. 

From the equation 
(\ref{otherope1}),
the next operator product expansion between the spin-$3$ and the
spin-$\frac{5}{2}$ has the following form
\bea
&&   \widetilde{(-1)\frac{1}{2} [D, \overline{D}] W}(z) \; (D W  \pm \overline{D} W)(w) =
\frac{1}{(z-w)^4} \,\, \frac{3(-12+5c)}{2c} \left[ DT  \pm \overline{D} T \right](w)
\nonu \\&& +\frac{1}
{(z-w)^3} \left[ -\frac{(6+c)}{c} \,\, \alpha \, (D W  \mp \overline{D} W)  + 
\frac{(-12+5c)}{(-1+c)} \,\, \pa ( D T
  \pm \overline{D} T) \right. \nonu \\
&& \left. - \frac{3(-12+5c)}{(-1+c)c} \,\, T (  D T \mp  \overline{D} T) 
\right](w) \nonu \\
&& + \frac{1}{(z-w)^2} \left[ -\frac{3(-108+39c-2c^2+c^3)}{c(3+c)(-12+5c)}
\alpha \, \pa (D W \mp \overline{D} W)
  + 
\frac{3(-3+c)^2(-12+5c)}{4(-1+c)(6+c)(-3+2c)} \,\, \pa^2  D T
   \right. \nonu \\
&& \pm \frac{3(-756+882c-303c^2-29c^3+10c^4)}{8(-1+c)c(6+c)(-3+2c)} \,\, \pa^2 
\overline{D} T +
\frac{9(-36+21c+c^2)}{c(3+c)(-12+5c)} \,\, \alpha \, T ( D W \pm  \overline{D} W )  
\nonu \\
&& +\frac{45(18-18c+7c^2)}{2(-1+c)c(6+c)(-3+2c)} \,\, ( T T D T \pm
T T \overline{D} T) + \frac{3(108-75c+23c^2)}{c(3+c)(-12+5c)} \,\, \alpha \, ( 
D T  \pm \overline{D} T) W
\nonu \\
&& -\frac{3(-12+5c)(-15+11c)}{4(-1+c)(6+c)(-3+2c)} \,\,
( \widetilde{[D, \overline{D}] T} 
D T  \pm  \overline{D} T \widetilde{[D, \overline{D}] T})  \nonu \\
&& -\frac{9(-3+c)(-12+5c)}{2(-1+c)(6+c)(-3+2c)} \,\, 
\pa D T  T \pm \frac{9(252-270c+83c^2+5c^3)}{2(-1+c)c(6+c)(-3+2c)} \,\,
\pa \overline{D} T T  \nonu \\
&&  - \frac{9(-144+234c-137c^2+5c^3)}{4(-1+c)c(6+c)(-3+2c)} \,\, \pa T 
D T \pm  \frac{9(-12+5c)(-30+19c+c^2)}{4(-1+c)c(6+c)(-3+2c)} \,\,  \pa T
\overline{D} T   \nonu \\
&& \left. +\frac{(-12+7c)}{2c} (D V \pm \overline{D} V) \right](w)
\nonu \\
&& +
\frac{1}{(z-w)} \left[  -\frac{(-15+c) c}{(3+c)(-12+5c)} \,\,
\alpha \, \pa^2 (D W  \mp \overline{D} W)
  + 
\frac{(-3+c)(9-3c+c^2)}{(-1+c)(6+c)(-3+2c)} \,\,
\pa^3  D T \right. \nonu \\
&& -\frac{6}{c} \,\, T ( D V \mp \overline{D} V) 
-\frac{54(6+c)}{c(3+c)(-12+5c)} \,\, \alpha T T ( D W \mp \overline{D}
W)  
\nonu \\
&& -\frac{27(-15+c)}{(-1+c)c(6+c)(-3+2c)} T T T (D T \mp \overline{D} T) 
+ \frac{36(-15+c)}{c(3+c)(-12+5c)} \,\, \alpha T ( D T \mp \overline{D} T ) W
\nonu \\
&& +\frac{81}{(6+c)(-3+2c)} T (\widetilde{[D, \overline{D}] T} D T \mp
\overline{D} T \widetilde{[D, \overline{D}] T} )
-\frac{6}{c} \,\, W ( D W  \mp \overline{D} W)
\nonu \\
&& +\frac{6(54-6c+c^2)}{c(3+c)(-12+5c)} \,\, \alpha  \, T ( \pa D W \pm
\pa \overline{D} W) -\frac{3}{(3+c)} \,\, \alpha ( D T  
\mp \overline{D} T ) \widetilde{[D, \overline{D}] W}  \nonu \\
&& +\frac{3(24+11c)}{(3+c)(-12+5c)} \,\, \alpha \, ( D T 
  \pm \overline{D} T) \pa W  +
\frac{12(72-27c+4c^2)}{c(3+c)(-12+5c)} \,\, \alpha \, ( \pa 
D T  \pm \pa \overline{D} T) W  \nonu \\
&& -\frac{27c}{(6+c)(-3+2c)} \,\, (\pa 
D T  \pm \pa \overline{D} T) \widetilde{[D, \overline{D}] T}  \pm
\frac{(-3+c)^2(-9+2c^2)}{2(-1+c)c(6+c)(-3+2c)} \,\,
\pa^3  \overline{D} T
\nonu \\
&&  \pm \frac{54(-9+2c^2)}{(-1+c)c(6+c)(-3+2c)} \,\,  \pa
\overline{D} T T T
  -\frac{9(-3+c)}{(-1+c)(-3+2c)} 
\,\,  ( \pa D T  \mp  \pa \overline{D} T) \pa T \nonu \\
&& \pm
\frac{9(-3+c)(3+c)(-9+2c)}{2(-1+c)c(6+c)(-3+2c)} \,\, 
\pa^2 \overline{D} T T + \frac{3}{(3+c)} \,\, \alpha \, \widetilde{[D, \overline{D}] T}
(D W \mp  \overline{D} W) \nonu \\
&& \nonu \\
&& -\frac{3(18-18c+7c^2)}{(-1+c)(6+c)(-3+2c)}  \,\, \pa \widetilde{[D,
  \overline{D}] T} 
( D T
\pm  \overline{D} T)  \nonu \\
&& + \frac{9(36-81c+17c^2)}{(-1+c)c(6+c)(-3+2c)} \,\, \pa T T D T  
 +\frac{54(-9+2c)}{(-1+c)(6+c)(-3+2c)} \,\,  \pa
D T T T
\nonu \\
&& -\frac{9(36-9c+c^2)}{(-1+c)(6+c)(-3+2c)} \pa^2 D T T
+ \frac{3(-36+c)(-6+c)}{c(3+c)(-12+5c)} \,\, \alpha \, (\pa T D W \pm
\pa T \overline{D} W)
\nonu \\
&&  - \frac{3(-6+c)(9-3c+c^2)}{(-1+c)c(6+c)(-3+2c)}  \,\, \pa^2 T
  ( D T \mp
 \overline{D} T) 
\pm \frac{9(-18-27c+17c^2)}{(-1+c)c(6+c)(-3+2c)} \,\, \pa T T \overline{D} T 
\nonu \\
&& \left. + 2 \,\, \pa (D V \pm \overline{D}
 V)\right](w) +\cdots
\nonu \\
&& \nonu \\
&& \longrightarrow
\frac{1}{(z-w)^4} \,\, \frac{15}{2} \left[ DT  \pm \overline{D} T \right](w)
\nonu \\&& +\frac{1}
{(z-w)^3} \left[ -\alpha \, (D W  \mp \overline{D} W)  + 
5 \,\, \pa ( D T
  \pm \overline{D} T)  - \frac{15}{c}  T (  D T \mp  \overline{D} T) 
\right](w) \nonu \\
&& + \frac{1}{(z-w)^2} \left[ -\frac{3}{5}
\alpha \, \pa (D W \mp \overline{D} W)
  + 
\frac{15}{8} \,\,  \pa^2  ( D T
    \pm  \,\,  
\overline{D} T) +
\frac{9}{5c} \,\, \alpha \, T ( D W \pm  \overline{D} W )  
\right. \nonu \\
&& +\frac{315}{4c^2} \,\,  T T ( D T \pm
 \overline{D} T) + \frac{69}{5c} \,\, \alpha \, ( 
D T  \pm \overline{D} T) W
 -\frac{33}{8c} \,\,
 \widetilde{[D, \overline{D}] T} 
( D T  \pm  \overline{D} T )  \nonu \\
&& \left. -\frac{45}{4c} \,\, 
(\pa D T   \mp  \,\,
\pa \overline{D} T) T    - \frac{45}{8c} \,\, \pa T 
( D T \mp   \,\,  
\overline{D} T)   +\frac{7}{2} (D V \pm \overline{D} V) \right](w)
\nonu \\
&& +
\frac{1}{(z-w)} \left[  -\frac{1}{5} \,\,
\alpha \, \pa^2 (D W  \mp \overline{D} W)
  + 
\frac{1}{2} \,\,
\pa^3  ( D T \pm \overline{D} T ) \right.  \nonu \\
&& -\frac{6}{c} \,\, T ( D V \mp \overline{D} V) 
-\frac{54}{5c^2} \,\, \alpha T T ( D W \mp \overline{D}
W)  
\nonu \\
&& -\frac{27}{2c^3} T T T (D T \mp \overline{D} T) 
+ \frac{36}{5c^2} \,\, \alpha T ( D T \mp \overline{D} T ) W
 +\frac{81}{2c^2} T \widetilde{[D, \overline{D}] T} (D T \mp
\overline{D} T  )
\nonu \\
&& -\frac{6}{c} \,\, W ( D W  \mp \overline{D} W)
+\frac{6}{5c} \,\, \alpha  \, T ( \pa D W \pm
\pa \overline{D} W) -\frac{3}{c} \,\, \alpha ( D T  
\mp \overline{D} T ) \widetilde{[D, \overline{D}] W}  \nonu \\
&& +\frac{33}{5c} \,\, \alpha \, ( D T 
  \pm \overline{D} T) \pa W  +
\frac{48}{5c} \,\, \alpha \, ( \pa 
D T  \pm \pa \overline{D} T) W   -\frac{27}{2c} \,\, (\pa 
D T  \pm \pa \overline{D} T) \widetilde{[D, \overline{D}] T} 
\nonu \\
&&   \frac{54}{c^2} ( \pa D T  
\pm \pa \overline{D} T) T T
  -\frac{9}{2c} 
\,\,  ( \pa D T  \mp  \pa \overline{D} T) \pa T  
 + \frac{3}{c} \,\, \alpha \, \widetilde{[D, \overline{D}] T}
(D W \mp  \overline{D} W) \nonu \\
&& -\frac{21}{2c}  \,\, \pa \widetilde{[D,
  \overline{D}] T} 
( D T
\pm  \overline{D} T)   + \frac{153}{2c^2} \,\, \pa T T ( D T
\pm \overline{D} T)  
 -\frac{9}{2c} (\pa^2 D T \mp \pa^2 \overline{D} T ) T
\nonu \\
&& + \frac{3}{5c} \,\, \alpha \, \pa T (D W \pm
\overline{D} W)
 - \frac{3}{2c}  \,\, \pa^2 T
  ( D T \mp
 \overline{D} T) 
\nonu \\
&& \left. + 2 \,\, \pa (D V \pm \overline{D}
 V)\right](w) +\cdots.
\label{rem3}
\eea
The relative coefficients $\frac{15}{2}, 5, \frac{15}{8}$,
and $\frac{1}{2}$ in the spin-$\frac{3}{2}$ field and its descendant
fields can be obtained from the coefficient formula in [48,1]. 
The relative coefficients $1, \frac{3}{5}$,
and $\frac{1}{5}$ in the spin-$\frac{5}{2}$ field and its descendant
fields can be obtained similarly.

Finally, one has the following operator product expansion between the
spin-$3$ and the spin-$3$  
\bea
&& \widetilde{(-1)\frac{1}{2} [D, \overline{D}] W}(z) \;
 \widetilde{(-1)\frac{1}{2} [D, \overline{D}] W}(w) 
=
 \frac{1}{(z-w)^6} \,\, \frac{1}{2} (-12+5c) \nonu \\
&& +\frac{1}{(z-w)^4} \left[  \frac{3(-8+c)}{c} \alpha
  \, W
  -\frac{3(-12+5c)}{2(-1+c)} \,\, \widetilde{[D, \overline{D}] T}
  +\frac{9(-4+5c)}{2(-1+c)c} \,\, T^2 \right](w) 
\nonu \\
&& + \frac{1}{(z-w)^3} \left[ \frac{3(-8+c)}{2c} \alpha \, \pa W
  -\frac{3(-12+5c)}{4(-1+c)} \,\, \pa \widetilde{[D, \overline{D}] T} +
\frac{9(-4+5c)}{2(-1+c)c} \pa T T \right](w) \nonu \\
&& +
\frac{1}{(z-w)^2} \left[ \frac{3(-12-19c+3c^2)}{4(3+c)(-12+5c)} \alpha
  \, \pa^2 W
  -\frac{9c(30-11c+2c^2)}{8(-1+c)(6+c)(-3+2c)} \,\, 
\pa^2 \widetilde{[D, \overline{D}] T} \right. \nonu \\
&& -\frac{36}{c} \,\, T V -\frac{18(24+11c)}{c(3+c)(-12+5c)} \,\,
\alpha T \widetilde{[D, \overline{D}] W}
+\frac{72(6+c)(6+5c)}{c^2(3+c)(-12+5c)} 
\,\, \alpha T T W
\nonu \\
&&  -
\frac{18(-54-3c+8c^2)}{(-1+c)c(6+c)(-3+2c)} \,\,  T T \widetilde{[D, \overline{D}] T}
\nonu \\
&& + \frac{108(-3+c)(3+4c)}{(-1+c)c^2(6+c)(-3+2c)} \,\, T^4
-\frac{12}{c} \,\, W^2  +\frac{63(-6+c)}{(-1+c)(6+c)(-3+2c)} \,\, \pa
\widetilde{ [D,
\overline{D}] T} T  \nonu \\
&& +\frac{252(-6+c)}{(-1+c)(6+c)(-3+2c)} \,\, T \overline{D} T D T
+ \frac{6(24+11c)}{(3+c)(-12+5c)} \,\,
\alpha ( \overline{D} T D W - D T \overline{D} W ) \nonu \\
&& + \frac{42(-6+c)c}{(-1+c)(6+c)(-3+2c)} \,\, 
(\pa \overline{D} T D T+  \pa D T \overline{D} T ) -
\frac{12(3+4c)}{(3+c)(-12+5c)} \,\, \alpha  \, \widetilde{[D, \overline{D}] T} W  \nonu \\
&& +\frac{3(18-27c+16c^2)}{2(-1+c)(6+c)(-3+2c)} \,\, \widetilde{[D, \overline{D}] T} 
\widetilde{[D, \overline{D}] T} 
+ \frac{54(48+c)}{(-1+c)c(6+c)(-3+2c)} \,\, \pa T T T \nonu \\
&& -\frac{9(-360+630c-353 c^2+6c^3)}{4(-1+c)c(6+c)(-3+2c)} \,\, \pa^2 T
 T - \frac{9(6-19c+6c^2)}{4(-1+c)(6+c)(-3+2c)} \,\, \pa T \pa T  \nonu \\
&& \left. + \frac{6(3+4c)}{(-1+c)(6+c)(-3+2c)} \,\,  
\pa^3 T -2 [D, \overline{D}] V  \right](w)
\nonu \\
&& + \frac{1}{(z-w)} \left[  \frac{(-15+c)c}{2(3+c)(-12+5c)} \,\, \alpha \,
  \pa^3 W
  -\frac{c(30-11c+2c^2)}{4(-1+c)(6+c)(-3+2c)} \,\,
\pa^3 \widetilde{[D, \overline{D}] T}  \right. \nonu \\
&& -\frac{9(24+11c)}{c(3+c)(-12+5c)} \,\, \alpha \, \pa (T  
[D, \overline{D}] W) +\frac{3(24+11c)}{(3+c)(-12+5c)} \,\, \alpha \, 
(\overline{D} T \pa D W +  \pa \overline{D} T D W) \nonu \\
&& + \frac{126(-6+c)}{(-1+c)(6+c)(-3+2c)} \,\, ( \pa   \overline{D} T
 T D T -\pa D T T \overline{D} T)
 \nonu \\
&& +\frac{36(6+c)(6+5c)}{c^2(3+c)(-12+5c)} \,\, \alpha  T T \pa W
+\frac{72(6+c)(6+5c)}{c^2(3+c)(-12+5c)} \,\, \alpha  \pa T T W 
-\frac{18}{c} \,\, \pa (T  V)  -\frac{12}{c} \pa W W
\nonu \\
&& +\frac{21(-6+c)(-3+c)}{(-1+c)(6+c)(-3+2c)} \,\, \pa^2 \overline{D} T
D T -\frac{6(3+4c)}{(3+c)(-12+5c)} \,\, \alpha \, \pa ( \widetilde{[D, \overline{D}] T}  W )
 \nonu \\
&&   +\frac{3(18-27c+16c^2)}{2(-1+c)(6+c)(-3+2c)} \,\,
\pa \widetilde{[D, \overline{D}] T} \widetilde{[D, \overline{D}] T} 
-\frac{3(24+11c)}{(3+c)(-12+5c)} \,\, \alpha \, \pa (D T  \overline{D} W )
\nonu \\
&& -\frac{9(-54-3c+8c^2)}{(-1+c)c(6+c)(-3+2c)} \,\, \pa \widetilde{[D, \overline{D}] T} T T +
\frac{63(-6+c)}{2(-1+c)(6+c)(-3+2c)} \,\, \pa \widetilde{[D, \overline{D}] T} \pa
T \nonu
 \\
&& 
+\frac{126(-6+c)}{(-1+c)(6+c)(-3+2c)} \,\, \pa T \overline{D} T D T 
+ \frac{21(-6+c)(-3+c)}{(-1+c)(6+c)(-3+2c)} \,\, \pa^2 D T \overline{D} 
T  \nonu \\
&&  
- \frac{18(-54-3c+8c^2)}{(-1+c)c(6+c)(-3+2c)} \,\, \pa T T \widetilde{[D, \overline{D}] 
T} 
-\frac{9(30-11c+2c^2)}{2(-1+c)(6+c)(-3+2c)} \,\, \pa^2 T \pa T
\nonu \\
&& +\frac{216(-3+c)(3+4c)}{(-1+c)c^2(6+c)(-3+2c)} \,\, \pa T T T T
+\frac{27(48+c)}{(-1+c)c(6+c)(-3+2c)} \,\, 
\pa T \pa T T
\nonu \\
&&  \left. -\frac{3(-144+258c-123c^2+2c^3)}{2(-1+c)c(6+c)(-3+2c)} \,\, \pa^3 T T  
+\frac{3(3+4c)}{2(-1+c)(6+c)(-3+2c)} \,\, \pa^4 T  
 - \pa [D, \overline{D}] V
\right](w) \nonu \\
&& +\cdots
\nonu \\
&& \nonu \\
&& \longrightarrow
 \frac{1}{(z-w)^6} \,\, \frac{5c}{2}  \nonu \\
&& +\frac{1}{(z-w)^4} \left[ 3 \alpha
  \, W
  -\frac{15}{2} \,\, \widetilde{[D, \overline{D}] T}
  +\frac{45}{2c} \,\, T^2 \right](w) 
\nonu \\
&& + \frac{1}{(z-w)^3} \frac{1}{2} \pa \left[ 3 \alpha \,  W
  -\frac{15}{2} \,\,  \widetilde{[D, \overline{D}] T} +
\frac{45}{2c}  T^2  \right](w) \nonu \\
&& +
\frac{1}{(z-w)^2} \left[ \frac{9}{20} \alpha
  \, \pa^2 W
  -\frac{9}{8} \,\, 
\pa^2 \widetilde{[D, \overline{D}] T}  -\frac{36}{c} \,\, T V 
+\frac{72}{c^2} 
\,\, \alpha T T W  -
\frac{72}{c^2} \,\,  T T \widetilde{[D, \overline{D}] T}
\right. \nonu \\
&& + \frac{216}{c^3} \,\, T^4
-\frac{12}{c} \,\, W^2  +\frac{126}{c^2} \,\, T \overline{D} T D T
+ \frac{66}{5c} \,\,
\alpha ( \overline{D} T D W - D T \overline{D} W ) \nonu \\
&& + \frac{21}{c} \,\, 
(\pa \overline{D} T D T+  \pa D T \overline{D} T ) -
\frac{48}{5c} \,\, \alpha  \, \widetilde{[D, \overline{D}] T} W  
+\frac{12}{c} \,\, \widetilde{[D, \overline{D}] T} 
\widetilde{[D, \overline{D}] T} 
 \nonu \\
&& \left. -\frac{27}{8c} \,\, \pa^2 T^2
   -2 [D, \overline{D}] V  \right](w)
\nonu \\
&& + \frac{1}{(z-w)} \left[  \frac{1}{10} \,\, \alpha \,
  \pa^3 W
  -\frac{1}{4} \,\,
\pa^3 \widetilde{[D, \overline{D}] T}   +\frac{33}{5c} \,\, \alpha \, 
\pa (\overline{D} T  D W - D T \overline{D} W) 
\right. \nonu \\
&&   +\frac{36}{c^2} \,\, \alpha  \pa (T T  W)
-\frac{18}{c} \,\, \pa (T  V)  -\frac{6}{c} \pa (W W)
\nonu \\
&& +\frac{21}{2c} \,\, \pa (\overline{D} T 
D T +D T \overline{D} T) -\frac{24}{5c} \,\, \alpha \, \pa ( \widetilde{[D, \overline{D}] T}  W )
+\frac{6}{c} \,\,
\pa (\widetilde{[D, \overline{D}] T} \widetilde{[D, \overline{D}] T}) 
\nonu \\
&& -\frac{36}{c^2} \,\, \pa (T T \widetilde{[D, \overline{D}] T})  
+\frac{63}{c^2} \,\, \pa  ( T  \overline{D} T D T)  
-\frac{3}{4c} \,\, \pa^3 T^2
\nonu \\
&& \left. +\frac{108}{c^3} \,\, \pa T^4    
 - \pa [D, \overline{D}] V
\right](w)  +\cdots.
\label{rem4}
\eea
Due to the presence of $T^2(w)$ in the classical limit, 
its descendant fields, $\pa T^2(w)$, $\pa^2 T^2(w)$, and 
$\pa^3 T^2 (w)$ appear and the field $T
\overline{D} T D T(w)$ occurs. Also the nonlinear term $T^4(w)$ and
its descendant field $\pa T^4 (w)$ appear in the right hand side.
The relative coefficients of various spin-$4$ fields to its
descendants, $1$ and $\frac{1}{2}$, can be obtained from the conformal
symmetry.
Also note that the relative coefficients, $1, \frac{1}{2},
\frac{3}{20}$,
and $\frac{1}{30}$ in the nonlinear $T^2$ term and its descendant
fields are the same as the ones in linear spin-$2$ terms [16]. 
All of these do not appear in the bulk theory.
Other remaining nonlinear terms appear in [13].  

We also present the other operator product expansions
\bea
&&  (D W \pm \overline{D} W)(z) \; W(w)  =  
-\frac{1}{(z-w)^3} 3 \left[ D T \mp \overline{D} T \right](w)
\nonu \\
&&  +\frac{1}{(z-w)^2} \left[
\alpha \, (D W \pm \overline{D} W) -  \frac{(-3+2c)}{(-1+c)} 
\,\, 
\pa (D T \mp \overline{D} T) -\frac{3}{(-1+c)} \,\, T ( D T \pm  \overline{D} T ) \right](w) 
\nonu \\
&& + \frac{1}{(z-w)} \left[ -( D V \mp \overline{D} V) +
\frac{3(-6+c)(-1+c)}{(3+c)(-12+5c)} \,\,
  \alpha \,
\pa (D W  \pm \overline{D} W) \right. \nonu \\
&&  \pm \frac{3(27-18c+3c^2+2c^3)}{4(-1+c)(6+c)(-3+2c)} \,\, 
\pa^2 \overline{D} T
+ \frac{6(-15+c)}{(3+c)(-12+5c)} \,\, \alpha \, T ( D W \mp  \overline{D} W) 
\nonu \\
&&  -\frac{9(-15+c)}{2(-1+c)(6+c)(-3+2c)} \,\, T T (D T \mp  \overline{D}
  T) -\frac{54(-1+c)}{(3+c)(-12+5c)} \,\, \alpha \, (D T  \mp \overline{D}
  T) W  \nonu \\
&&  + \frac{27c}{2(6+c)(-3+2c)} \,\, ( \widetilde{[D, \overline{D}]T} D T  \mp
\overline{D} T \widetilde{[D, \overline{D}]
  T})
  -\frac{3c(9-3c+c^2)}{2(-1+c)(6+c)(-3+2c)} \,\, \pa^2 D T \nonu \\
&&  -\frac{3(-18+24c+c^2)}{(-1+c)(6+c)(-3+2c)} \,\, \pa D T T 
  \mp \frac{3(9-3c+c^2)}{(-1+c)(6+c)(-3+2c)} \,\,\pa \overline{D} T T \nonu \\
&& \left. -\frac{9(12+c)}{2(6+c)(-3+2c)}  \,\, \pa T 
  D T \mp \frac{9(-6+c)}{2(6+c)(-3+2c)} \,\, \pa T \overline{D} T  \right] (w)
+  \cdots
\nonu \\
&& \nonu \\
&& \longrightarrow
-\frac{1}{(z-w)^3} 3 \left[ D T \mp \overline{D} T \right](w)
\nonu \\
&&  +\frac{1}{(z-w)^2} \left[
\alpha \, (D W \pm \overline{D} W) -2  
\,\, 
\pa (D T \mp \overline{D} T) -\frac{3}{c} \,\, T ( D T \pm  \overline{D} T ) \right](w) 
\nonu \\
&& + \frac{1}{(z-w)} \left[ -( D V \mp \overline{D} V) +
\frac{3}{5} \,\,
  \alpha \,
\pa (D W  \pm \overline{D} W) 
  \pm \frac{3}{4} \,\, 
\pa^2 \overline{D} T
+ \frac{6}{5c} \,\, \alpha \, T ( D W \mp  \overline{D} W) 
\right. \nonu \\
&&  -\frac{9}{4c^2} \,\, T T (D T \mp  \overline{D}
  T) -\frac{54}{5c} \,\, \alpha \, (D T  \mp \overline{D}
  T) W  
  + \frac{27}{4c} \,\, ( \widetilde{[D, \overline{D}]T} D T  \mp
\overline{D} T \widetilde{[D, \overline{D}]
  T})
  \nonu \\
&&  \left. -\frac{3}{2c} \,\, (\pa D T 
  \pm \pa \overline{D} T) T  -\frac{9}{4c}  \,\, \pa T (
  D T \pm  \overline{D} T ) - \frac{3}{4} \pa^2 D T \right] (w)
+  \cdots,
\nonu 
\eea
and 
\bea
&& (-1)\frac{1}{2} \widetilde{[D, \overline{D}] W}(z) \; W(w) =  
\frac{1}{(z-w)^4} 3 T(w) +\frac{1}{(z-w)^3} 3 \pa T(w) \nonu \\
&& +\frac{1}{(z-w)^2} \left[-\frac{3(-8+c)}{2(-12+5c)} \,\, \alpha \, \widetilde{[D, 
\overline{D} ] W} -\frac{9c(-12+5c)}{4(-1+c)(6+c)(-3+2c)} \,\, \pa \widetilde{[D,
\overline{D}] T} \right.
\nonu \\
&&  +\frac{12}{c} \,\, \alpha \, T W-
\frac{3(36-9c+8c^2)}{2(-1+c)(6+c)(-3+2c)} \,\, T \widetilde{[D, \overline{D}]T}  
\nonu \\
&&  -\frac{27(-12+5c)}{2(-1+c)c(6+c)(-3+2c)} \,\, T^3 -
\frac{9c(-12+5c)}{(-1+c)(6+c)(-3+2c)} \,\,
\overline{D} T D T \nonu \\
&& \left. -\frac{3(18-15c+2c^2+2c^3)}{2(-1+c)(6+c)(-3+2c)} \,\, \pa^2 T
+ \frac{27(-12+5c)}{2(-1+c)(6+c)(-3+2c)} \,\, \pa T T + 3 V \right](w) 
\nonu \\
&& +  
\frac{1}{(z-w)} \left[ -\frac{(-15+c)c}{2(3+c)(-12+5c)} 
\,\, \alpha \, \pa \widetilde{[D, 
\overline{D} ] W} + \frac{6}{(3+c)}
\,\, \alpha \, T \pa W  \right. \nonu \\
&&  +
\frac{6}{(3+c)} \,\, \alpha \, ( \overline{D} T D W +  D T \overline{D} W )-
\frac{6c(-12+5c)}{(-1+c)(6+c)(-3+2c)} \,\, ( \pa \overline{D} T D T - 
\pa D T \overline{D} T  ) \nonu \\
&& -\frac{6(9-3c+c^2)}{(-1+c)(6+c)(-3+2c)} \,\, \pa \widetilde{[D, \overline{D}] T}
T  +
\frac{12}{(3+c)} \,\, \alpha \, \pa T W \nonu \\
&&   
+\frac{3c(3+4c)}{(-1+c)(6+c)(-3+2c)} \,\, \pa T \widetilde{[D, \overline{D}] 
T} -\frac{27(-12+5c)}{(-1+c)c(6+c)(-3+2c)} \,\, \pa T T T  \nonu \\
&& \left. +\frac{(18-15c+2c^2+2c^3)}{2(-1+c)(6+c)(-3+2c)} \,\, \pa^3 T 
+ 2 \pa V \right](w) 
+\cdots
\nonu \\
&& \nonu \\
&& \longrightarrow 
\frac{1}{(z-w)^4} 3 T(w) +\frac{1}{(z-w)^3} 3 \pa T(w) \nonu \\
&& +\frac{1}{(z-w)^2} \left[-\frac{3}{10} \,\, \alpha \, \widetilde{[D, 
\overline{D} ] W}    +\frac{12}{c} \,\, \alpha \, T W-
\frac{6}{c} \,\, T \widetilde{[D, \overline{D}]T}  
 -
\frac{45}{2c} \,\,
\overline{D} T D T -\frac{3}{2} \,\, \pa^2 T
+ 3 V \right](w) 
\nonu \\
&& +  
\frac{1}{(z-w)} \left[ -\frac{1}{10} 
\,\, \alpha \, \pa \widetilde{[D, 
\overline{D} ] W} + \frac{6}{c}
\,\, \alpha \, T \pa W    +\frac{6}{c} \,\, \alpha \, ( \overline{D} T
D W +  
D T \overline{D} W ) \right. \nonu \\
&&  -
\frac{15}{c} \,\, ( \pa \overline{D} T D T - 
\pa D T \overline{D} T  )  -\frac{3}{c} \,\, \pa \widetilde{[D, \overline{D}] T}
T  -
\frac{12}{c} \,\, \alpha \, \pa T W    
 +\frac{6}{c} \,\, \pa T \widetilde{[D, \overline{D}] T} 
\nonu \\
&& \left. -\frac{18}{c^2} \,\, \pa T T T +\frac{1}{2} \pa^3 T  + 2
\pa V \right](w) 
\nonu \\
&& +\cdots,
\nonu
\eea
where the former can be obtained from $(4.3)$ while the latter can be
obtained from $(4.4)$.



\begin{thebibliography}{99}

\bibitem{BS} 
  P.~Bouwknegt and K.~Schoutens,
  ``W symmetry in conformal field theory,''  
Phys.\ Rept.\  {\bf 223}, 183 (1993)  [hep-th/9210010].  

\bibitem{Zamolodchikov} 
  A.~B.~Zamolodchikov,
  ``Infinite Additional Symmetries in Two-Dimensional 
Conformal Quantum Field Theory,''  
Theor.\ Math.\ Phys.\  {\bf 65}, 1205 (1985)  
[Teor.\ Mat.\ Fiz.\  {\bf 65}, 347 (1985)].  

\bibitem{FL1} 
  V.~A.~Fateev and S.~L.~Lukyanov,
  ``The Models of Two-Dimensional 
Conformal Quantum Field Theory with Z(n) Symmetry,''  
Int.\ J.\ Mod.\ Phys.\ A {\bf 3}, 507 (1988).  

\bibitem{FL2} 
  S.~L.~Lukyanov and V.~A.~Fateev,
  ``Conformally Invariant Models Of Two-dimensional Qft With Z(n)
  Symmetry,''  
Sov.\ Phys.\ JETP {\bf 67}, 447 (1988).  

\bibitem{FL3} 
  S.~L.~Lukyanov and V.~A.~Fateev,
  ``Physics reviews: Additional symmetries 
and exactly soluble models in two-dimensional 
conformal field theory,''  Chur, Switzerland: 
Harwood (1990) 117 p. (Soviet Scientific Reviews A, Physics: 15.2)

\bibitem{BBSS1} 
  F.~A.~Bais, P.~Bouwknegt, M.~Surridge and K.~Schoutens,
  ``Extensions of the Virasoro Algebra 
Constructed from Kac-Moody Algebras Using Higher Order Casimir
  Invariants,''  
Nucl.\ Phys.\ B {\bf 304}, 348 (1988).  

\bibitem{BBSS2} 
  F.~A.~Bais, P.~Bouwknegt, M.~Surridge and K.~Schoutens,
  ``Coset Construction for Extended Virasoro Algebras,''  
Nucl.\ Phys.\ B {\bf 304}, 371 (1988).  

\bibitem{GG} 
  M.~R.~Gaberdiel and R.~Gopakumar,
  ``An $AdS_3$ Dual for Minimal Model CFTs,''  
Phys.\ Rev.\ D {\bf 83}, 066007 (2011)  
[arXiv:1011.2986 [hep-th]].  

\bibitem{GG1} 
  M.~R.~Gaberdiel and R.~Gopakumar,
  ``Triality in Minimal Model Holography,''  JHEP {\bf 1207}, 127
  (2012)  
[arXiv:1205.2472 [hep-th]].  

\bibitem{GG2} 
  M.~R.~Gaberdiel and R.~Gopakumar,
  ``Minimal Model Holography,''  arXiv:1207.6697 [hep-th].  

\bibitem{CHR} 
  T.~Creutzig, Y.~Hikida and P.~B.~Ronne,
  ``Higher spin $AdS_3$ supergravity and its dual CFT,''  
JHEP {\bf 1202}, 109 (2012)  [arXiv:1111.2139 [hep-th]].  

\bibitem{CG} 
  C.~Candu and M.~R.~Gaberdiel,
  ``Supersymmetric holography on $AdS_3$,''  
arXiv:1203.1939 [hep-th].  

\bibitem{HP} 
  K.~Hanaki and C.~Peng,
  ``Symmetries of Holographic Super-Minimal Models,''  
arXiv:1203.5768 [hep-th].  

\bibitem{KSNPB} 
  Y.~Kazama and H.~Suzuki,
  ``New N=2 Superconformal 
Field Theories and Superstring Compactification,''  
Nucl.\ Phys.\ B {\bf 321}, 232 (1989).  

\bibitem{KSPLB} 
  Y.~Kazama and H.~Suzuki,
  ``Characterization of 
N=2 Superconformal Models Generated by 
Coset Space Method,''  
Phys.\ Lett.\ B {\bf 216}, 112 (1989).  

\bibitem{Ahn1206} 
  C.~Ahn,
  ``The Large N 't Hooft Limit of Kazama-Suzuki Model,''  JHEP {\bf 1208}, 047 (2012)  [arXiv:1206.0054 [hep-th]].  

\bibitem{BHMS} 
  S.~Banerjee, S.~Hellerman, J.~Maltz and S.~H.~Shenker,
  ``Light States in Chern-Simons Theory 
Coupled to Fundamental Matter,''  arXiv:1207.4195 [hep-th].  

\bibitem{GHKSS} 
  R.~Gopakumar, A.~Hashimoto, I.~R.~Klebanov, S.~Sachdev and K.~Schoutens,
  ``Strange Metals in One Spatial Dimension,''  Phys.\ Rev.\ D {\bf 86}, 066003 (2012)  [arXiv:1206.4719 [hep-th]].  

\bibitem{Vasiliev} 
  M.~A.~Vasiliev,
  ``Holography, Unfolding and Higher-Spin Theory,''  
arXiv:1203.5554 [hep-th].  

\bibitem{GHJ} 
  M.~R.~Gaberdiel, T.~Hartman and K.~Jin,
  ``Higher Spin Black Holes from CFT,''  JHEP {\bf 1204}, 103 (2012)  
[arXiv:1203.0015 [hep-th]].  

\bibitem{HGPR} 
  M.~Henneaux, G.~Lucena Gomez, J.~Park and S.~-J.~Rey,
  ``Super- W(infinity) Asymptotic Symmetry of 
Higher-Spin $AdS_3$ Supergravity,''  JHEP {\bf 1206}, 037 (2012)  
[arXiv:1203.5152 [hep-th]].  

\bibitem{Ahn2012} 
  C.~Ahn,
  ``The Primary Spin-4 Casimir Operators in the Holographic SO(N)
  Coset Minimal Models,''  
JHEP {\bf 1205}, 040 (2012)  [arXiv:1202.0074 [hep-th]].  

\bibitem{CY} 
  C.~-M.~Chang and X.~Yin,
  ``Correlators in $W_N$ Minimal Model Revisited,''  JHEP {\bf 1210}, 050 (2012)  [arXiv:1112.5459 [hep-th]].  

\bibitem{AKP} 
  M.~Ammon, P.~Kraus and E.~Perlmutter,
  ``Scalar fields and three-point functions in D=3 higher spin
  gravity,''  JHEP {\bf 1207}, 113 (2012)  [arXiv:1111.3926 [hep-th]].  

\bibitem{CGGR} 
  A.~Castro, R.~Gopakumar, M.~Gutperle and J.~Raeymaekers,
  ``Conical Defects in Higher Spin Theories,''  
JHEP {\bf 1202}, 096 (2012)  [arXiv:1111.3381 [hep-th]].  

\bibitem{Ahn2011} 
  C.~Ahn,
  ``The Coset Spin-4 Casimir Operator and Its Three-Point Functions
  with Scalars,''  
JHEP {\bf 1202}, 027 (2012)  [arXiv:1111.0091 [hep-th]].  

\bibitem{PR} 
  K.~Papadodimas and S.~Raju,
  ``Correlation Functions in Holographic Minimal Models,''  
Nucl.\ Phys.\ B {\bf 856}, 607 (2012)  [arXiv:1108.3077 [hep-th]].  

\bibitem{CFP} 
  A.~Campoleoni, S.~Fredenhagen and S.~Pfenninger,
  ``Asymptotic W-symmetries in 
three-dimensional higher-spin gauge theories,''  
JHEP {\bf 1109}, 113 (2011)  [arXiv:1107.0290 [hep-th]].  

\bibitem{GV} 
  M.~R.~Gaberdiel and C.~Vollenweider,
  ``Minimal Model Holography for SO(2N),''  
JHEP {\bf 1108}, 104 (2011)  [arXiv:1106.2634 [hep-th]].  

\bibitem{CY1} 
  C.~-M.~Chang and X.~Yin,
  ``Higher Spin Gravity with Matter in $AdS_3$ and Its CFT Dual,''  JHEP {\bf 1210}, 024 (2012)  [arXiv:1106.2580 [hep-th]].  

\bibitem{GGHR} 
  M.~R.~Gaberdiel, R.~Gopakumar, T.~Hartman and S.~Raju,
  ``Partition Functions of Holographic Minimal Models,''  
JHEP {\bf 1108}, 077 (2011)  [arXiv:1106.1897 [hep-th]].  

\bibitem{Ahn1106} 
  C.~Ahn,
  ``The Large N 't Hooft Limit of Coset Minimal Models,''  JHEP {\bf
  1110}, 125 (2011)  [arXiv:1106.0351 [hep-th]].  

\bibitem{CG1} 
  C.~Candu and M.~R.~Gaberdiel,
  ``Duality in N=2 minimal model holography,''  
arXiv:1207.6646 [hep-th].  

\bibitem{HS} 
  C.~M.~Hull and B.~J.~Spence,
  ``N=2 Current Algebra And Coset Models,''  
Phys.\ Lett.\ B {\bf 241}, 357 (1990).  

\bibitem{Romans} 
  L.~J.~Romans,
  ``The N=2 super W(3) algebra,''  
Nucl.\ Phys.\ B {\bf 369}, 403 (1992).  

\bibitem{Odake}
S. Odake, ``Superconformal Algebras and Their Extensions,''
Soryushiron Kenkyu(Kyoto) {\bf 78} (1989) 201 (in Japanese).

\bibitem{Ahn94} 
  C.~Ahn,
  ``Explicit construction of N=2 W(3) current in the N=2 coset 
SU(3) / SU(2) x U(1) model,''  
Phys.\ Lett.\ B {\bf 348}, 77 (1995)  [hep-th/9410170].  

\bibitem{BW} 
  R.~Blumenhagen and A.~Wisskirchen,
  ``Extension of the 
N=2 virasoro algebra by two primary fields of dimension 2 and 3,''  
Phys.\ Lett.\ B {\bf 343}, 168 (1995)  [hep-th/9408082].  

\bibitem{Wisskirchen} 
  A.~Wisskirchen,
  ``Construction of N=2 superW algebras,''  BONN-IB-94-21.  

\bibitem{KT} 
  S.~Krivonos and K.~Thielemans,
  ``A Mathematica package for computing 
N=2 superfield operator product expansions,''  
Class.\ Quant.\ Grav.\  {\bf 13}, 2899 (1996)  
[hep-th/9512029].  

\bibitem{Thielemans} 
  K.~Thielemans,
  ``A Mathematica package for 
computing operator product expansions,''  
Int.\ J.\ Mod.\ Phys.\ C {\bf 2}, 787 (1991).  

\bibitem{Bowcock} 
  P.~Bowcock,
  ``Quasi-primary Fields And 
Associativity Of Chiral Algebras,''  
Nucl.\ Phys.\ B {\bf 356}, 367 (1991).  

\bibitem{Ahn93} 
  C.~Ahn,
  ``Free superfield realization of N=2 quantum super W(3) algebra,''  
Mod.\ Phys.\ Lett.\ A {\bf 9}, 271 (1994)  [hep-th/9304038].  

\bibitem{BW1} 
  P.~Bowcock and G.~M.~T.~Watts,
  ``On the classification of quantum W algebras,''  
Nucl.\ Phys.\ B {\bf 379}, 63 (1992)  [hep-th/9111062].  

\bibitem{AKS} 
  C.~Ahn, S.~Krivonos and A.~S.~Sorin,
  ``The Full structure of quantum N=2 superW(3)**2 algebra,''  
Mod.\ Phys.\ Lett.\ A {\bf 10}, 1299 (1995)  [hep-th/9501100].  

\bibitem{AIKS} 
  C.~Ahn, E.~Ivanov, S.~Krivonos and A.~S.~Sorin,
  ``Quantum N=2 superW(2)(3) algebra in superspace,''  
Mod.\ Phys.\ Lett.\ A {\bf 11}, 1705 (1996)  [hep-th/9512214].  

\bibitem{IK} 
  E.~Ivanov and S.~Krivonos,
  ``Superfield realizations of N=2 superW(3),''  
Phys.\ Lett.\ B {\bf 291}, 63 (1992)  
[Erratum-ibid.\ B {\bf 301}, 454 (1993)]  [hep-th/9204023].  

\bibitem{LPRetal} 
  H.~Lu, C.~N.~Pope, L.~J.~Romans, X.~Shen and X.~J.~Wang,
  ``Polyakov construction of the N=2 superW(3) algebra,''  
Phys.\ Lett.\ B {\bf 264}, 91 (1991).  

\bibitem{GH} 
  M.~R.~Gaberdiel and T.~Hartman,
  ``Symmetries of Holographic Minimal Models,''  
JHEP {\bf 1105}, 031 (2011)  [arXiv:1101.2910 [hep-th]].  

\bibitem{Ito} 
  K.~Ito,
  ``Free field realization of N=2 superW(3) algebra,''  
Phys.\ Lett.\ B {\bf 304}, 271 (1993)  [hep-th/9302039]. 

\bibitem{Ozer} 
  H.~T.~Ozer,
  ``On the superfield realization of superCasimir WA(n) algebras,''
  Int.\ J.\ Mod.\ Phys.\ A {\bf 17}, 317 (2002)  [hep-th/0102203].  

\bibitem{BFKNRV}
  R.~Blumenhagen, M.~Flohr, A.~Kliem, W.~Nahm, A.~Recknagel and R.~Varnhagen,
  ``W algebras with two and three generators,''
  Nucl.\ Phys.\  B {\bf 361}, 255 (1991).

\bibitem{KW}
  H.~G.~Kausch and G.~M.~T.~Watts,
  ``A Study of W algebras using Jacobi identities,''
  Nucl.\ Phys.\  B {\bf 354}, 740 (1991).

\bibitem{Hornfeck1} 
  K.~Hornfeck,
  ``The Minimal supersymmetric extension of WA(n-1),''  
Phys.\ Lett.\ B {\bf 275}, 355 (1992).  

\bibitem{Hornfeck2} 
  K.~Hornfeck,
  ``W algebras with set of primary fields of dimensions (3, 4, 5) and
  (3, 4, 5, 6),''  Nucl.\ Phys.\ B {\bf 407}, 237 (1993)
  [hep-th/9212104].  



\end{thebibliography}
\end{document}